\documentclass[aps,prd,showpacs,eqsecnum,twocolumn,superscriptaddress,a4paper,nofootinbib]
{revtex4-2}
\usepackage{amsmath,amssymb,graphicx,color}
\usepackage{bm}
\usepackage{epsf}
\usepackage{mathrsfs}

\begin{document}

\title{Long-term evolution of a merger-remnant neutron star in general relativistic magnetohydrodynamics I: Effect of magnetic winding}

\author{Masaru Shibata} 
\affiliation{Max Planck Institute for
  Gravitational Physics (Albert Einstein Institute), Am M\"uhlenberg 1,
  Potsdam-Golm 14476, Germany}
\affiliation{Center for Gravitational Physics, Yukawa Institute for Theoretical
  Physics, Kyoto University, Kyoto, 606-8502, Japan}

\author{Sho Fujibayashi} 
\affiliation{Max Planck Institute for
  Gravitational Physics (Albert Einstein Institute), Am M\"uhlenberg 1,
  Potsdam-Golm 14476, Germany}


\author{Yuichiro Sekiguchi} \affiliation{Department of Physics, Toho
  University, Funabashi, Chiba 274-8510, Japan}
\affiliation{Center for Gravitational Physics, Yukawa Institute for Theoretical
  Physics, Kyoto University, Kyoto, 606-8502, Japan}

\date{\today}
\newcommand{\beq}{\begin{equation}}
\newcommand{\eeq}{\end{equation}}
\newcommand{\beqn}{\begin{eqnarray}}
\newcommand{\eeqn}{\end{eqnarray}}
\newcommand{\pa}{\partial}
\newcommand{\vp}{\varphi}
\newcommand{\varep}{\varepsilon}
\newcommand{\ep}{\epsilon}
\newcommand{\comp}{(M/R)_\infty}
\def\cL{\mathscr{L}}
\newcommand{\cE}{{\cal{E}}}
\newcommand{\cB}{{\cal{B}}}
\newcommand{\cb}{{\cal{H}}}
\newcommand{\cJ}{{\cal{J}}}
\newcommand{\cC}{{\cal{C}}}
\newcommand{\cI}{{\cal{I}}}
\begin{abstract}

Long-term ideal and resistive magnetohydrodynamics (MHD) simulations
in full general relativity are performed for a massive neutron star
formed as a remnant of binary neutron star mergers.  Neutrino
radiation transport effects are taken into account as in our previous
papers.  The simulation is performed in axial symmetry and without
considering dynamo effects as a first step.  In the ideal MHD, the
differential rotation of the remnant neutron star amplifies the
magnetic-field strength by the winding in the presence of a seed
poloidal field until the electromagnetic energy reaches $\sim 10\%$ of
the rotational kinetic energy, $E_{\rm kin}$, of the neutron star. The
timescale until the maximum electromagnetic energy is reached depends
on the initial magnetic-field strength and it is $\sim 1$\,s for the
case that the initial maximum magnetic-field strength is $\sim
10^{15}$\,G. After a significant amplification of the magnetic-field
strength by the winding, the magnetic braking enforces the initially
differentially rotating state approximately to a rigidly rotating
state. In the presence of the resistivity, the amplification is
continued only for the resistive timescale, and if the maximum
electromagnetic energy reached is smaller than $\sim 3$\% of $E_{\rm
  kin}$, the initial differential rotation state is approximately
preserved. In the present context, the post-merger mass ejection is
induced primarily by the neutrino irradiation/heating and the magnetic
winding effect plays only a minor role for the mass ejection.

\end{abstract}
\maketitle


\section{Introduction}\label{sec1}

Theoretical exploration for the evolution of the merger remnants of
binary neutron stars has become a hot topic in this
decade~\cite{SH19}, because such systems can shine as an
electromagnetic counterpart of gravitational waves emitted in their
inspiral stage and the signals bring us a variety of information for
the nature of the merger process and neutron stars that cannot be
obtained only by the gravitational-wave observation, as the first 
observation of the binary neutron star merger
shows~\cite{GW170817,GW170817a}. There are a variety of the
possibilities for the remnants of the binary neutron star
mergers~\cite{SH19} and for the corresponding electromagnetic
counterparts~\cite{MB2012,HP2015}.  Irrespective of the possibilities,
the key for the strong electromagnetic emission is the mass ejection
from the remnant. Thus, one important aspect of the theoretical study
is to clarify the properties of the ejecta such as total mass, typical
velocity, and typical elements.

The promising component of the remnant that induces the mass ejection
is a disk (or torus) surrounding the central compact object, which is
either a massive neutron star or a black hole. The reason for this is
that the disk is differentially rotating, approximately with the
Keplerian rotational profile, and is likely to be magnetized because
the disk matter stems from neutron stars. In this situation, the
magnetorotational instability (MRI)~\cite{BH98} is activated, and as a
result, a turbulence would be developed, enhancing the turbulence
viscosity.  The resulting viscous heating and angular momentum
transport inevitably enhance the activity of the disk, and eventually,
the mass ejection is
induced~\cite{MF2013,MF2014,Just2015,SM17,FTQFK18,Fujiba18,Janiuk19,FTQFK19,Miller19,Fujiba20,Fujiba2020,FFL20}.
In addition, the amplified magnetic field could eventually constitute
a aligned poloidal magnetic field along the rotational axis and this
could further enhance the mass ejection
efficiency~\cite{SM17,FTQFK19,FTQFK19,Miller19}. With these
considerations, numerical simulations have been extensively performed
in the last decade to clarify the details of the mass ejection from
the accretion disks.


Although many works have been done for exploring the evolution of the
accretion disks formed as remnants of the neutron-star merger, there
are only a few works for directly evolving the remnant massive neutron
star for a long timescale of $\agt 1$\,s~\cite{Fujiba18,Fujiba2020}
(more specifically, no long-term magnetohydrodynamics (MHD) simulation
has been done: but see \cite{moesta} for a short-term simulation).  A
unique property of the remnant massive neutron stars is found in their
angular velocity profile, $\Omega(\varpi)$, where $\varpi$ is the
cylindrical radius: Reflecting the nearly irrotational velocity field
of the pre-merger stage of two neutron stars, the inner region of the
remnant neutron stars should be differentially rotating with the
angular velocity that increases with the cylindrical radius
($d\Omega/d\varpi >0$; e.g., Ref.~\cite{STU2005}).  This situation is
in contrast to that for the disk rotating around a central object,
which approximately has a Keplerian profile, i.e., the angular
velocity decreases with the increase of the cylindrical radius, and
thus, unstable to the MRI.  For the inner region of the remnant
neutron stars (in particular for the central part of it) for which
$d\Omega/d\varpi > 0$, the MRI does not play any role.

However, irrespective of the sign of $d\Omega/d\varpi$, the winding of
the magnetic-field lines occurs in the presence of a seed poloidal
magnetic field together with differential rotation. In contrast to the
MRI for which the growth of the magnetic-field strength exponentially
proceeds, the (toroidal) magnetic-field strength increases only
linearly with time by the winding. Nevertheless, the growth rate is not
still negligible in the presence of the rapid rotation because the
amplification factor is approximately written as $\Omega t$ (e.g.,
Refs.~\cite{Shapiro2000,SLSS2006,Sun2019}) and $\Omega \sim 5000\,{\rm rad/s}$
(see Sec.~\ref{sec3}) for the merger remnants. Thus in $0.2$\,s, the
magnetic field can be amplified by $\sim 10^3$ times unless strong
resistive processes are present.  That is, even in the presence of a
seed poloidal magnetic field typically of $10^{13}$--$10^{14}$\,G
(which may be a conservative value for the merger
remnants~\cite{Kiuchi}), the (toroidal) magnetic field could be
amplified to the order of $10^{16}$--$10^{17}$\,G in $\sim 0.2$\,s
after the merger. The resulting magnetic pressure could be a
substantial fraction of the matter pressure, and hence, the MHD effect
could play an important role for the late time evolution of the
remnant neutron star.

Motivated by this consideration, we perform general relativistic MHD
(GRMHD) simulations for a massive neutron star formed as a remnant of
equal-mass binary neutron star mergers: As in our series of
papers~\cite{Fujiba17,Fujiba18,Fujiba2020}, we employ a numerical
result of a simulation for binary neutron star mergers as the initial
condition and perform an axisymmetric simulation. We solve Einstein's
equation, GRMHD equations, and neutrino-radiation hydrodynamics
equations altogether. Both the ideal and resistive MHD simulations are
performed. We here note that the origin of the resistivity of the
neutron stars is not certain. The purpose here is to simply show that
in the presence of a hypothetical resistive MHD process, the evolution
of the remnant massive neutron stars can be significantly modified.

In the assumption of axial symmetry, the poloidal magnetic field is
not amplified even when the toroidal-field strength is significantly
enhanced, due to the anti-dynamo property~\cite{Cowling}.
As a first step toward more detailed exploration of this topic, we do
not consider any dynamo effects for the growth of the magnetic-field
strength in this paper. Thus, we focus only on the growth of the
toroidal magnetic field and its effect on the evolution of the remnant
neutron star. The effect of the dynamo is planned to be studied in the
subsequent work.

The paper is organized as follows: In Sec.~\ref{sec2}, we summarize
the basic equations employed in the present numerical simulation
paying particular attention to the GRMHD equations and our method for
solving resistive MHD equations.  Section~\ref{sec3} presents
numerical results of the simulations paying particular attention to
the growth of the magnetic-field energy by the winding and the
resulting mass ejection.  Section~\ref{sec4} is devoted to a
summary. In Appendix~\ref{app1}, a number of the results of test
simulations for our newly developed resistive MHD implementation are
presented.  Throughout this paper, we use the geometrical units of
$c=1=G$ where $c$ and $G$ denote the speed of light and the
gravitational constant, respectively (but $c$ is often recovered to
clarify the units in the following sections). Latin and Greek indices
denote the spacetime and space components, respectively. In
Sec.~\ref{sec2}, we suppose to use Cartesian coordinates for the
spatial components whenever equations are written. 

\section{Basic equations and method for numerical computations}\label{sec2}

\subsection{Brief summary}\label{sec2-1}

In this work, we numerically solve Einstein's equation, ideal or
resistive MHD equations, evolution equations for the lepton fractions
including the electron fraction, and (approximate) neutrino-radiation
transfer equations. Except for the MHD parts (see the following
subsections for details), the numerical implementation is the same as
that in our latest viscous-hydrodynamics
work~\cite{Fujiba20,Fujiba2020}: Einstein's equation is solved using
the original version of the Baumgarte-Shapiro-Shibata-Nakamura
formalism~\cite{BSSN} together with the puncture
formulation~\cite{puncture}, Z4c constraint propagation
prescription~\cite{Z4c}, and 5th-order Kreiss-Oliger dissipation.  The
axial symmetry for the geometric variables is imposed using the
cartoon method~\cite{cartoon,cartoon2} with the 4th-order accuracy in
space.  For evolving the lepton fractions, we take into account
electron and positron capture, electron-positron pair annihilation,
nucleon-nucleon bremstrahlung, and plasmon
decay~\cite{Fujiba18,Fujiba2020}.  We employ a tabulated equation of
state based on the DD2 equation of state~\cite{DD2} for a relatively
high-density part and the Timmes (Helmhoitz) equation of state for the low-density
part~\cite{Timmes}.  In this tabulated equation of state,
thermodynamics quantities such as $\varep$, $P$, and $h$ are written
as functions of $\rho$, $Y_e$, and $T$ where $\varep$, $P$,
$h(=c^2+\varep+P/\rho)$, $\rho$, $Y_e$ and $T$ are the specific
internal energy, pressure, specific enthalpy, rest-mass density,
electron fraction, and matter temperature, respectively.  We choose
the lowest rest-mass density to be $0.1\,{\rm g/cm^3}$ in the table,
and the atmosphere density for $\rho_*:=\rho u^t \sqrt{-g}$ in the
hydrodynamics simulation is chosen to be $10^3\,{\rm g/cm^3}$ in the
central region of $r \leq 100$\,km and it is decreased down to
$1\,{\rm g/cm^3}$ with the dependence of $\propto r^{-3}$ in the outer
region (i.e., for the far region it is $1\,{\rm g/cm^3}$).  Here
$u^\mu$ and $g$ denote the four velocity of the fluid and the
determinant of the spacetime metric $g_{\mu\nu}$,
respectively. 
$\rho_*$ obeys the continuity equation in the form of
\beq
\pa_t \rho_* + \pa_i (\rho_* v^i)=0,\label{contin}
\eeq
where $v^i=u^i/u^t$ is the three velocity, $dx^i/dt$.
From Eq.~(\ref{contin}), the conserved rest mass is defined by
\beq
M_*:=\displaystyle \int \rho_* d^3x.
\eeq

\subsection{Maxwell's equations}\label{sec2-2}

First we write down the equations for the electromagnetic fields, which are 
derived from Maxwell's equations:
\beqn
&&\nabla_\mu F^{\mu\nu} = -4\pi j^\nu, \label{eq:maxcell1}\\
&&\nabla_\mu {}^\ast \hspace{-0.8mm}F^{\mu\nu} = 0.
\eeqn
Here $\nabla_\mu$ is the covariant derivative with the respect to
$g_{\mu\nu}$, $F^{\mu\nu}$ is the electromagnetic tensor, $j^\mu$ is
the current four vector, and ${}^\ast \hspace{-0.8mm}F^{\mu\nu}$ is
the dual of $F^{\mu\nu}$ defined by
\beqn
{}^\ast \hspace{-0.8mm}F_{\mu\nu}:={1 \over 2}\epsilon_{\mu\nu\alpha\beta}F^{\alpha\beta},
\eeqn
with $\epsilon_{\mu\nu\alpha\beta}$ the Levi-Civita tensor of
$\epsilon_{txyz}=\sqrt{-g}$.  Using the unit timelike vector normal to
the spacelike hypersurface, $n^\mu$, we define the electric and
magnetic fields by
\beqn
E^\mu &:=& F^{\mu\nu} n_\nu,\\
B^\mu &:=& {1 \over 2} n_\alpha \epsilon^{\alpha\mu\nu\beta} F_{\nu\beta},
\eeqn
and thus, the electromagnetic tensor and its dual are written as
\beqn
F^{\mu\nu}&=&n^{\mu} E^{\nu} - n^{\nu} E^{\mu}+\epsilon^{\mu\nu\alpha} B_{\alpha}, \label{eq:Fab} \\
{}^\ast \hspace{-0.8mm}F_{\mu\nu}&=&\epsilon_{\mu\nu\alpha}E^\alpha-n_\mu B_\nu + n_\nu B_\mu, \label{eq:Fab1}
\eeqn
where $\epsilon^{\nu\alpha\beta}=n_\mu\epsilon^{\mu\nu\alpha\beta}$.
Since $E^\mu n_\mu=B^\mu n_\mu=0$, $E^t=B^t=0$. Note the definition of
$n_\mu=-\alpha \nabla_\mu t$ with $\alpha$ the lapse function.


Then Maxwell's equations are written in terms of $E^\mu$ and $B^\mu$
as follows: For the constraint equations,
\beqn
&& D_k E^k=4\pi \rho_{\rm e}, \label{eq:emcon01}\\
&& D_k B^k=0, \label{eq:emcon02}
\eeqn
and for the evolution equations, 
\beqn
&& \pa_t E^i-\cL_{\beta} E^i = \alpha K E^i - D_k \left(\alpha
\epsilon^{kij}B_j \right)-4\pi \alpha \bar j^i,~~~~ 
\label{eq:emevol01} \\
&& \pa_t B^i-\cL_{\beta} B^i = \alpha K B^i + D_k \left(\alpha
\epsilon^{kij}E_j \right), 
\label{eq:emevol02}
\eeqn
where $\rho_{\rm e}:=-j^a n_a$ and $\bar j^i := \gamma^{i}_{~k}j^k$
with $\gamma_{ij}$ the spatial metric defined by
$\gamma_{\mu\nu}:=g_{\mu\nu}+n_\mu n_\nu$.  $D_k$ denotes the
covariant derivative with respect to $\gamma_{ij}$, $\beta^i$ the
shift vector, $K$ the trace of the extrinsic curvature $K_{ij}$, and
$\cL_\beta$ denotes the Lie derivative with respect to $\beta^i$.  We
numerically solve the evolution equations in the
forms~\cite{Shibata16}
\beqn
&&\pa_t \cE^i = - \pa_k \left(\beta^i \cE^k - \beta^k \cE^i
+ \alpha \epsilon^{kij} \cB_j \right) \nonumber \\
&& ~~~~~~~~~~~~-4\pi \left(\cJ^i-Q \beta^i \right), \label{eq:emevo1}\\
&&\pa_t \cB^i = - \pa_k \left( \beta^i \cB^k - \beta^k \cB^i
- \alpha \epsilon^{kij} \cE_j \right), \label{eq:emevo2}
\eeqn
where $\cE^i :=\sqrt{\gamma}\, E^i$, $\cB^i := \sqrt{\gamma}\, B^i$,
$\cJ^i :=\sqrt{-g}\, \bar j^i$, and $Q:=\sqrt{\gamma}\, \rho_{\rm e}$
with $\gamma={\rm det}(\gamma_{ij})$ (in curvilinear coordinates, the
definition should be appropriately modified by excluding the
contribution of the coordinates in $\gamma$). We note
$(-g)=\alpha^2\gamma$.  In this notation, the constraint equations are
written in the simple forms as
\beqn
\pa_i \cE^i &=&4\pi Q,\label{pacE}\\
\pa_i \cB^i &=&0. \label{pacB}
\eeqn

To close the equations, we in general need the Ohm's law, for which we
here write as
\beqn
j^{\mu}=\tilde \rho_{\rm e} u^{\mu}+\sigma_{\rm c} \left( F^{\mu\nu} u_\nu 
+ \alpha_{\rm d} {}^\ast \hspace{-0.8mm}F^{\mu\nu}u_\nu \right), \label{eq:emohm}
\eeqn
where $\tilde \rho_{\rm e}:=-j^\mu u_\mu=w^{-1}( \rho_{\rm
  e}-\sigma_{\rm c}E^\mu u_\mu+\sigma_{\rm c}\alpha_{\rm d} B^\mu
u_\mu)$ is the charge density observed in the frame comoving with
the fluid, and $\sigma_{\rm c}$ is the conductivity with $w:=-n_\mu
u^\mu=\alpha u^t$. Note that the resistivity is defined by
$\eta:=1/(4\pi\sigma_{\rm c})$.  The third term in the right-hand side
of Eq.~(\ref{eq:emohm}) denotes a dynamo term (in the simplest
version) with $\alpha_{\rm d}$ being the so-called $\alpha$-dynamo
parameter (see, e.g.,~Ref.~\cite{B2005} for the dynamo theory and
Ref.~\cite{BDZ2013} for the relativistic formulation). The dynamo term
is a phenomenological one and we do not consider it in this paper
(i.e., $\alpha_{\rm d}=0$) except for Appendix~\ref{app:dynamo}.  Note
that $F^{\mu\nu} u_\nu$ and $-{}^\ast \hspace{-0.8mm}F^{\mu\nu}u_\nu
(=:b^\mu)$ denote the electric and magnetic fields in the frame
comoving with the fluid.

In the ideal MHD case for which we suppose $\sigma_{\rm c} \rightarrow
\infty$, it is convenient to write the electromagnetic tensor in terms
of the magnetic field in the frame comoving with the fluid, $b^\mu$, as
\beqn
F^{\mu\nu}&=&\epsilon^{\mu\nu\alpha\beta} u_\alpha b_{\beta},  \\
{}^\ast \hspace{-0.8mm}F^{\mu\nu}&=& b^\mu u^\nu- b^\nu u^\mu. 
\eeqn
Then, it becomes trivial that the corresponding electric field,
$F^{\mu\nu}u_\nu$, vanishes.  Then, the condition of 
$F^{\mu\nu}u_\nu=0$ with Eq.~(\ref{eq:Fab}) gives the electric field, $\cE^i$, as
\beq
\cE^i=-{1 \over w} \epsilon^{ijk} u_j \cB_k, \label{ideal2}
\eeq
and thus, $\cE^\mu u_\mu(=\cE^i u_i)=0$. 

In the ideal MHD case, the current is not determined from
Eq.~(\ref{eq:emohm}) due to the fact that $\sigma_{\rm c}\rightarrow
\infty$. Instead, it has to be determined from Eq.~(\ref{eq:emevol01})
for $\cE^i$ given by Eq.~(\ref{ideal2}).  Thus, only
Eq.~(\ref{eq:emevol02}) or (\ref{eq:emevo2}) becomes the evolution
equation for the electromagnetic field, which determines the
magnetic-field evolution. Using Eq.~(\ref{ideal2}), this equation is
written in the well-known form as
\beqn
\pa_t \cB^i = \pa_k \left(\cB^k v^i - \cB^i v^k\right). \label{induction0}
\eeqn

\subsection{Ideal MHD}\label{sec2-3}

In the ideal MHD, the energy-momentum tensor is written as
\beq
T_{\mu\nu}=\left(\rho h + {b^2 \over 4\pi}\right) u_\mu u_\nu + \left(P + {b^2 \over 8\pi} \right)g_{\mu\nu}
-b_\mu b_\nu, \label{idealEMtensor}
\eeq
where $b^2=b^\mu b_\mu$. Here, we have the relations as
\beqn
 \alpha b^t &=& B^k u_k,\\
 w b_i &=& B_i + B^k u_k u_i,\\
 w^2 b^2 &=& B^2 +(B^k u_k)^2,
\eeqn
with $B^2 = B^k B_k$, and thus, the energy-momentum tensor is also
written in terms of $B^k$. 

For evolving the MHD equations, we define $S_0:=\sqrt{\gamma}\,
T^{\mu\nu} n_\mu n_\nu$ and $S_i:=-\sqrt{\gamma}\, T^{\mu\nu} n_\mu
\gamma_{\nu i}$, which are written as
\beqn
S_0&=&\rho_* h w - \sqrt{\gamma}P + {\sqrt{\gamma} \over 4\pi} \left[B^2 -{1 \over 2w^2}\left\{B^2
+(B^k u_k)^2\right\} \right] ,\nonumber \\
&& \label{eqS0} \\
S_i&=&\rho_* h u_i + {\sqrt{\gamma} \over 4\pi} \left(B^2 u_i - B^k u_k B_i \right). \label{eqSi}
\eeqn
Their evolution equations are
\beqn
&&\pa_t S_0 + \pa_j \Big[S_0 v^j +\sqrt{\gamma} P_{\rm tot}(v^j + \beta^j)
-{\sqrt{-g} \over 4\pi w}B^k u_k B^j\Big] \nonumber \\
&&~~~=\sqrt{-g}K_{ij} S^{ij}  - S_k D^k \alpha, \\
&&\pa_t S_i + \pa_j \Big[S_i v^j +\sqrt{-g} P_{\rm tot} \delta_i^{~j} \nonumber \\
&&~~~~~~~~~~~~~~- {\sqrt{-g} \over 4\pi w^2} B^j(B_i + u_i B^k u_k) \Big] \nonumber \\
&&~~~=-S_0 \pa_i \alpha + S_k \pa_i \beta^k - {1 \over 2} \sqrt{\gamma} S_{jk} \pa_i \gamma^{jk},
\eeqn
where $P_{\rm tot}:=P + b^2/8\pi$ and $S_{ij}:=\gamma_i^{~\mu}\gamma_j^{~\nu} T_{\mu\nu}$. 

Multiplying $B^i$ to Eq.~(\ref{eqSi}), we obtain $B^i S_i=\rho_* h B^i u_i$. Thus, 
Eq.~(\ref{eqSi}) is rewritten as
\beq
S_i + {\sqrt{\gamma} \over 4\pi \rho_* h} B^k S_k B_i
=\left( \rho_* h + {\sqrt{\gamma} \over 4\pi} B^2 \right) u_i . \label{eqSi2}
\eeq
Using this, the normalization relation of $u^\mu$ is written in terms
of $\gamma_{ij}$, $\rho_*$, $S_i$, and $B^i$ as 
\beqn
&&\gamma^{ij}\left(S_i + {\sqrt{\gamma} \over 4\pi \rho_* h} B^k S_k B_i\right)
\left(S_j + {\sqrt{\gamma} \over 4\pi \rho_* h} B^k S_k B_j\right) \nonumber \\
&& \times \left( \rho_* h + {\sqrt{\gamma} \over 4\pi} B^2 \right)^{-2} +1=w^2. 
\label{eqw2}
\eeqn
Hence, for given (evolved) values of $\gamma_{ij}$, $\rho_*$, $S_i$,
$S_0$, and $B^i$ together with a given equation of state, this
equation together with Eq.~(\ref{eqS0}) constitute simultaneous
equations for $h$ and $w$, and thus, they are used for the so-called
primitive recovery procedure in the ideal MHD. Our numerical approach
for the primitive recovery when employing tabulated equations of state
is essentially the same as that in the viscous hydrodynamics
case~\cite{Fujiba17}.

\subsection{MHD in general cases} \label{sec2-4}

In the general MHD case, not only the magnetic field but also
the electric field are explicitly included in the energy-momentum tensor as
\beqn
T_{\mu\nu} &=& \rho h u_\mu u_\nu + P g_{\mu\nu} \nonumber \\
&+&{1 \over 4\pi} \Biggl[{E^2 + B^2 \over 2}
\left(\gamma_{\mu\nu}+n_{\mu} n_{\nu}\right) - E_{\mu}E_{\nu} -B_{\mu}B_{\nu} \nonumber \\
&&~~~~~+\left(n_{\mu} \epsilon_{\nu\alpha\beta} +n_{\nu} \epsilon_{\mu\alpha\beta} \right)
E^{\alpha} B^{\beta}\Biggr],
\eeqn
where $E^2=E_k E^k$. We note that in the ideal MHD case, using
Eq.~(\ref{ideal2}) together with the fact that $E^i/\sqrt{E_k E^k}$,
$B^i/\sqrt{B_k B^k}$, and $\gamma^{ij}u_j/\sqrt{\gamma^{kl}u_ku_l}$ can
constitute the orthonormal bases of the three space, the
energy-momentum tensor of Eq.~(\ref{idealEMtensor}) is derived in a
straightforward manner.

For the general MHD case, $S_0$ and $S_i$ are written as
\beqn
S_0&=& \rho_* h w - \sqrt{\gamma} P + {1 \over 8\pi}\sqrt{\gamma}\left(E^2 + B^2\right),\label{eqS03}\\
S_i&=& \rho_* h u_i  + {\sqrt{\gamma} \over 4\pi}\epsilon_{ijk} E^j B^k, \label{eqSi3}
\eeqn
and their evolution equations are
\beqn
&&\pa_t S_0 + \pa_j \Bigg[S_0 v^j +\sqrt{\gamma} \left(P-{E^2 + B^2 \over 8\pi} \right)(v^j + \beta^j) \nonumber \\
&&~~~~~~~~~~~~~~ + {1 \over 4\pi} \sqrt{-g} \epsilon^{jkl} E_k B_l \Bigg] \nonumber \\
&&~~~=\sqrt{-g}K_{ij} S^{ij}  - S_k D^k \alpha, \\
&&\pa_t S_i + \pa_j \Bigg[S_i v^j +\sqrt{-g} \left(P + {E^2 + B^2 \over 8\pi}\right) \delta_i^{~j} \nonumber \\
&&~~ - {\sqrt{-g} \over 4\pi} \Big( E_i E^j + B_i B^j \Big) 
-{\sqrt{\gamma} \over 4\pi} (v^j + \beta^j)\epsilon_{ikl} E^k B^l \Bigg] \nonumber \\
&&~~~=-S_0 \pa_i \alpha + S_k \pa_i \beta^k - {1 \over 2} \sqrt{-g} S_{jk} \pa_i \gamma^{jk},
\eeqn
where
\beqn
S_{ij}&=&\rho h u_i u_j + P\gamma_{ij} \nonumber \\
&& + {1 \over 4\pi} \left[-E_i E_j -B_i B_j + {1\over 2}\gamma_{ij}(E^2 + B^2)\right].
~~~
\eeqn

Using Eq.~(\ref{eqSi3}), the normalization relation for $u^\mu$ is written as
\beqn
&&\gamma^{il}\left(S_i-{\sqrt{\gamma} \over 4\pi}\epsilon_{ijk} E^j B^k\right)
\left(S_l-{\sqrt{\gamma} \over 4\pi}\epsilon_{ljk} E^j B^k\right)(\rho_* h)^{-2}
\nonumber \\
&& + 1=w^2. \label{eq:recover}
\eeqn
Thus, for the resistive MHD, this equation together with
Eq.~(\ref{eqS03}) constitute the simultaneous equations for $h$ and
$w$ for given (evolved) values of $\gamma_{ij}$, $\rho_*$, $S_i$,
$S_0$, $B^i$, and $E^i$, and are used for the primitive recovery
procedure. 

Practically, we solve the MHD equations (both in ideal and resistive
MHD) using the cylindrical coordinates $(\varpi, \varphi, z)$ in this
work.  For this procedure, we write the equations for $S_0$, $S_y$,
$\rho_*$, and the conservation equations for the lepton fraction in a
conservative form as in our viscous hydrodynamics
simulations~\cite{Fujiba20,Fujiba2020}. This is in particular
important to guarantee the conservation of the rest mass and angular
momentum in numerical computation.

\subsection{Numerical methods for solving Maxwell's equations}\label{sec2-5}

We solve the ideal MHD equations mostly in the same method as
described in Ref.~\cite{SS2005}: We evolve the hydrodynamics equations
for $S_0$, $S_i$, $\rho_*$, and the lepton fraction using a
high-resolution shock capturing scheme (a 3rd-order upwind scheme).
In the new implementation, one improvement is made for solving the
induction equation on the evolution of the magnetic field, $\cB^k$.
The previous implementation solved the equations for $\cB^x$ and
$\cB^z$ using a constraint transport scheme~\cite{CT88} and that for
$\cB^y$ by an upwind scheme, which is the same as for evolving $S_0$,
$S_i$, and $\rho_*$. However, we were aware of the fact that this
method was so diffusive that the magnetic-field energy was spuriously
decreased with a short timescale within 100\,ms with the typical grid
resolution of the grid spacing $\sim 200$\,m (we use the grid
resolution of DD2-135M of Ref.~\cite{Fujiba2020} for most of the
present simulations and that of DD2-135L of Ref.~\cite{Fujiba2020} for
4 test simulations).  Thus, for a long-term simulation with the
duration more than seconds, this scheme is practically useless
(although a sophisticated high-order scheme may solve this problem).
In the new implementation, we solve the induction equation for $\cB^x$
and $\cB^z$ by a high-resolution upwind scheme, which is the same as
for evolving $S_0$, $S_i$, and $\rho_*$, and the induction equation
for $\cB^y$ by the 4th-order centered finite differencing with the
operation of the 5th-order Kreiss-Oliger dissipation.  To
approximately preserve the divergence-free condition of
Eq.~(\ref{pacB}), a divergence cleaning is introduced (in the same
manner as in the resistive MHD scheme: see below).  We find that in
this method, the stable numerical evolution is feasible and also
spurious oscillation is suppressed with the reasonable choice for the
coefficient of the Kreiss-Oliger term.  By this procedure, the
diffusive evolution of the magnetic field is suppressed in a much
better manner than in our previous implementation.  The maximum error
size of the divergence-free condition, which is defined by $|(\pa_i
\cB^i)\varDelta x /\cB^z|$ where $\varDelta x$ is the grid spacing
that covers remnant neutron stars, remains to be $\sim 10^{-5}$ for
the entire simulation time in the present numerical resolution.

In the resistive MHD, we also employ the same shock capturing scheme
as that employed in the ideal MHD for the hydrodynamics part that
evolve $S_0$, $S_i$, $\rho_*$, and the lepton fraction. On the other
hand, Maxwell's equations are solved in a different procedure.  In
addition, we change the method of the time evolution depending on the
magnitude of the conductivity. In the following, we describe the
methods for the low- and high-conductivity cases separately, focusing
only on the case that the timescale of $1/(4\pi \sigma_{\rm c})$ is
shorter than the time step interval of the numerical simulation,
$\varDelta t$: A partially implicit scheme is employed for appropriately
handling the equation of the electric field for the high values of
$\sigma_{\rm c}$.


For the low-conductivity case, we evolve the electromagnetic equations
using the method that is used for evolving geometrical variables and
incorporating a divergence cleaning prescription.
Specifically, we rewrite Eqs.~(\ref{eq:emevo1}) and
(\ref{eq:emevo2}) in the forms:
\beqn
&&\pa_t \cE^i = - \pa_k \left(\beta^i \cE^k - \beta^k \cE^i
+ \alpha \epsilon^{kij} \cB_j \right) \nonumber \\
&& ~~~~~~~~~~~~-4\pi \left(\cJ^i-Q \beta^i \right), \label{eq:emevo1a}\\
&&\pa_t \cB^i = - \pa_k \left( \beta^i \cB^k - \beta^k \cB^i
- \alpha \epsilon^{kij} \cE_j \right), \nonumber \\
&& ~~~~~~~~~~~-\alpha\gamma^{1/2} \gamma^{ij} \pa_j \phi_B\label{eq:emevo2a} \\
&&\pa_t \phi_B=\beta^k \pa_k \phi_B - \alpha \kappa \phi_B-\alpha \gamma^{-1/2} \pa_k \cB^k,
\label{eq:emevo3a}
\eeqn
where $\phi_B$ is a new auxiliary variable associated with the
divergence cleaning, $\kappa$ is a constant, and $\cJ^i-Q\beta^i$ is
written as
\beqn
\cJ^i-Q\beta^i&=&Q v^i
+\sigma_{\rm c} \alpha_{\rm d} \cB^k u_k (v^i + \beta^i) \nonumber \\
&+&\alpha \sigma_{\rm c} \Big[w A^i_{~j}\cE^j + \epsilon^{ijk}u_j \cB_k \nonumber \\
&&~~~~~~+\alpha_{\rm d}\left(-w \cB^i + \epsilon^{ijk}u_j \cE_k\right)\Big], \label{current}
\eeqn
with $A^i_{~j}=\delta^i_{~j}-w^{-2}\bar u^i u_j$ and $\bar u^i=\gamma^{ij}u_j$. 
We notice that in our scheme, we always evaluate $Q$ by $\pa_k \cE^k /4\pi$. 

In numerical implementation, the transport terms in
Eqs.~(\ref{eq:emevo1a}) and (\ref{eq:emevo2a}) are evaluated by the
4th-order centered finite differencing, and the transport term in
Eq.~(\ref{eq:emevo3a}) is evaluated by the 4th-order upwind scheme as
done for the geometric variables. Whenever we evolve $\cE^i$, $\cB^i$,
and $\phi_B$, we operate the Kreiss-Oliger dissipation procedure in
the same manner as we do for the geometric variables.

To evolve the system, we basically use the 4th-order Runge-Kutta
scheme but for $\cE^i$ we partially employ an implicit scheme. This is
implemented in the following manner: In Eq.~(\ref{eq:emevo1a}), we
move the term of $4\pi \sigma_{\rm c} \alpha w A^i_{~j} \cE^j$ that
appears in the current term to the left-hand side and write the
left-hand side of the evolution equation in the form
\beqn
&&\pa_t \cE^i + 4\pi \sigma_{\rm c} \alpha w A^i_{~j} \cE^j \nonumber \\
&&={1 \over \varDelta t}\left(\cE^i + 4\pi \sigma_{\rm c} \alpha w A^i_{~j}
\cE^j \varDelta t- \cE^i_0\right)
\nonumber \\
&&:={1 \over \varDelta t}\left(M^i_{~j}\cE^j  - \cE^i_0\right), 
\eeqn
where $\cE^i_0$ denotes the values of $\cE^i$ in a previous time-step and 
$M^i_{~j}=(1 + \zeta) \delta^i_{~j} - \zeta \bar u^i u_j/w^2$ 
with $\zeta=4\pi \sigma_{\rm c}\alpha w \varDelta t$. 
Here, the inverse of $M^i_{~j}$ is simply written as
\beq
(M^{-1})^i_{~j} = {1 \over 1+\zeta} \delta^i_{~j} + {\zeta \over (1+\zeta)(w^2+\zeta)}\bar u^i u_j. 
\eeq
Thus, the updated values of $\cE^i$ is easily obtained in each Runge-Kutta time step as
\beqn
\cE^i&=&(M^{-1})^i_j \biggl[\cE^j_0  - \varDelta t\, F_T^j
\nonumber \\
&&-4\pi \varDelta t \Big\{
Q v^j-\sigma_{\rm c} (- \alpha_{\rm d} \cB^k u_k)(v^j + \beta^j) \nonumber \\
&&~~~+\alpha \sigma_{\rm c} \Big(\epsilon^{jkl} u_k \cB_l 
 +\alpha_{\rm d}(-w \cB^j + \epsilon^{jkl}u_k \cE_{0l})\Big) \label{cE2}
 \Big\} \biggr], \nonumber \\
\label{implE}
\eeqn
where $F_T^j$ denotes the term associated with the transport term,
which is evaluated in the explicit manner.  We suppose that
$\alpha_{\rm d}$ is small (or zero) and thus the last term of
Eq.~(\ref{cE2}) is not included for the implicit prescription,
although it is straightforward to include it.

As we illustrate in Appendix~\ref{app1}, this implementation enables
to solve several test-bed problems successfully even for the case that
$\sigma_{\rm c}$ is large such that $\zeta \gg 1$. The electromagnetic
fields are also evolved stably even in the absence of the upwind
scheme.


For the high-conductivity case close to the ideal MHD case
(specifically for $\zeta \geq 100$ in our present implementation), we
modify this procedure. The reason for the modification is that the
equation of the magnetic field approximately reduces to the induction
equation (with a weak diffusion term: see below) as
\beqn
\pa_t \cB^i \approx \pa_k (v^i \cB^k - v^k \cB^i) + O(\sigma_{\rm c}^{-1}).
\label{induction99}
\eeqn
Here in Eq.~(\ref{induction99}) we omit the divergence cleaning term
for simplicity, although we include it in the practical simulations.
As a natural consequence of reducing approximately to a transport
equation, for the case of the long resistive dissipation timescale,
the centered finite differencing could introduce a long-term growing
unstable mode in numerical computation, in particular for the
evolution of the poloidal magnetic field leading to the violation of
divergence-free condition of $\pa_k \cB^k=0$ (note that this condition
is written only by the poloidal-field component in the axially
symmetric case). Below, we first derive Eq.~(\ref{induction99}) from
Eq.~(\ref{eq:emevo2a}).

For the case that $\zeta(=4\pi\sigma_{\rm c}\alpha w \varDelta t) \gg 1$,
the expression of the solution by the implicit method of Eq.~(\ref{implE})
should be mathematically modified by replacing $(M^{-1})^i_{~j}$ by
\beq
(M^{-1})^i_{~j} = {1 \over \zeta} \delta^i_{~j} + {1 \over w^2+\zeta}\bar u^i u_j. 
\eeq
This is easily understood that the stiff term associated with
$\sigma_{\rm c}$ in the right-hand side of the evolution equation for
$\cE^i$ enforces the exponential decay of the initial electric field.
Then, Eq.~(\ref{implE}) is rewritten into the form 
\beqn
\cE^i&=&(M^{-1})^i_j \left(\cE^j_0  - \varDelta t\, F^j -4\pi \varDelta t (Q v^j + \alpha_{\rm d} \sigma_{\rm c} d^j)\right)
\nonumber \\ 
&&~~~~~~ -{1 \over w}\epsilon^{jkl} u_k \cB_l \nonumber \\
&=:&\tilde \cE^i-{1 \over w}\epsilon^{jkl} u_k \cB_l,\label{implE2}
\eeqn
where $d^j=\cB^ku_k(v^j+\beta^j)+\alpha (-w\cB^j + \epsilon^{jkl}u_k\cE_{0l})$
denotes the term associated with the dynamo. 
Hence, the terms associated with the transport for $\cE^i$
(see Eq.~(\ref{eq:emevo2a})) is rewritten as
\beqn
&& -\beta^i \cB^k + \beta^k \cB^i +\alpha \epsilon^{kij} \cE_j \nonumber \\
&& =v^i \cB^k - v^k \cB^i+\alpha \epsilon^{kij} \tilde \cE_j.\label{implE3}
\eeqn
Thus, Eq.~(\ref{induction99}) is derived from Eq.~(\ref{eq:emevo2a})
(here we suppose again that $\alpha_{\rm d}$ is small).

As in the ideal MHD case, the transport part of
Eq.~(\ref{induction99}) should be evaluated by using an upwind scheme
for the numerical stability.  On the other hand, the last term of
Eq.~(\ref{implE3}) constitutes a diffusion term in
Eq.~(\ref{eq:emevo2a}), which can be evaluated by the centered finite
differencing.  In our numerical implementation, the transport term is
handled in the same manner as in the ideal MHD case, while the
diffusion term is handled by the 4th-order centered finite
differencing. In addition, the divergence cleaning and 5th-order
Kreiss-Oliger dissipation are implemented as in the relatively low
conductivity case.

Before closing this subsection, we briefly comment on our artificial
prescription for handling the region in which the electromagnetic
energy density is much larger than the rest-mass energy density,
because such regions often cannot be solved accurately in the MHD
simulation. In the present paper, if the ratio of $B^2/(4\pi \rho
c^2)$ (in the ideal MHD case) or $(E^2 + B^2)/(8\pi \rho c^2)$ (in the
resistive MHD case) is larger than $10^3$, we set $u_i=0$
artificially. In addition, if $w(=\alpha u^t)$ exceeds $5$ or $\varep$
exceeds $10^3$ after the primitive-recover process, we also
artificially set $u_i=0$ and $\varep$ is determined from the condition
that the temperature is $k T=0.1$\,MeV where $k$ is the Boltzmann
constant. The reason for this is that in these 
high-electromagnetic-energy cases, the primitive recovery often fails.
The region with these extreme conditions appears for the case that the
magnetic winding proceeds significantly and as a result a strong
outflow is driven from the neutron-star surface toward a low-density
atmosphere. In the absence of these artificial prescriptions, the
computation often crashed in our current implementation.  Since we
give up resolving the region with $w \geq 5$, we cannot explore a
very-high Lorentz-factor jet in the present simulations.

\subsection{Initial condition and relevant timescales}\label{sec2-6}

As in our series of the papers~\cite{Fujiba17,Fujiba18,Fujiba2020},
the initial condition for the matter field is supplied from the result
of a simulation for binary neutron star mergers.  Specifically, we
employ the DD2-135M model of Ref.~\cite{Fujiba2020}: a merger remnant
of binary neutron stars with each neutron-star mass $1.35M_\odot$. As
touched in Sec.~\ref{sec2-5}, four simulations are also performed using a
lower-resolution model, DD2-135L, to check the dependence of the
results on the grid resolution.


We then superimpose electromagnetic fields for which the energy
density is much smaller than the internal and kinetic energy density
of the fluid. In this work, we initially give a poloidal magnetic
field written in the form
\beqn
\cB^\varpi&=&-{1 \over \varpi} \pa_z A_\varphi,\\
\cB^z &=& {1 \over \varpi} \pa_\varpi A_\varphi,
\eeqn
where 
\beq
A_\varphi=A_0 \varpi^2\,
{\rm max}\left({P \over P_{\rm max}} - \delta_A, 0\right), \label{eqAA}
\eeq
with $\delta_A=10^{-5}$ and $P_{\rm max}$ is the maximum pressure.
$A_0$ determines the initial field strength.
With this setting, the magnetic field is present for a high-density
region with $\rho \agt 10^{11.5}\,{\rm g/cm^3}$. 
The toroidal magnetic field is set to be zero ($\cB^y=0$) initially.
The electric field is determined by the ideal MHD condition
(\ref{ideal2}).  We also performed two ideal MHD simulations with
$\delta_A=10^{-2}$ (for which the settings agree approximately with
Bh-R0-Y and Bh-R0-N in Table~\ref{table1}).  Irrespective of the
values of $\delta_A$, the magnetic-field is strong only inside the
remnant massive neutron star for which the magnetic-field effect
appears in the most remarkable manner in the present context (i.e., in
the axisymmetric simulation with no dynamo effect). Indeed, it is
found that the results on the evolution of the electromagnetic energy
and angular velocity profile depend only weakly on the value of
$\delta_A$. Thus in the following, we focus only on the results of
$\delta_A=10^{-5}$.

As Eq.~(\ref{eqAA}) shows, we pay attention to the MHD effect only for
the high-density region, i.e., for the remnant neutron star, and do
not pay attention to the MHD effect in the disk. Because the angular
velocity of the disk decreases along the cylindrical-radius direction,
the MRI would play a crucial role for its evolution. For exploring the
evolution, we need a long-term non-axisymmetric simulation that can
capture the effects of turbulent motion and resulting effective
viscosity induced. This is beyond the scope of this paper.

We also performed simulations initially with a purely toroidal
magnetic field of
\beqn
\cB^y=A_0 \varpi z \,{\rm max}\left({P \over P_{\rm max}} - 10^{-5}, 0\right).
\label{initoro}
\eeqn
Here the dependence on the coordinates, $\varpi z$, stems from the
required symmetry for $\cB^y$ in the axially and planely (with respect
to the $z=0$ plane) symmetric assumption.  In axisymmetric simulations
with this initial condition, the magnetic field should be simply
preserved or decay with the timescale determined by $\sigma_{\rm c}$
(cf.~Eq.~(\ref{taudis}) below). In Appendix~\ref{app2}, we show that
our implementation can derive the expected numerical results.

Table~\ref{table1} lists the initial conditions (with the purely
poloidal magnetic field). $\hat B_{\rm init}$ and $E_{\rm B}$ denote the
maximum value of $\cB^z \gamma^{-1/3}(=B^z \gamma^{-1/6})$ and the
electromagnetic energy at the initial state. Here, the electromagnetic
energy is defined by
\beq
E_{\rm B}={1 \over 8\pi} \int (B^2 + E^2) \sqrt{-g}\,d^3x. 
\eeq
Note that $E^2$ is much smaller than $B^2$. Since the volume of the
neutron star is of the order of $10^{19}\,{\rm cm}^3$, the volume
averaged magnetic-field strength is initially $\sim 10^{15}$\,G for
$E_{\rm B} \sim 10^{48}$\,erg.  The labels, Bv, Bh, Bm, and Bl, in the
model name of Table~\ref{table1} refer to the initial magnetic-field
strength (very high, high, medium, and low).

Because the remnant neutron star which we employ is differentially
rotating, with this setting, the toroidal magnetic-field strength is
initially increased linearly with time by the winding effect as
(e.g.,~Refs.~\cite{Shapiro2000,SLSS2006,Sun2019})
\beqn
B^T \approx B^\varpi_0 {d\Omega \over d\ln\varpi} t,
\eeqn
where $B^\varpi_0$ is the initial local magnitude of the $\varpi$
component of the magnetic field.  This growth continues until the
magnetic-field energy increases to $\sim 10\%$ of the rotational
kinetic energy, $E_{\rm kin}$, of the neutron star.  Thus the
winding timescale is defined approximately by
\beqn
\tau_{\rm wind}&:=&\sqrt{{E_{\rm kin} \over E_{B,0}}} (\Omega_{\rm max}-\Omega_0)^{-1}
\nonumber \\
&\approx& 0.2\,{\rm s}\left({E_{\rm kin} \over 10^{53}\,{\rm erg}}\right)^{1/2}
\left({E_{B,0} \over 10^{47}\,{\rm erg}}\right)^{-1/2}
\nonumber \\ &&~~~~~\times
\left({\Omega_{\rm max}-\Omega_0 \over 5000\,{\rm rad/s}}\right)^{-1},
\label{tauwind}
\eeqn
where $\Omega_{\rm max}$ and $\Omega_0$ denote the maximum value of
the angular velocity and angular velocity at the center,
respectively. $E_{B,0}$ is the initial magnetic-field energy, and
$E_{\rm kin}$ is defined by
\beq 
E_{\rm kin}:={1 \over 2} \int \rho_* h u_\varphi v^\varphi d^3x.
\eeq 
For the model employed in this paper, the initial value of $E_{\rm
  kin}$ is $\approx 1.16 \times 10^{53}$\,erg.  In the estimate of
Eq.~(\ref{tauwind}), we supposed that the region of the maximum angular
velocity would be located in an outer region of the remnant massive neutron
star~(see, e.g.,~Ref.~\cite{Fujiba20}) and it governs the winding and
the growth of the electromagnetic energy. Also, we assumed that
a part of the poloidal magnetic field for which the electromagnetic
energy is $\sim 10\%$ of $E_{B,0}$ contributes to the winding. 

In the case of the finite value of $\sigma_{\rm c}$, the electromagnetic 
fields are dissipated. The dissipation timescale is approximately written as
\beqn
\tau_{\rm dis} &\approx& {4\pi \ell^2 \sigma_{\rm c} \over c^2} \nonumber \\
&=& 0.13\,{\rm s} \left({\ell \over 3\,{\rm km}}\right)^2
\left({\sigma_{\rm c} \over 10^8\,{\rm s}^{-1}}\right), \label{taudis}
\eeqn
where $\ell$ denotes the typical variation scale of the magnetic
field.  Because we do not know the typical size of $\sigma_{\rm c}$,
we pay attention to the cases for which $\tau_{\rm dis}$ is comparable
to $\tau_{\rm wind}$, i.e., $\sigma_{\rm c}=10^7$--$10^9\,{\rm
  s}^{-1}$.

For the case of $\tau_{\rm dis} \gg \tau_{\rm wind}$, obviously, the
resistive dissipation does not play a role. By contrast, for
$\tau_{\rm dis} \ll \tau_{\rm wind}$, the magnetic field is diffused
out before the magnetic winding significantly amplifies the
magnetic-field strength. Also, the winding effect does not directly
determine the final magnetic-field profile for $\tau_{\rm dis} \ll
\tau_{\rm wind}$.  This point is understood by the following model for
the evolution of $\cB^y$, for which the evolution equation is
approximately written as
\beq
\pa_t \cB^y \approx \varpi \cB^i \pa_i \Omega + \eta \Delta \cB^y,
\eeq
where $\Delta$ denotes the Laplacian operator (different from
$\varDelta$).  Here, we picked up only the terms related to the
winding and resistive dissipation. For this equation, we consider a
toy model in which the first term in the right-hand side is not
time-varying as $\varpi\cB^i\pa_i\Omega=F(x^i)=:\eta\Delta F_2(x^i)$.
Then, it is easily found that for $t \ll \tau_{\rm dis}$, the magnetic
winding approximately determines the solution as $\cB^y \approx t
F(x^i)$. On the other hand, for $t \agt \tau_{\rm dis}$, the
asymptotic solution is $\cB^y \approx - F_2(x^i)$.  Thus, the winding
history is not reflected, although the order of magnitude of $F_2$ is
written as $\sim \Omega \tau_{\rm dis} |\cB^i|$, which reflects the
winding in $t \leq \tau_{\rm dis}$.  We note that near the rotational
axis, $F \propto \varpi^3$, and hence, $F_2 \propto \varpi^5$. Thus,
even if $\cB^y$ is positive during the winding, the final relaxed
value for it can be negative.

\begin{table}[t]
\caption{Initial condition and set-up for the numerical simulation.
  We list the maximum value of $\hat B_{\rm init}=\cB^z\gamma^{-1/3}$,
  the magnetic-field energy, and the value of $\sigma_{\rm c}^{-1}$ in
  units of milliseconds. The last column shows the on or off of the
  irradiation/heating and pair-annihilation of neutrinos. For all the
  initial conditions, the total baryon mass is $M_*=2.95M_\odot$, the
  gravitational mass is $M=2.64M_\odot$, the total rotational
  kinetic energy is $E_{\rm kin}\approx 1.16 \times 10^{53}$\,erg, and
  the total angular momentum is $J=4.65 \times 10^{49}$\,g\,cm$^2$/s.
}
\begin{tabular}{ccccc} \hline
~~Model~~ & ~~$\hat B_{\rm init}$\,(G)~~ &~~$E_{\rm B}$\,(erg)~~ & ~$\sigma_{\rm c}\,{\rm (s^{-1})}$~ & Pair \& Heat.~ \\
 \hline \hline
Bv-R0-Y & $4.4 \times 10^{15}$ & $9.4 \times 10^{47}$ & $\infty$ & On   \\
Bh-R0-Y & $2.2 \times 10^{15}$ & $2.4 \times 10^{47}$ & $\infty$ & On   \\
Bh-R0-N & $2.2 \times 10^{15}$ & $2.4 \times 10^{47}$ & $\infty$ & Off  \\
Bh-Rw-Y & $2.2 \times 10^{15}$ & $2.4 \times 10^{47}$ & $10^{11}$& On    \\
Bh-Rl-Y & $2.2 \times 10^{15}$ & $2.4 \times 10^{47}$ & $10^9$   & On   \\
Bh-Rm-Y & $2.2 \times 10^{15}$ & $2.4 \times 10^{47}$ & $10^8$   & On   \\
Bh-Rh-Y & $2.2 \times 10^{15}$ & $2.4 \times 10^{47}$ & $10^7$   & On   \\
Bm-R0-Y & $1.1 \times 10^{15}$ & $5.9 \times 10^{46}$ & $\infty$ & On   \\
Bm-R0-N & $1.1 \times 10^{15}$ & $5.9 \times 10^{46}$ & $\infty$ & Off  \\
Bm-Rl-Y & $1.1 \times 10^{15}$ & $5.9 \times 10^{46}$ & $10^9$   & On   \\
Bm-Rm-Y & $1.1 \times 10^{15}$ & $5.9 \times 10^{46}$ & $10^8$   & On   \\
Bm-Rh-Y & $1.1 \times 10^{15}$ & $5.9 \times 10^{46}$ & $10^7$   & On   \\
Bl-Rl-Y & $5.6 \times 10^{14}$ & $1.5 \times 10^{46}$ & $10^9$   & On   \\
 \hline
\end{tabular}
\label{table1}
\end{table}

\section{Evolution of a remnant neutron star}\label{sec3}

\subsection{Summary of the evolution}\label{sec3-1}

Table~\ref{table1} lists the models for which we performed
simulations.  The initial electromagnetic-field energy is varied among
$1.5 \times 10^{46}$--$9.4 \times 10^{47}$\,erg.  With this setting,
the magnetic winding timescale is approximately 0.1--1\,s.

The simulations are performed for the ideal MHD models (referred to as
R0 in the model name) and for the resistive MHD models with
$\sigma_{\rm c}=10^7$, $10^8$, $10^9$, and $10^{11}\,{\rm s}^{-1}$
(referred to as Rh, Rm, Rl, and Rw in the model name).  For
$\sigma_{\rm c}=10^{11}\,{\rm s}^{-1}$, the dissipation timescale by
the resistivity is $\sim 100$\,s, which is much longer than the
winding timescale in our setting. Thus, this model is employed to
check that the ideal MHD results are approximately reproduced in the
resistive MHD implementation.


For most of the models, neutrino effects including the
irradiation/heating and pair annihilation are taken into account as in
our previous papers~\cite{Fujiba17,Fujiba2020}.  To clarify the
importance of the neutrino effects on the evolution of the massive
neutron stars and on the mass ejection, for the ideal MHD case, we
perform two simulations in which the effects of both the neutrino
irradiation/heating and neutrino pair annihilation are switched off (models
Bh-R0-N and Bm-R0-N: in the presence of the neutrino effects, the
model is referred to with the label Y).

As we touched in Secs.~\ref{sec2-5} and \ref{sec2-6}, numerical
simulations were performed with two grid resolutions for two ideal MHD
(Bh-R0-Y and Bh-R0-N) and two resistive MHD (Bh-Rw-Y and Bh-Rl-Y)
cases. By comparing the results of two different resolutions, we find
that the numerical results depend weakly on the grid resolution (see
Appendix~\ref{app3}). Specifically, for the lower grid resolution, the
amplification of the magnetic-field strength is suppressed, and thus,
the maximum magnetic-field energy becomes smaller, in particular in
the presence of the neutrino irradiation/heating and pair-annihilation
effects.  For the resistive MHD simulation, we also find that the
dissipation is spuriously enhanced for the lower grid
resolution. However, besides these quantitative effects, the evolution
process of the neutron star is not modified qualitatively by changing
the grid resolution.

\begin{figure*}[t]
\includegraphics[width=88mm]{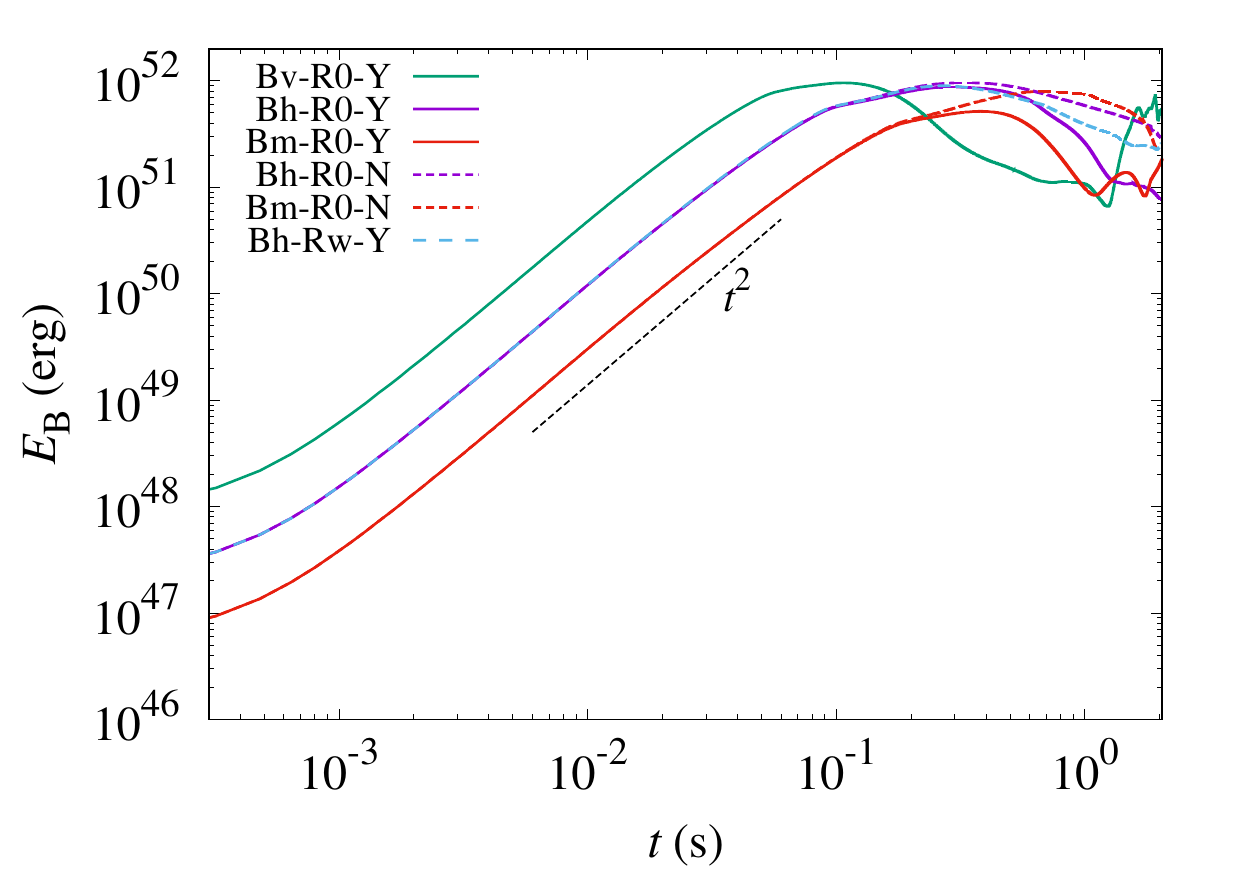}~~
\includegraphics[width=88mm]{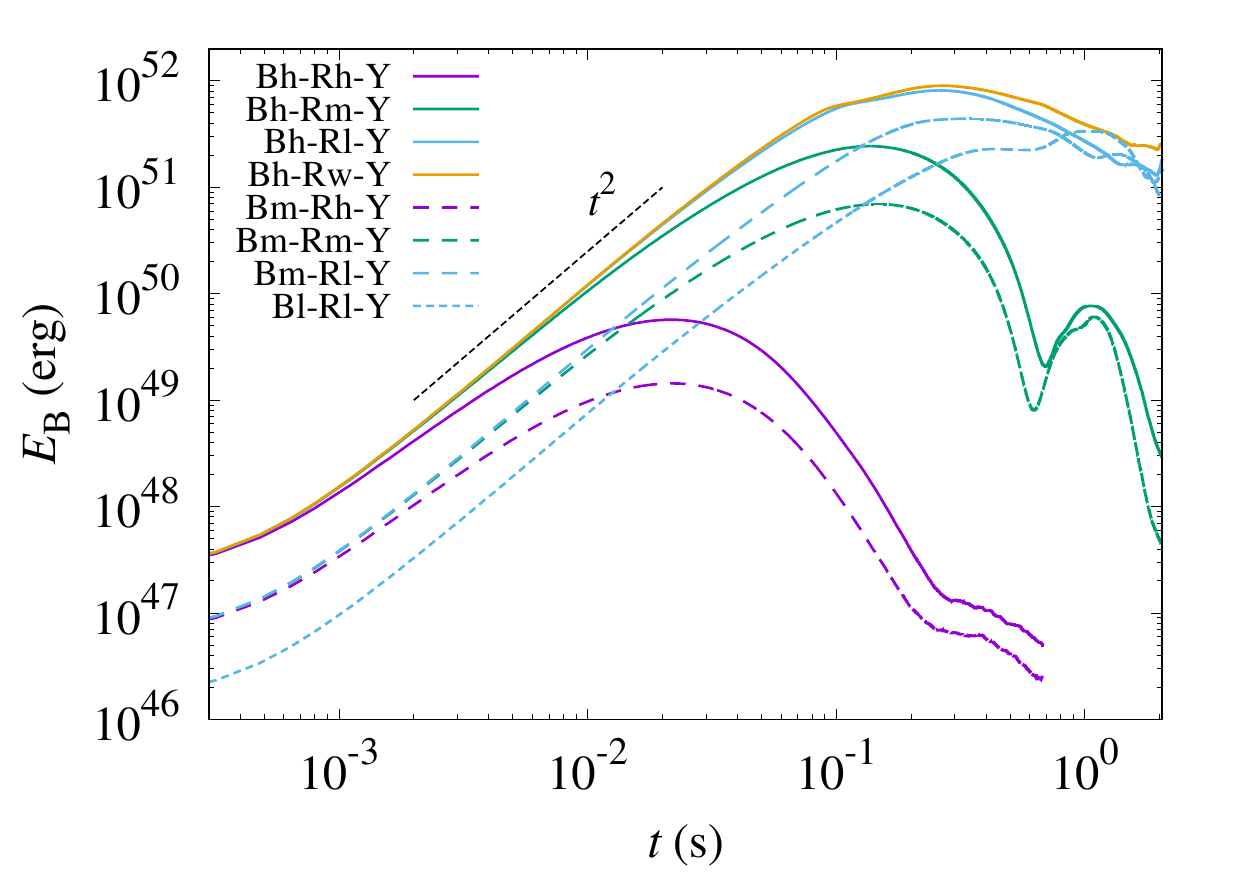} 
\caption{Evolution of the electromagnetic energy for all the ideal MHD models, 
Bv-R0-Y, Bh-R0-Y, Bh-R0-N, Bm-R0-Y, Bm-R0-N, and for a resistive
  model with a high conductivity, Bh-Rw-Y (left panel) and for all the
  resistive models employed in this paper (right panel).
\label{fig1}}
\end{figure*}

In the present context, the typical MHD process in the remnant neutron
star of neutron-star mergers in the presence of an initial seed
poloidal magnetic field is governed by the magnetic winding effect,
for which the growth timescale of the toroidal magnetic-field energy
is written approximately by Eq.~(\ref{tauwind}).  When the
electromagnetic energy exceeds a few~\% of the rotational kinetic
energy of the remnant neutron star (for this case $E_{\rm kin} \approx
1.16 \times 10^{53}$\,erg), the magnetic braking starts playing an
important role. As a consequence, the angular momentum in the neutron
star is redistributed and the differentially rotating configuration of
the angular velocity is enforced to approach a (approximately) rigidly
rotating one.  Then, the angular velocity in the central region
exceeds $\sim 5000$\,rad/s (i.e., the rotational period is $\sim
1.2$\,ms).  After the toroidal magnetic-field strength is
significantly increased by the winding, the mass outflow from the
neutron-star surface is enhanced by the increased magnetic pressure
(although the mass ejection in the present context is driven primarily
by the neutrino irradiation/heating: see Sec.~\ref{sec3-2}). After the
magnetic braking works significantly, the increase of the angular
velocity in the central region is decelerated. In addition, the
magnetic-field profile in the neutron star is modified significantly
by the mass outflow (and resulting magnetic-flux escape) process. The
detail on these findings is described in Sec.~\ref{sec3-1a}.

In the presence of only low resistivity (i.e., a high value of
$\sigma_{\rm c}$), the evolution process of the remnant neutron star is
essentially the same as in the ideal MHD case.  In the presence of a
high resistivity with which the dissipation timescale of the magnetic
field is shorter than the winding timescale, not only the winding but
also the magnetic-field dissipation plays an important role. For such
cases, the magnetic winding does not occur significantly and the
magnetic-field effect on the angular velocity profile is minor (see
Sec.~\ref{sec3-1b} for details).


\subsubsection{Ideal MHD}\label{sec3-1a}

\begin{figure*}[t]
\includegraphics[width=86mm]{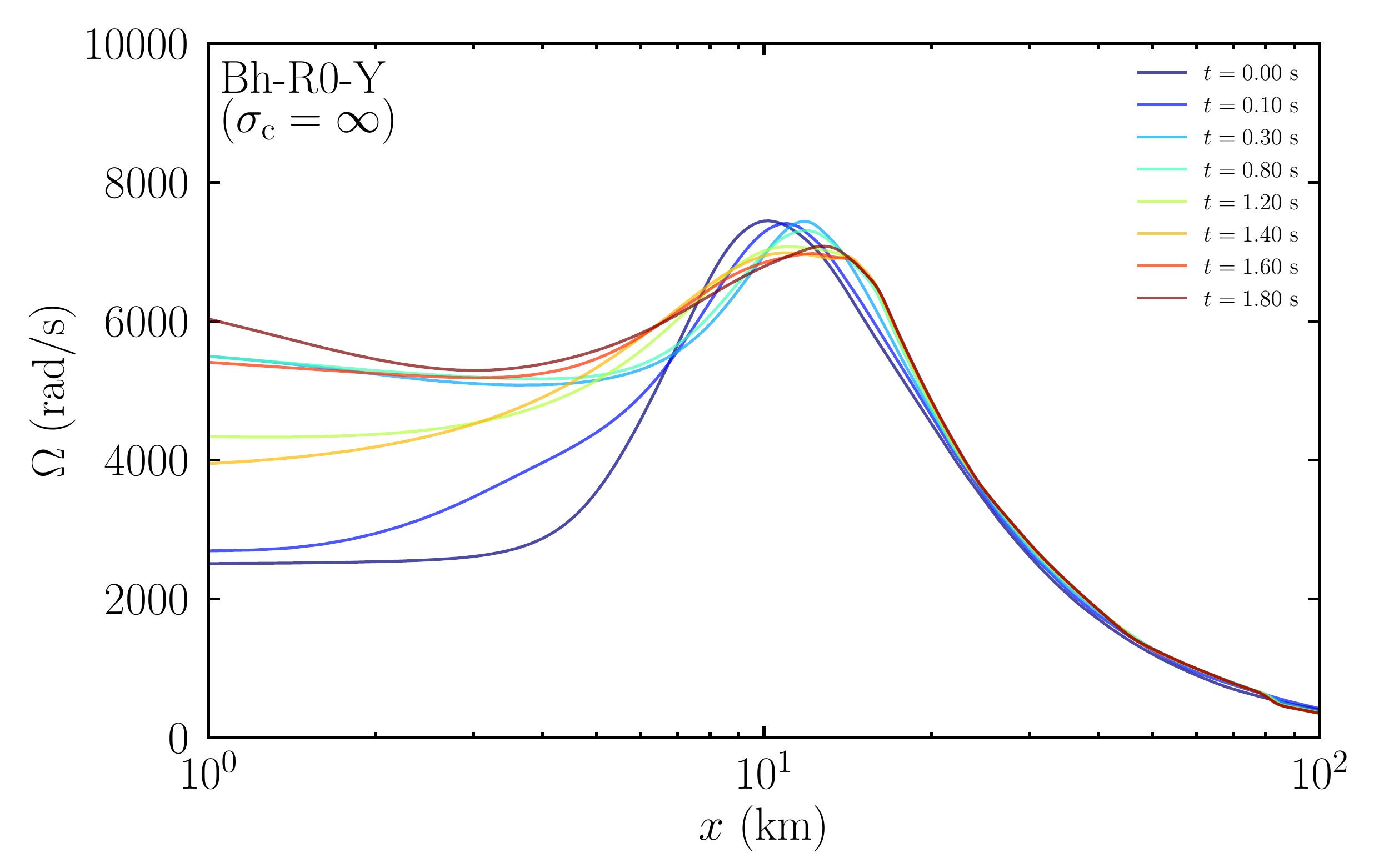} 
\includegraphics[width=86mm]{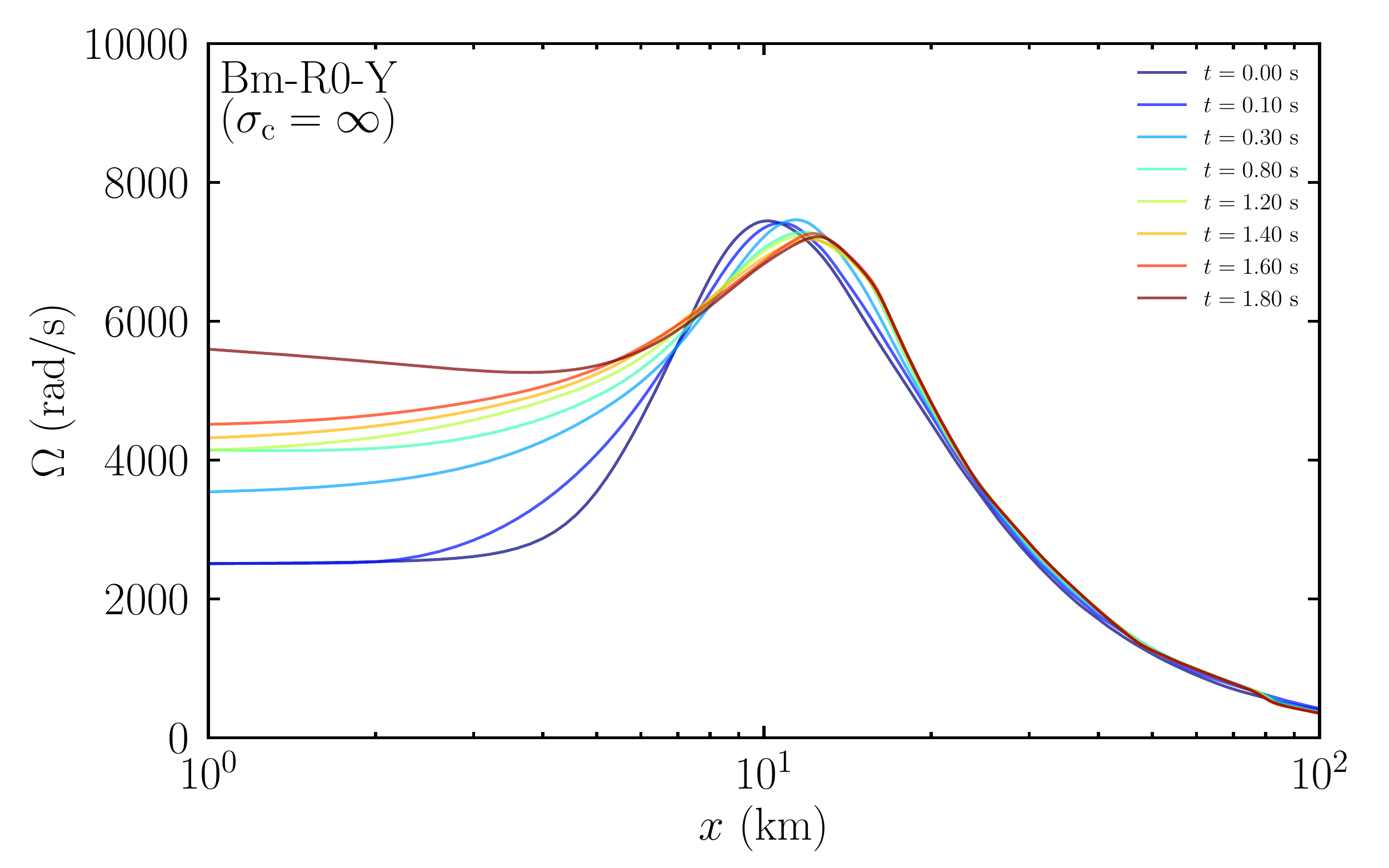} \\
\includegraphics[width=86mm]{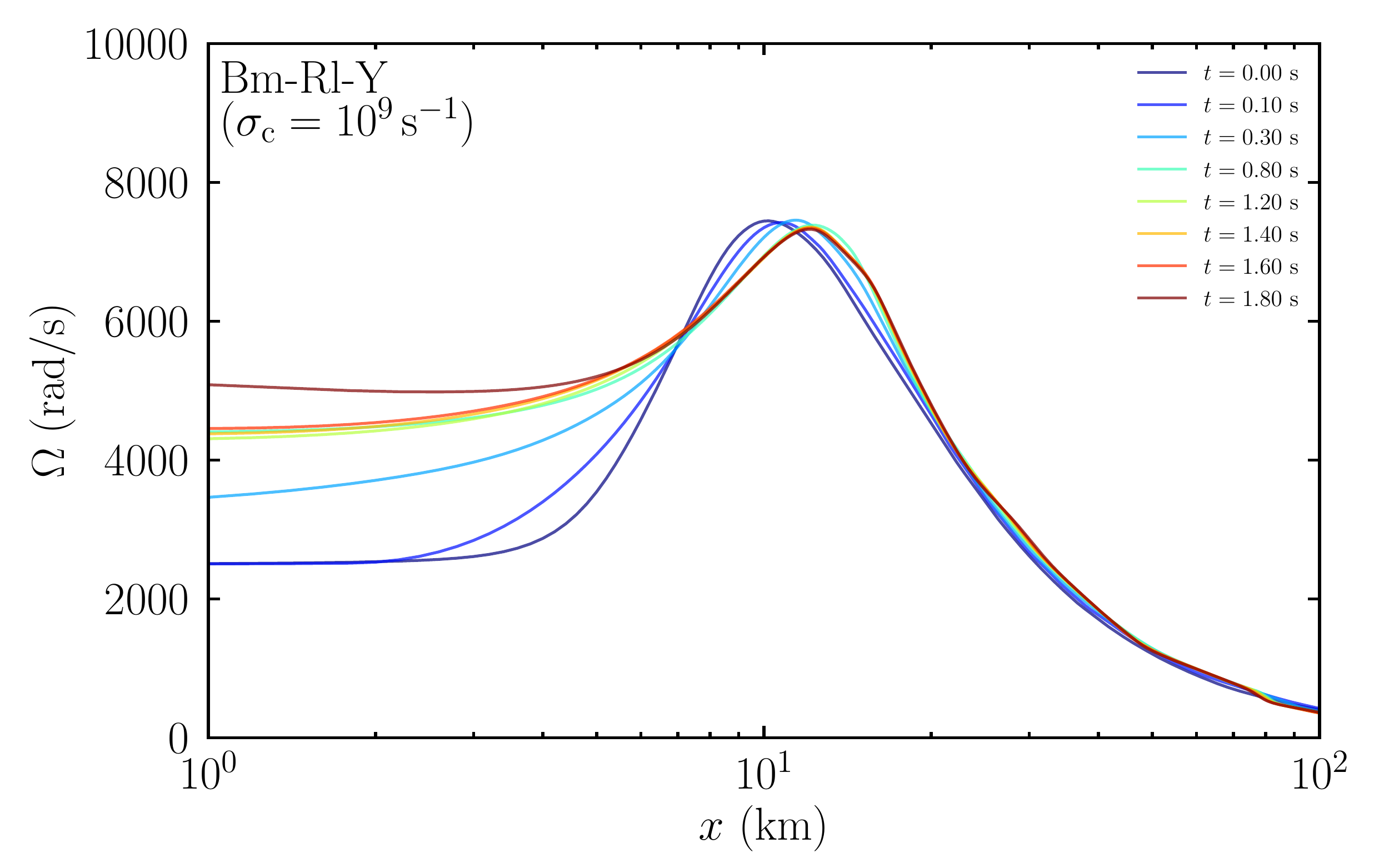} 
\includegraphics[width=86mm]{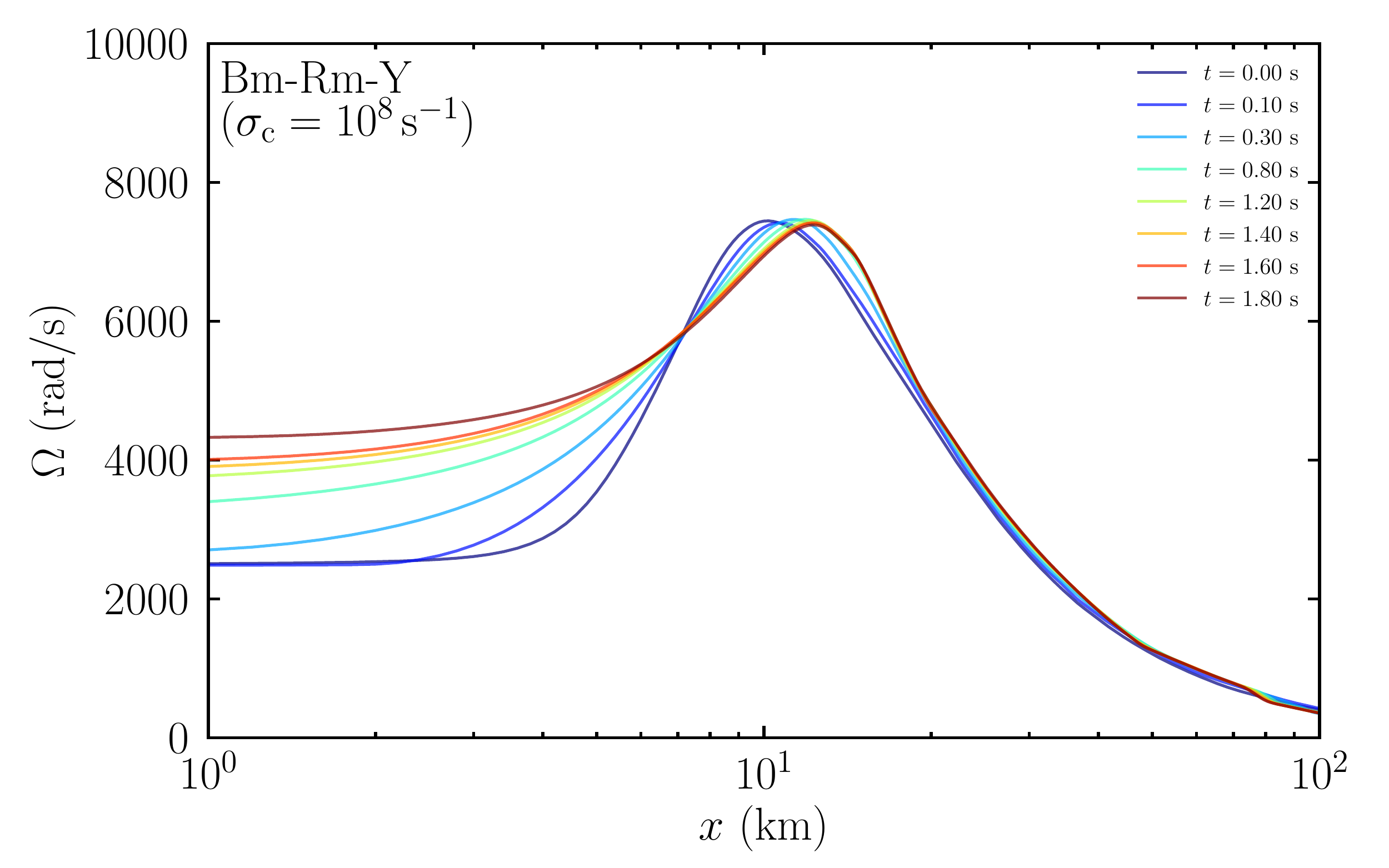} \\
\includegraphics[width=86mm]{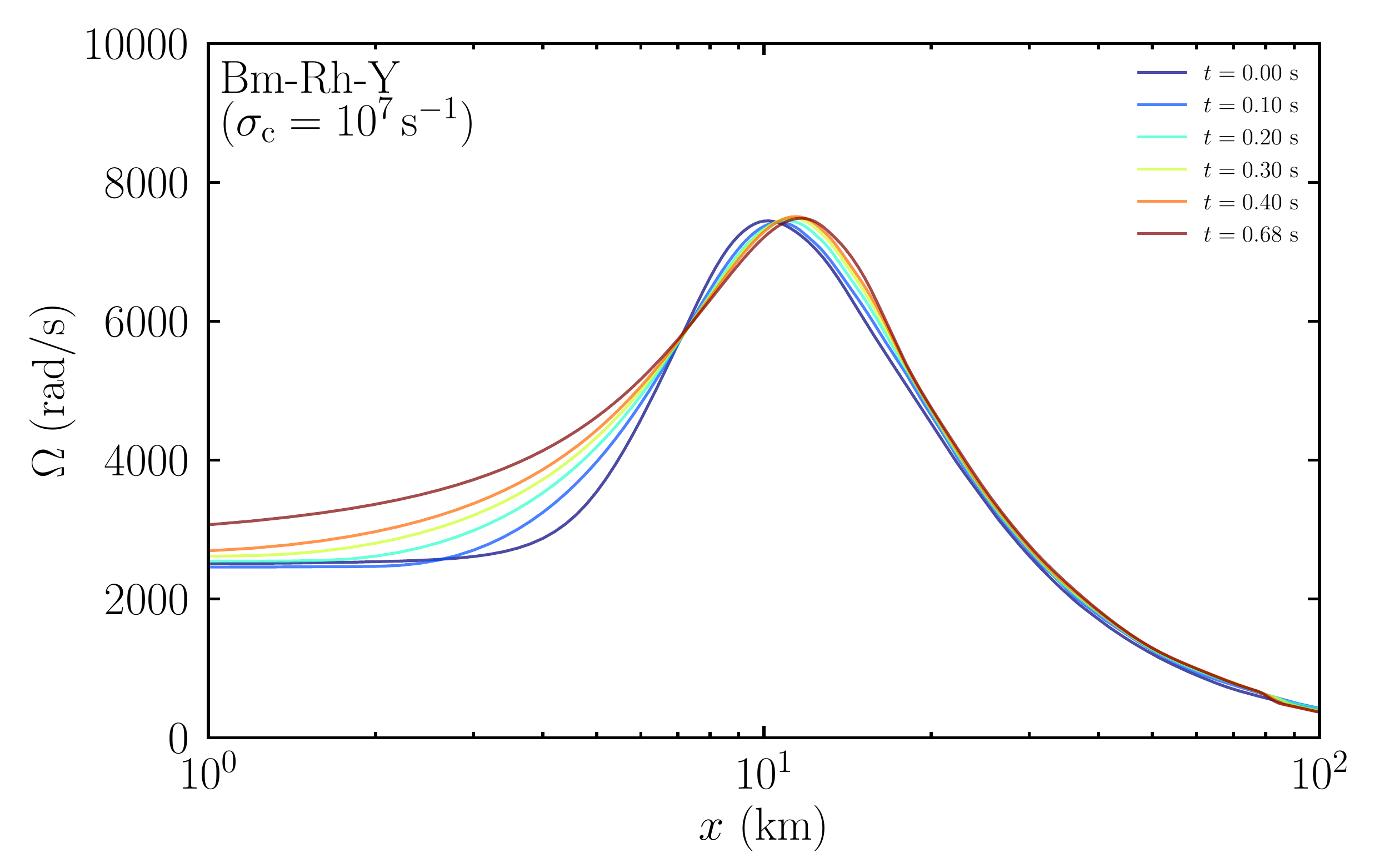}
\includegraphics[width=86mm]{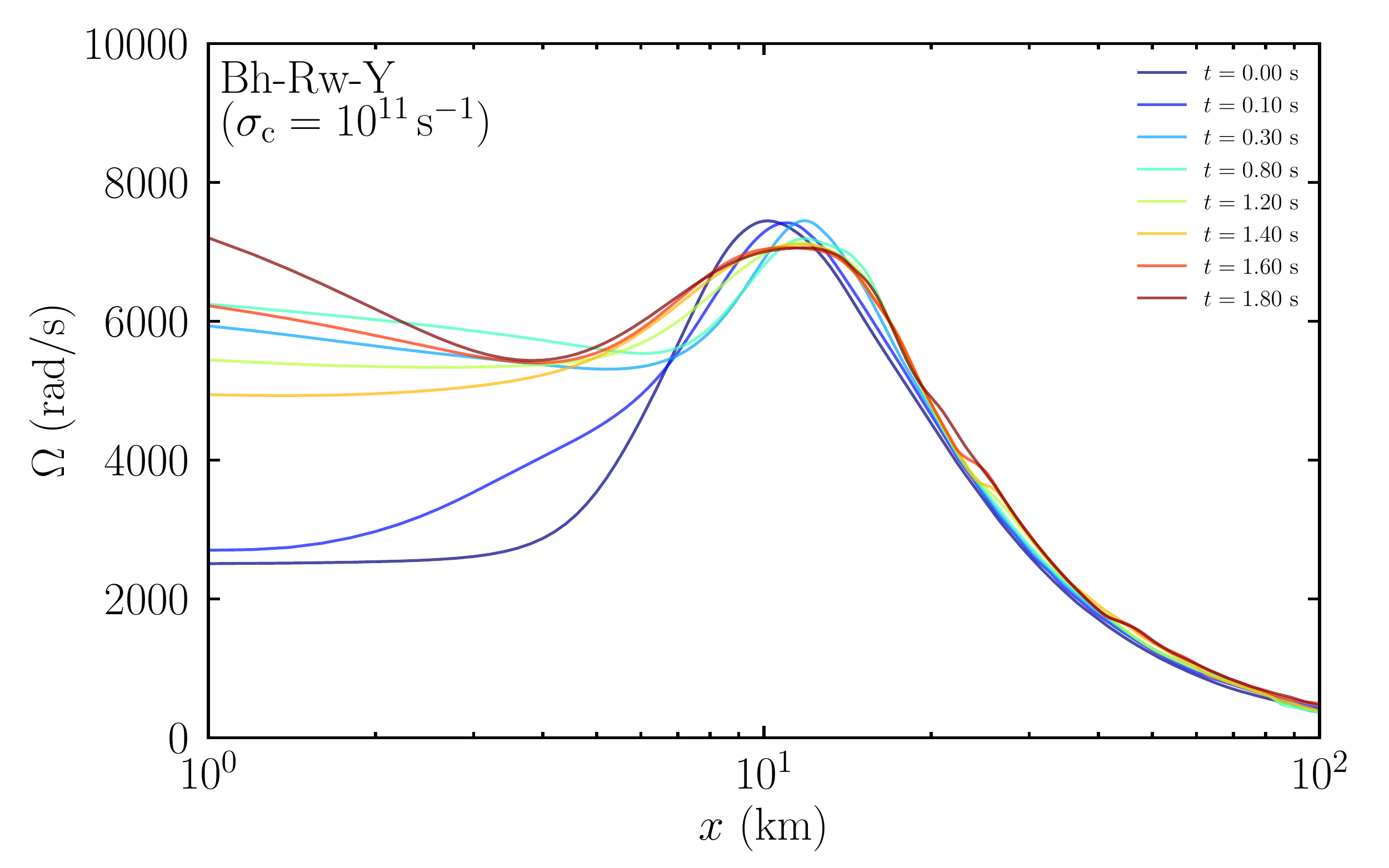}
\caption{Evolution of the angular velocity profile for models Bh-R0-Y
  (top left), Bm-R0-Y (top right), Bm-Rl-Y (middle left), Bm-Rm-Y
  (middle right), Bm-Rh-Y (bottom left), and Bh-Rw-Y (bottom right).
  We note that due to the neutrino cooling, the remnant neutron star
  contracts and its central region spins up. For model Bm-Rh-Y, the
  increase of the angular velocity is predominantly due to this
  effect.
\label{fig2}}
\end{figure*}

\begin{figure*}[p]
\includegraphics[width=82mm]{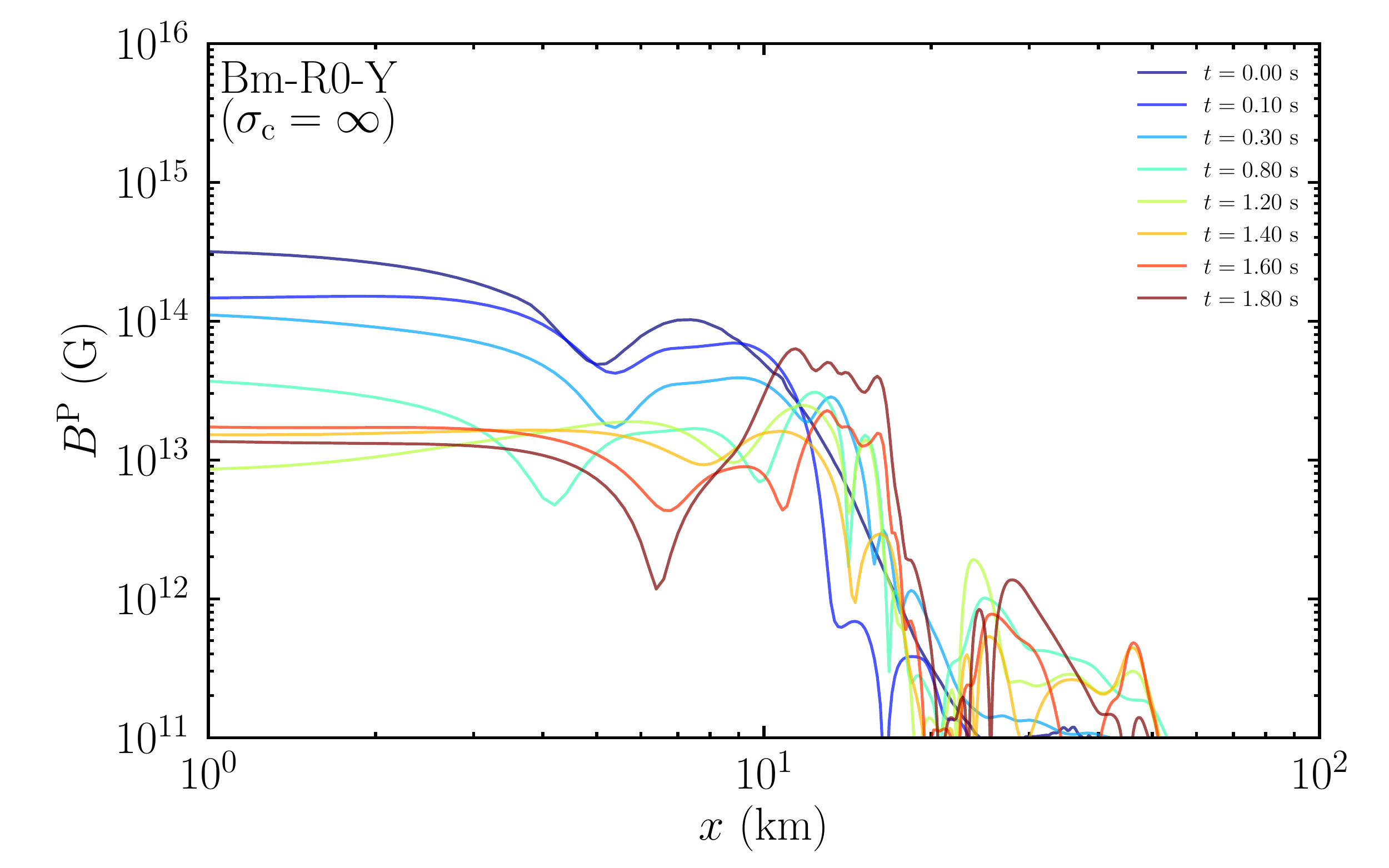} 
\includegraphics[width=82mm]{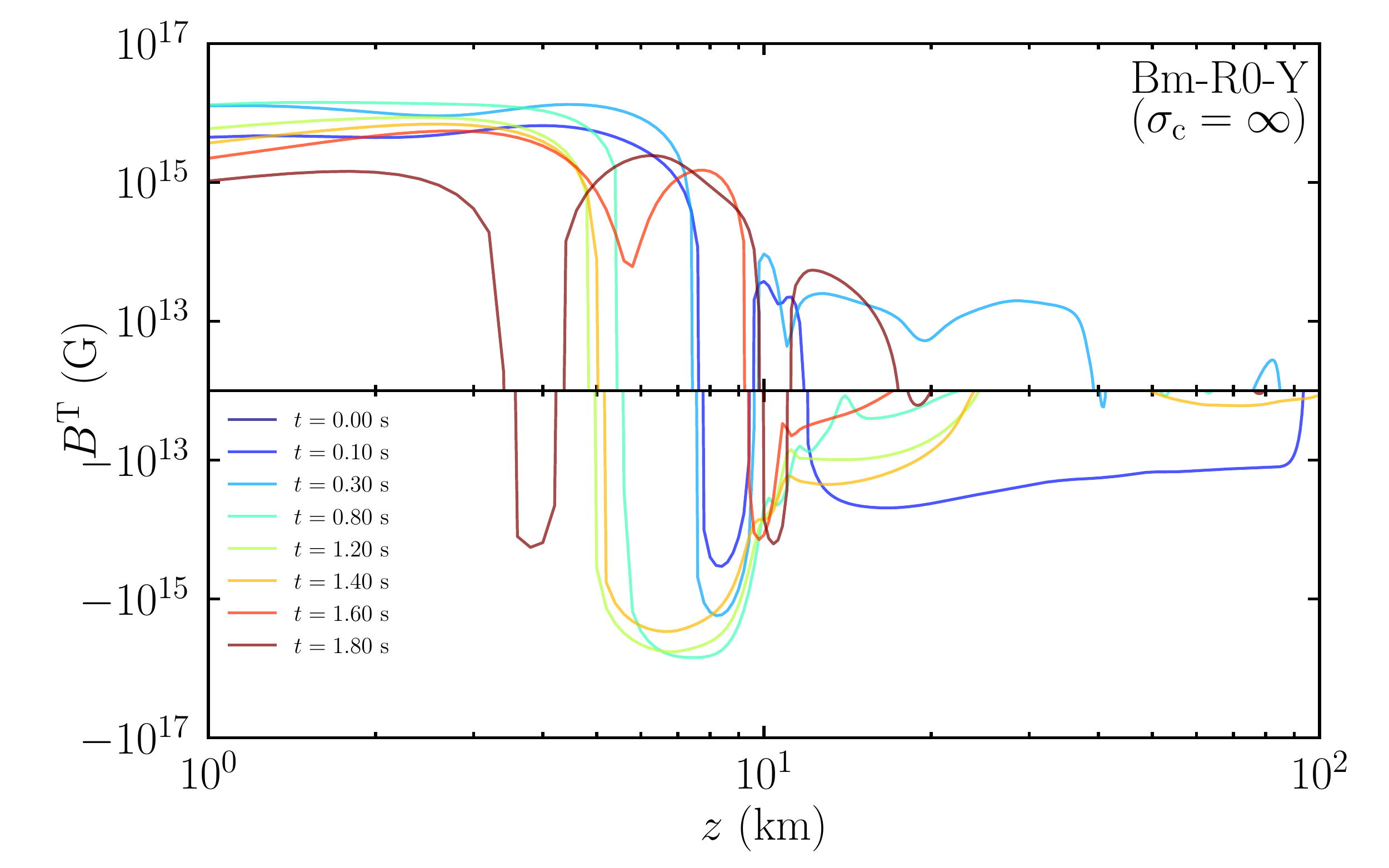} \\
\includegraphics[width=82mm]{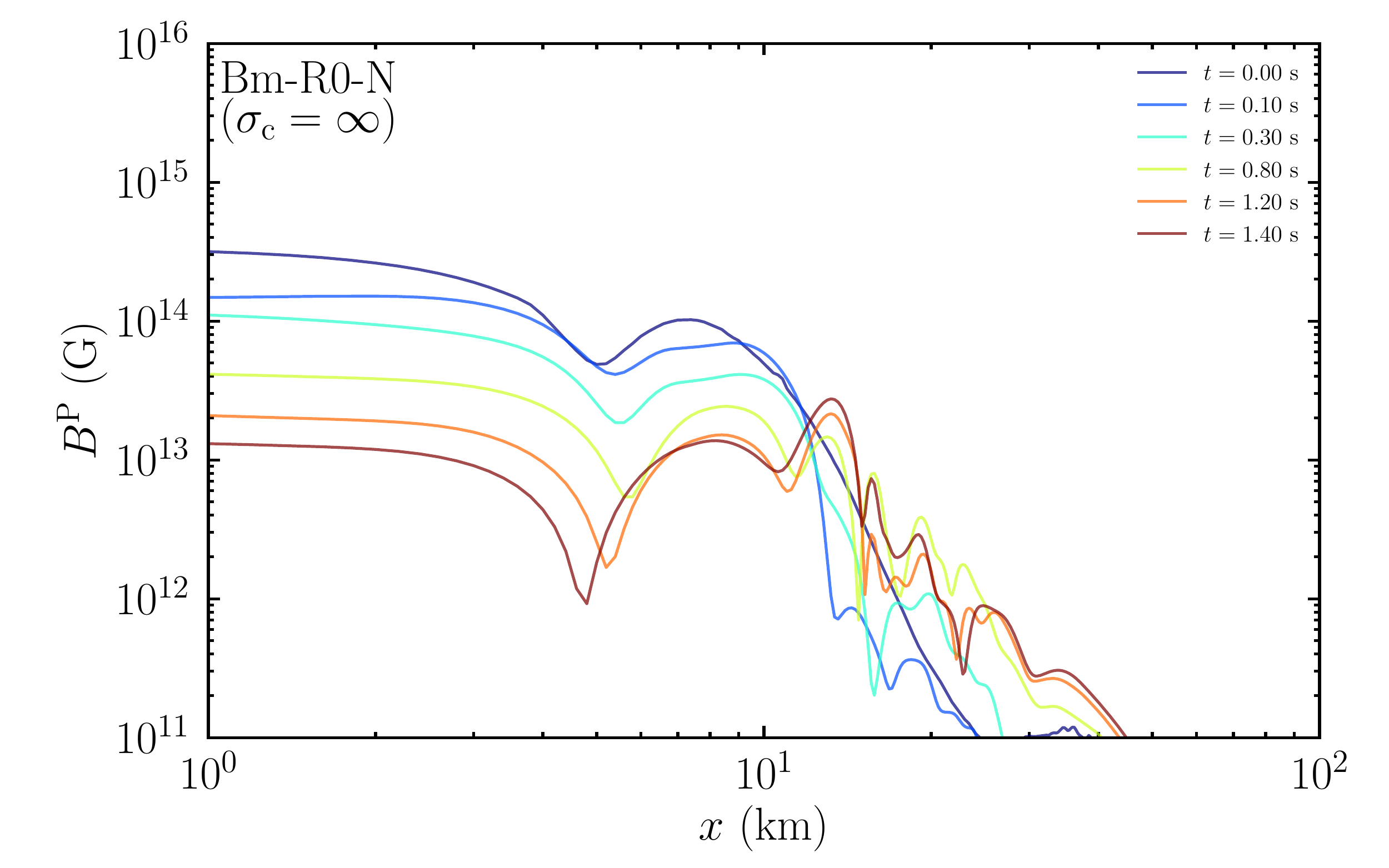} 
\includegraphics[width=82mm]{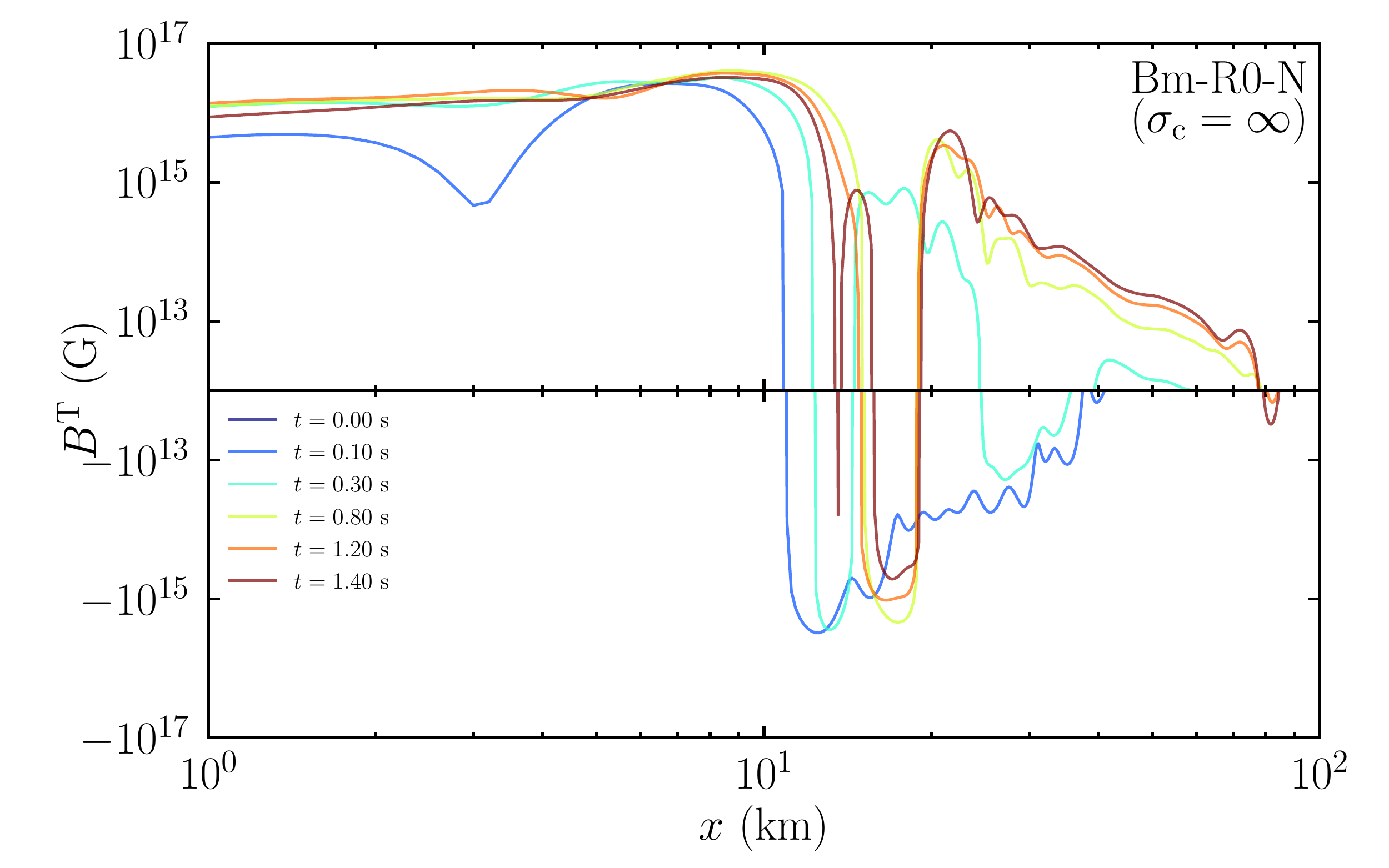} \\
\includegraphics[width=82mm]{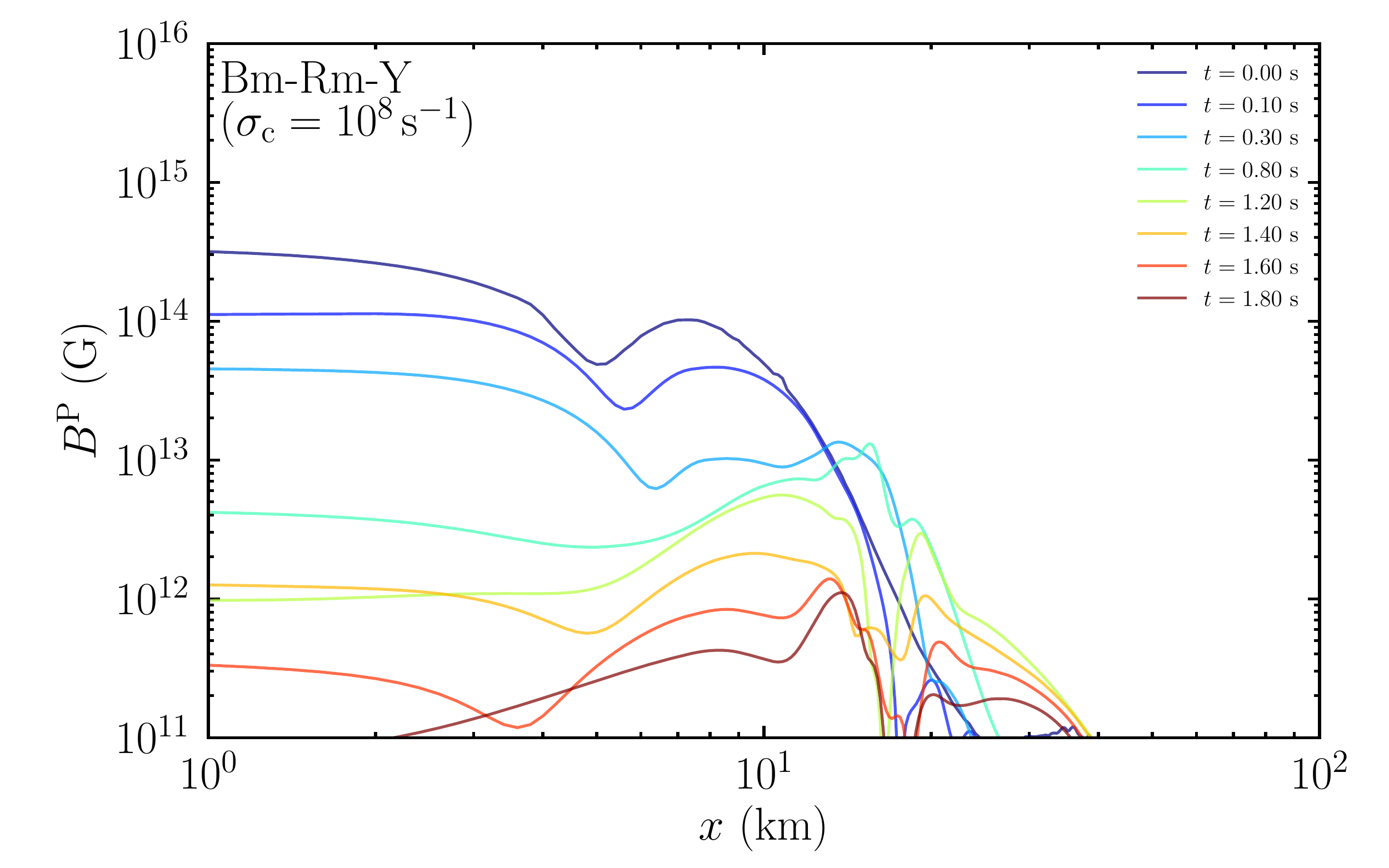} 
\includegraphics[width=82mm]{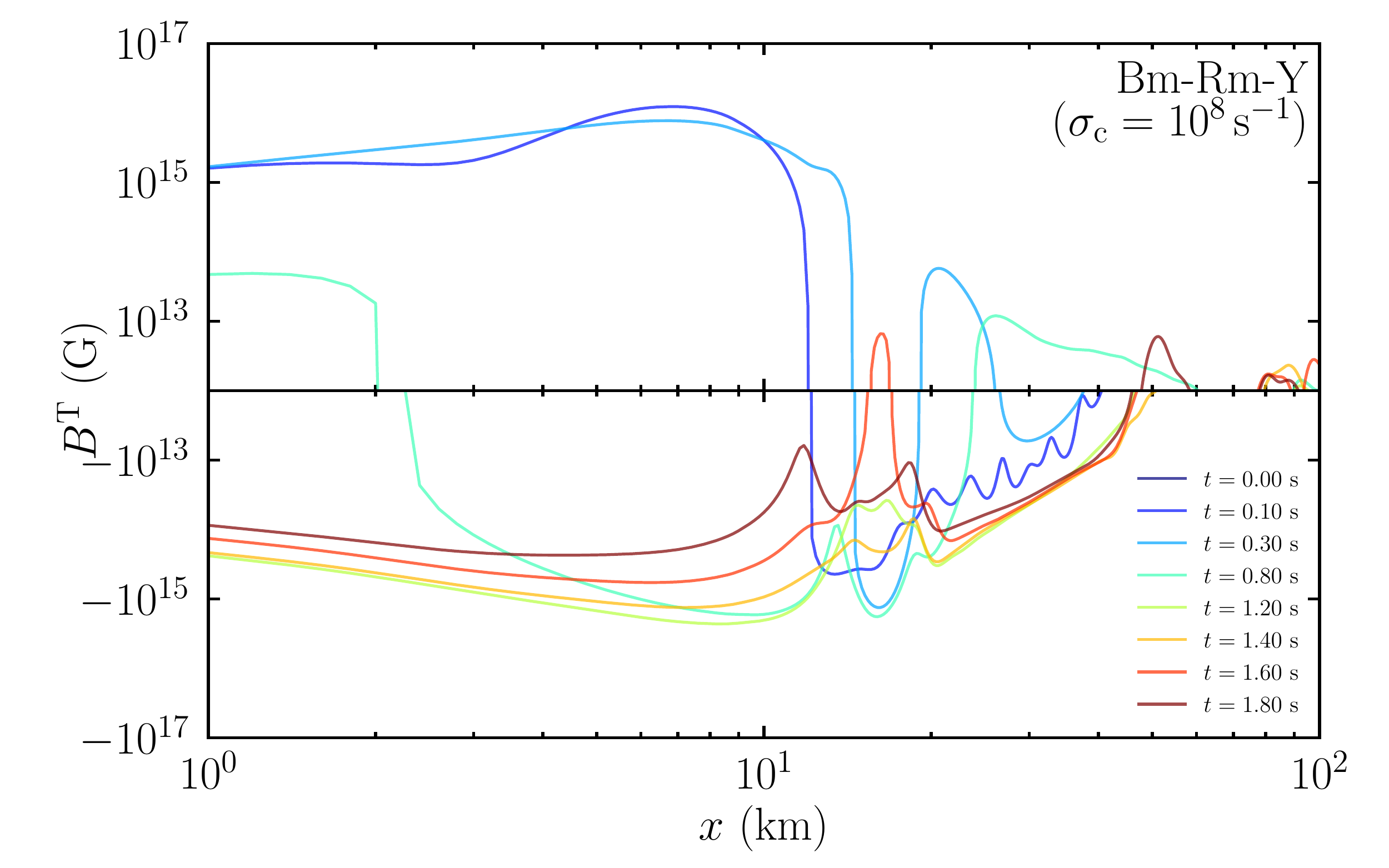} \\
\includegraphics[width=82mm]{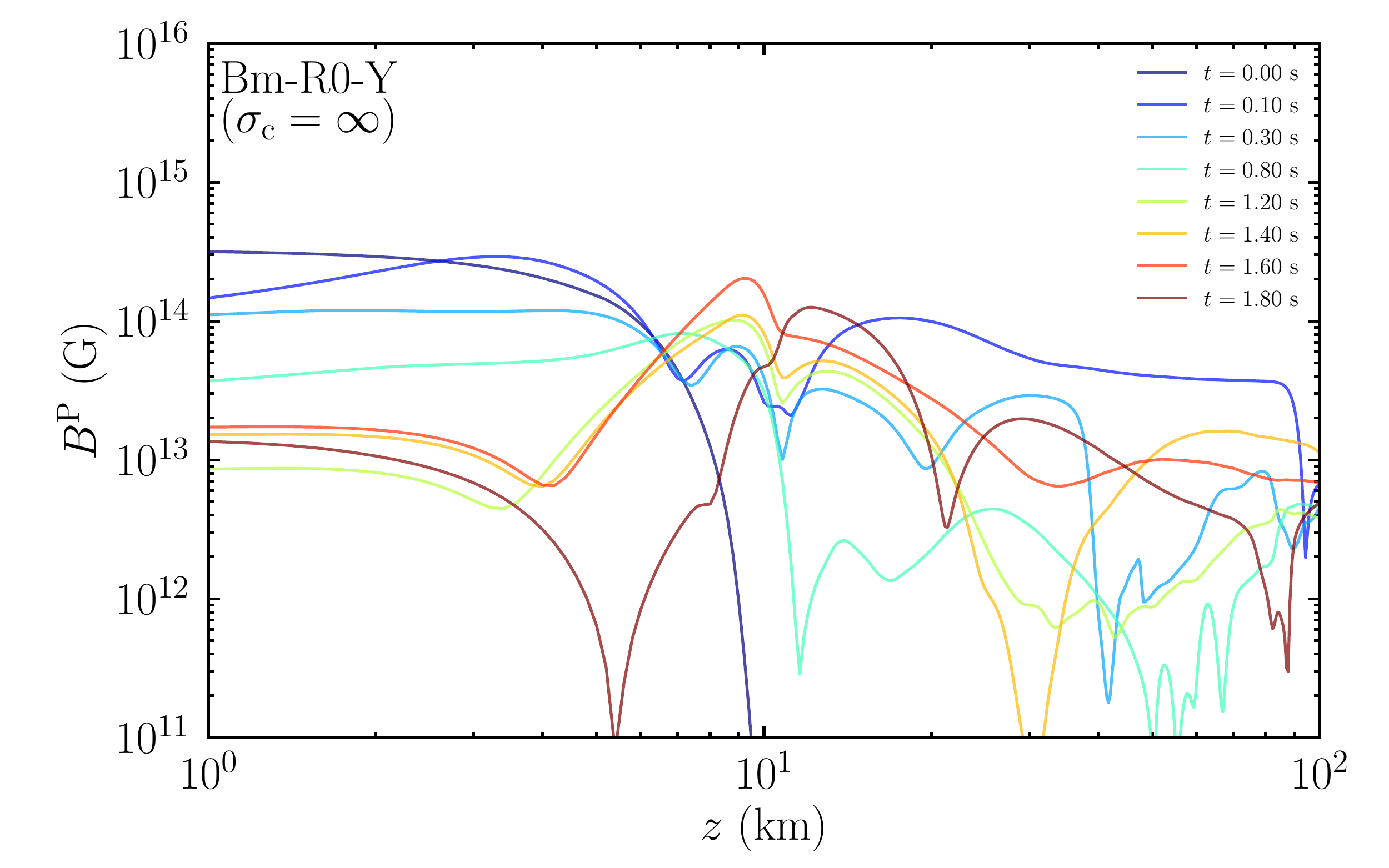} 
\includegraphics[width=82mm]{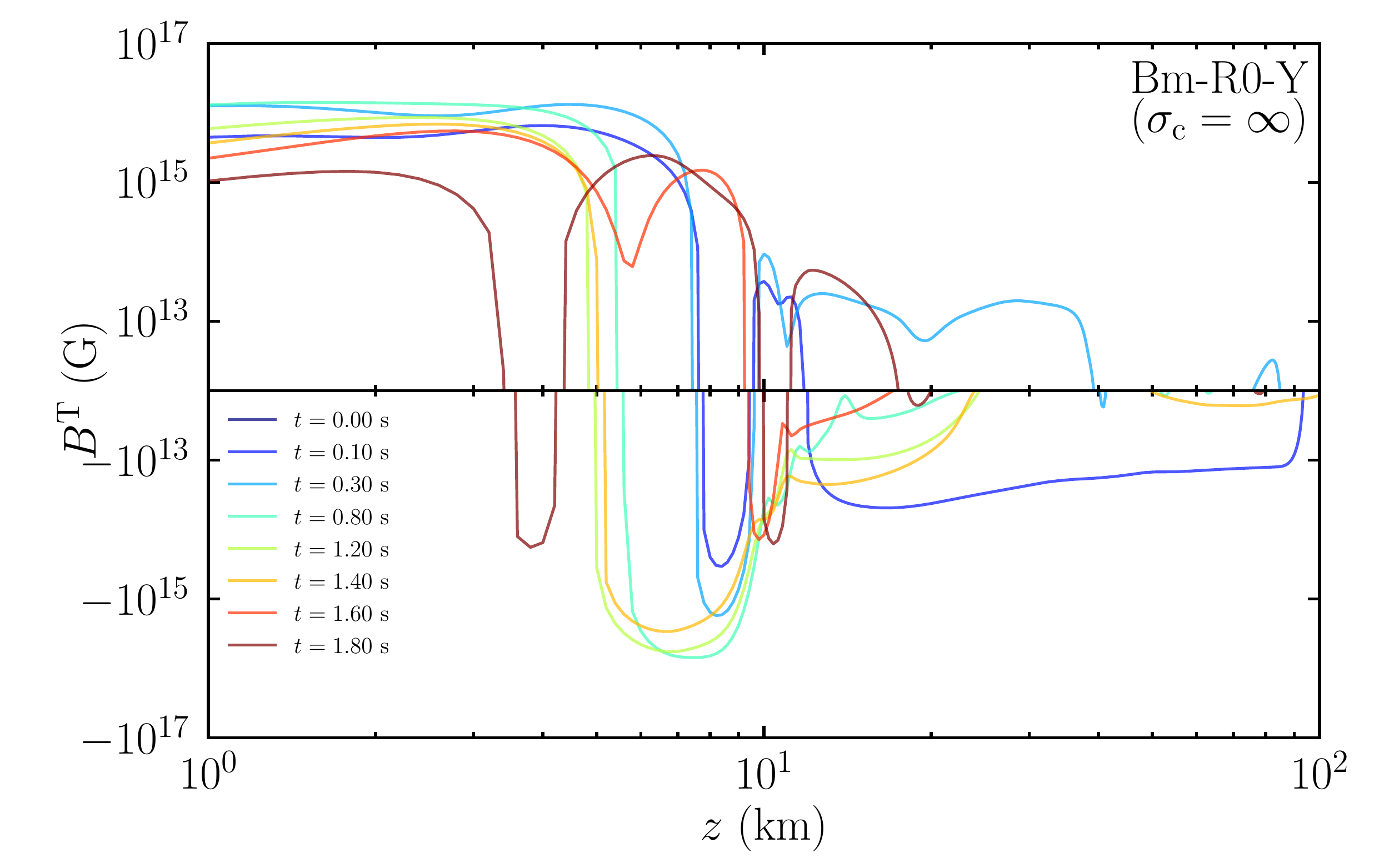}
\caption{Poloidal (left) and toroidal (right) magnetic-field strength
  along the $x$-direction with $z=1$\,km at selected time slices for models
  Bm-R0-Y (top panel), Bm-R0-N (2nd top panel), and Bm-Rm-Y (3rd
  panel). The bottom panel shows the poloidal and toroidal magnetic-field
  strengths along the $z$-direction with $x=1$\,km for model Rm-B0-Y. 
  The field strengths of the poloidal and toroidal
  components are defined by $B^P=\sqrt{\cB^x\cB_x + \cB^z
    \cB_z} \gamma^{-1/2}$ and $B^T={\rm sign}(\cB^y) \sqrt{\cB^y\cB_y}
  \gamma^{-1/2}(\approx\cB^y \gamma^{-1/3})$, respectively. 
\label{fig3}}
\end{figure*}

Figure~\ref{fig1} shows the electromagnetic energy as a function of
time.  The left panel plots the results for all the ideal MHD models
together with one resistive model of a high conductivity (Bh-Rw-Y) and
the right panel plots those for all the resistive MHD models. In the
ideal MHD cases, the electromagnetic energy, $E_{\rm B}$, is initially
increased by the magnetic winding effect by several orders of
magnitude. In this early stage, $E_{\rm B}$ is approximately
proportional to $t^2$ irrespective of the models (including the
resistive MHD models except for the case of $\sigma_{\rm c}=10^7\,{\rm
  s}^{-1}$ for which the resistive dissipation timescale is quite
short; the order of 10\,ms).

For the ideal MHD case, after $E_{\rm B}$ exceeds $\sim 3\%$ of
$E_{\rm kin}$, the increases of $E_{\rm B}$ is decelerated (i.e., the
slope of $E_{\rm B}$ becomes shallower than $t^2$), indicating that
the winding effect is weaken and the magnetic braking becomes
important. However, the increase of $E_{\rm B}$ still continues until
it reaches $\sim 10\%$ of $E_{\rm kin}$, and then, the growth of the
electromagnetic field is saturated. The saturation energy is slightly
smaller for the smaller initial field strength. This is primarily due
to the fact that more magnetic fluxes are escaped associated with the
mass outflow for longer-term evolution, and may be partly due to the
numerical dissipation, i.e., in the longer-term evolution the damage
by the numerical dissipation is more serious: We note that the
simulations are always performed for $\sim 10^3$ rotational period of
the neutron star.

The saturated energy is slightly higher in the absence of the neutrino
irradiation/heating and pair-annihilation effects. Our interpretation
for this is that in the absence of these effects, the mass outflow
from the neutron star is suppressed (cf.~Sec.~\ref{sec3-2}), resulting
escape of the magnetic flux (associated with the mass outflow) is
also suppressed, and as a result, the amplification of the
electromagnetic field proceeds to a higher level.

As we already mentioned briefly, the mass ejection is driven primarily
by the neutrino irradiation/heating (see Sec.~\ref{sec3-2} for
details).  However, with the increase of the electromagnetic energy
toward the saturation, the mass ejection from the neutron star is
slightly enhanced.  This is due to the magnetic pressure primarily
resulting from the significantly amplified toroidal magnetic field.
By this enhancement of the outflow, a part of the magnetic flux
escapes from the neutron star, and as a result, the total
electromagnetic energy decreases.  However, the decrease rate becomes
eventually low ($E_{\rm B}$ eventually shows the oscillatory
behavior) and the electromagnetic energy in average appears to relax
to $\sim 10^{51}$\,erg, i.e., $\sim 1\%$ of the rotational kinetic
energy of the neutron star.  We note that the rotational kinetic
energy does not change significantly by the magnetic braking effects
because it is always larger by more than one order of magnitude than
the electromagnetic energy, and thus, the magnetic-field effect is
minor. Instead, the rotational kinetic energy slightly increases with
time because the neutron star contracts due to the emission of
neutrinos of total energy of $\sim 10^{53}$\,erg, and hence, the
neutron star slightly spins up in the absence of the electromagnetic
effect, as shown in Ref.~\cite{Fujiba2020}.

For the nearly ideal MHD case, i.e., for the resistive MHD case with a
tiny resistivity (high conductivity, $\sigma_{\rm c}=10^{11}\,{\rm
  s}^{-1}$; model Bh-Rw-Y), the curve of $E_{\rm B}$ is similar to the
corresponding ideal MHD model (Bh-R0-Y) as found in the left panel of
Fig.~\ref{fig1}. This is natural because for this model, the resistive
dissipation timescale is quite long $\sim 100$\,s
(cf.~Eq.~(\ref{taudis})). In the late stage with $t \agt 1$\,s, the
disagreement between the results for models Bh-R0-Y and Bh-Rw-Y is not
negligible. Our interpretation for this is that the formulation and
the finite-differencing scheme are not completely the same for the
ideal and resistive MHD computations; e.g., in the resistive MHD we
solve the equations for the electric field but in the ideal MHD we do
not.  In particular, for regions in which the electromagnetic energy
density is much higher than the rest-mass energy density, we have to
introduce an artificial treatment to avoid the crush of the
computation, and for such regions the prescription becomes different
in the two MHD implementations.  Hence, in the long-term simulation,
the numerical error accumulated by such an artificial treatment could
cause this difference (accordingly, for $t \agt 1$\,s, the numerical
results are not very accurate).


Associated with the magnetic braking by the amplified toroidal
magnetic fields, the angular velocity profile of the neutron star is
always modified for the ideal MHD case (see top two panels of
Fig.~\ref{fig2}).\footnote{We note that due to the neutrino cooling,
  the remnant neutron star contracts and its central region spins
  up. The increase of the angular velocity is partly due to this
  effect. For the high-resistivity cases such as models Bh-Rh-Y and
  Bm-Rh-Y, the increase of the angular velocity in the central region
  is caused mainly by this effect: This was confirmed by comparing the
  numerical result with no magnetic fields~\cite{Fujiba2020}.}  As we
mentioned in Sec.~\ref{sec1}, initially, the angular velocity is an
increase function of $\varpi$ in the central region of the neutron
star. This profile is modified and approaches gradually a rigidly
rotating state for the region in which the strong magnetic field is
present. Figure~\ref{fig2} illustrates that the angular velocity in
the central region is increased by a factor of $\sim 2$ by this
effect. Thus, the degree of the differential rotation is reduced by
the winding. The timescale for this process is determined by the
Alfv{\' e}n timescale~\cite{Shapiro2000,SLSS2006}
\beqn
\tau_{\rm a}&=&{R \over B^T/\sqrt{4\pi \rho h}} \nonumber \\
&\approx & 0.11\,{\rm s}
\left({R \over 10\,{\rm km}}\right)
\left({B^T \over 10^{15}\,{\rm G}}\right)^{-1}
\left({\rho \over 10^{15}\,{\rm g/cm^3}}\right)^{1/2},
\eeqn
where $R$, $B^T$, and $\rho$ denote the typical radius, toroidal
magnetic-field strength, and rest-mass density of the neutron star,
and we set $h \approx 1$ for simplicity.
Unless the magnetic-field profile is significantly modified, after the
magnetic braking occurs, the angular velocity should basically show
oscillatory behavior with the timescale of $\tau_{\rm a}$, because the
angular velocity approximately obeys a hyperbolic
equation~\cite{Shapiro2000}. However, in the present context, the
poloidal magnetic flux is significantly escaped by the neutrino-driven
mass outflow and a simple oscillation is not seen.  We also note that
in the non-axisymmetric case, the turbulence and dynamo effects also
would modify the magnetic-field profile significantly~\cite{Sun2019}.


The top and bottom panels of Fig.~\ref{fig3} show the poloidal (left)
and toroidal (right) magnetic-field strength along $x$-direction with
$z=1$\,km (top) and along the $z$-direction with $x=1$\,km (bottom)
for model Rm-B0-Y. The poloidal and toroidal magnetic-field strengths
are defined, respectively, by $\sqrt{\cB^x\cB_x + \cB^z \cB_z}
\gamma^{-1/2}$ and ${\rm sign}(\cB^y)\sqrt{\cB^y\cB_y}
\gamma^{-1/2}(\approx\cB^y \gamma^{-1/3})$.  It is found that the
toroidal-field strength (absolute value) is increased far beyond
$10^{15}$\,G inside the neutron star quite uniformly. By contrast, the
maximum value of the poloidal field strength does not increase but
rather decreases in the late phase. This decrease is due to the fact
that associated with the matter outflow, the magnetic flux is ejected
from the neutron star. By this process, instead, a strong poloidal
field is produced in the polar direction outside the neutron star (see
the bottom panels of Fig.~\ref{fig3}): The typical poloidal-field
strength at the neutron-star pole is $\sim 10^{14}$\,G and $\sim
10^{13}$\,G at $z \sim 100$\,km.

Associated with the mass outflow from the neutron-star surface,
magnetic-fluxes escape toward a large $\varpi$ and large $z$
direction. Because the angular velocity of the escaped matter
decreases with the motion toward $\varpi$-direction, the generation of
a negative toroidal magnetic field is proceeded along the field lines
of the escaped matter. This effect subsequently makes the toroidal
magnetic field in the neutron star negative. For model Bm-R0-Y, the
mass outflow from the neutron-star surface is appreciably caused by 
the neutrino-driven wind. As a result, the region with the negative
toroidal field expands gradually with time (see the top-right panel of
Fig.~\ref{fig3}).  By contrast, for model Bm-R0-N, the neutrino-driven
wind is absent and the mass outflow from the neutron-star surface is
weak.  Consequently, the region with the negative toroidal field
appears only in the vicinity of the neutron star surface (see the
second-top-right panel of Fig.~\ref{fig3}).



As indicated in Fig.~\ref{fig3}, the toroidal magnetic-field energy
dominates over the poloidal magnetic-field one after the significant
winding: The poloidal-field energy is $< 0.1\%$ of the toroidal field
energy.  However, this may be an artifact in axisymmetric computations
in which no dynamo effect is present~\cite{Cowling}. In reality, the
poloidal-field strength may become much higher than the present
results.

\subsubsection{Resistive MHD}\label{sec3-1b}

\begin{figure*}[t]
\includegraphics[width=86mm]{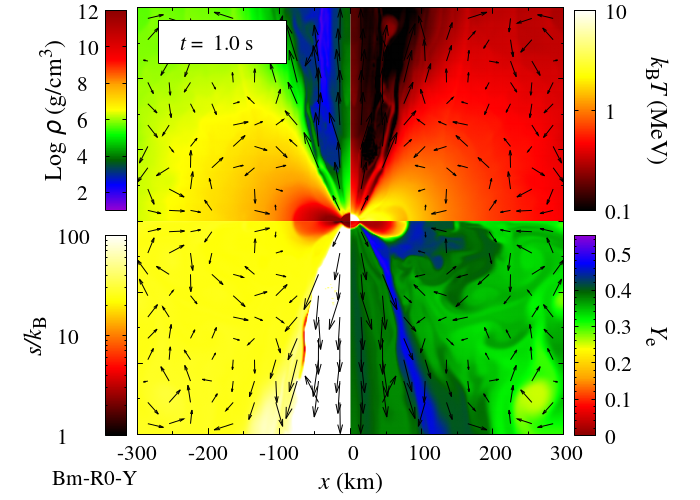}
\includegraphics[width=86mm]{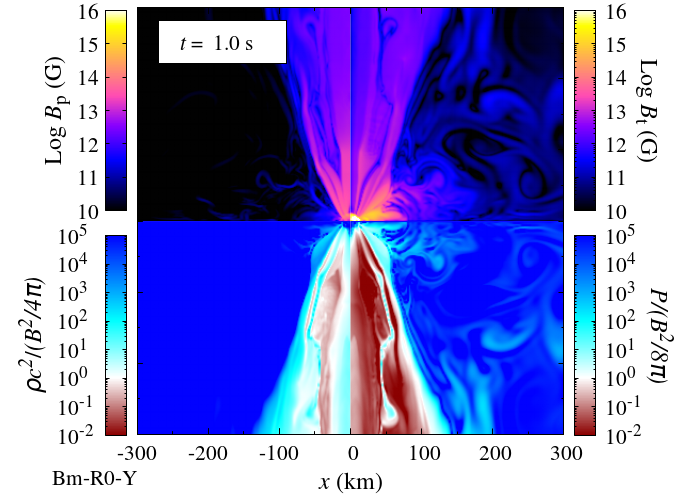}  \\ 
\includegraphics[width=86mm]{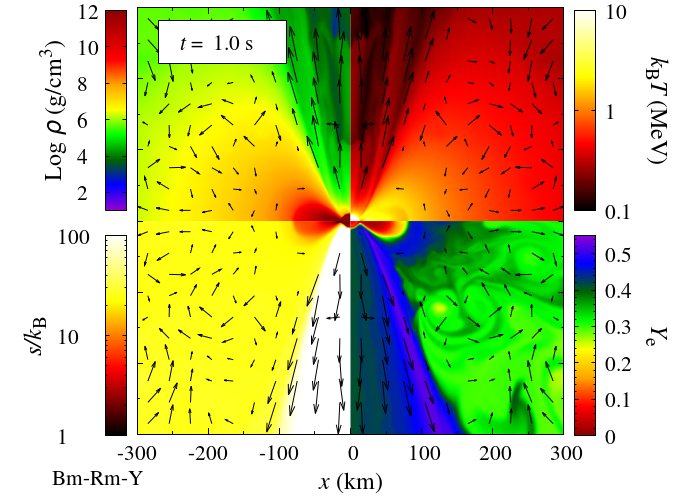} 
\includegraphics[width=86mm]{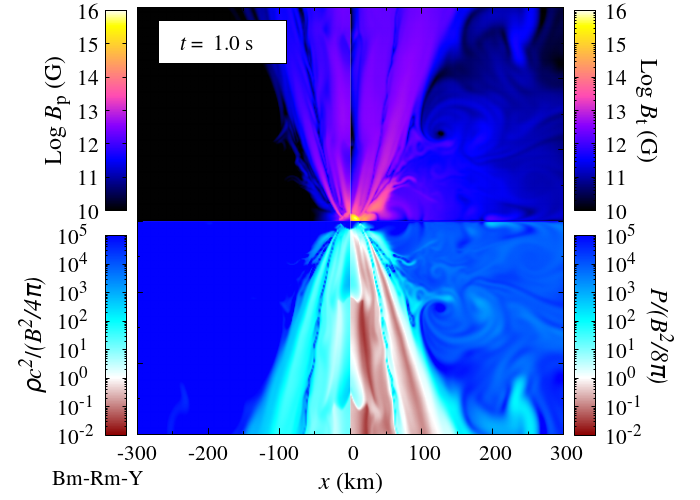} 
\caption{Profiles for the density, temperature, specific entropy,
  electron fraction (left for these four plots), poloidal and toroidal
  magnetic-field strength, the ratio of the rest-mass 
  to magnetic energy density, and the ratio of the gas
  pressure to the magnetic pressure (right for these four plots) after
  the saturation of the growth of the electromagnetic field for models
  Bm-R0-Y (top panels) and Bm-Rm-Y (bottom panels).
\label{fig4}}
\end{figure*}

For the resistive MHD models with low values of the conductivity
(i.e., high resistivity), the magnetic-field strength and the maximum
value of $E_{\rm B}$ reached by the magnetic winding depend on the
values of $\sigma_{\rm c}$, and thus, are different from those for the
ideal MHD case (see the right panel of Fig.~\ref{fig1} as well as
Figs.~\ref{fig2} and \ref{fig3}). Specifically, the maximum
magnetic-field strength depends on the ratio, $\tau_{\rm
  dis}/\tau_{\rm wind}=:R_{\tau}$: For smaller values of $R_\tau$
(i.e., $R_\tau \alt 1$), the maximum values of the magnetic-field
strength and $E_{\rm B}$ are smaller because the resistive dissipation
of the electromagnetic energy proceeds faster than the magnetic
winding. Thus, for the smaller values of the initial electromagnetic
energy, these maximum values are smaller for a given value of
$\sigma_{\rm c}$.  This is found clearly from the results for
$\sigma_{\rm c}=10^7$ and $10^8\,{\rm s}^{-1}$ (compare the results of
models Bh-Rm-Y and Bh-Rh-Y with Bm-Rm-Y and Bm-Rh-Y, respectively).
By contrast, for $R_\tau \agt 1$, the resistive effect is not very
remarkable: For example, the maximum value of $E_{\rm B}$ always
exceeds 1\% of the rotational kinetic energy, $E_{\rm kin}$, because
the magnetic winding sufficiently proceeds until the saturation of the
growth is reached: See the results with $\sigma_{\rm c} \geq
10^9\,{\rm s}^{-1}$.

We note that irrespective of the maximum value of $E_{\rm B}$ reached,
the toroidal magnetic-field energy dominates over the poloidal
magnetic-field one in the end as in the ideal MHD case.  Again, this
could be an artifact in axisymmetric computations with no dynamo
effect in which the poloidal field cannot be amplified~\cite{Cowling}.
In the non-axisymmetric case, this is unlikely to be the case, and
hence, the poloidal field may be also enhanced in the later stage.

For the case that the maximum value of $E_{\rm B}$ is smaller than
$\sim 3\%$ of $E_{\rm kin}$, the effect of the magnetic braking is
weak, and hence, the differential rotation is modified weakly. As
found from the middle-right and bottom-left panels of Fig.~\ref{fig2},
the angular velocity near the rotational axis ($x=0$) does not
increase appreciably for $\sigma_{\rm c} \leq 10^8\,{\rm s}^{-1}$ (see
also footnote 1). By contrast, the angular-velocity profile for
$\sigma_{\rm c} \geq 10^9\,{\rm s}^{-1}$ evolves in a similar manner
to that in the ideal MHD case (compare the top-left and bottom-right
panels or the top-right and middle-left panels of Fig.~\ref{fig2}).

The third row of Fig.~\ref{fig3} shows the evolution of the
magnetic-field profile near the equatorial plane for a resistive model
(Bm-Rm-Y). Figure~\ref{fig3} illustrates that the magnetic-field
strength depends strongly on $\sigma_{\rm c}$: For $\sigma_{\rm c}
\leq 10^8\,{\rm s}^{-1}$, the magnetic field is dissipated in $0.1$\,s
and the resulting field strength is by 1--2 orders of magnitude
smaller than that in the ideal MHD case (compare the top and third-top
panels of Fig.~\ref{fig3}).  For model Bm-Rm-Y, the toroidal field
component varies from the positive to negative values in the neutron
star for late times.  As we mentioned our interpretation for this by an
analysis at the end of Sec.~\ref{sec2}, this reflects the resistive
evolution of the toroidal magnetic field in axial symmetry.

For the case that the resistive dissipation timescale is short, $\alt
0.1$\,s, the thermal energy is generated in the phase in which the
neutrino-driven mass outflow is most efficient. However, for such
cases, $R_\tau$ is small, $< 1$, and hence, the electromagnetic energy
caused by the winding is much smaller than the rotational kinetic
energy and internal energy of the neutron star. Hence, the generated
thermal energy plays only a minor role for the mass outflow and
evolution of the neutron star (although this may contribute a bit to
enhancing the mass ejection).  For $R_\tau \agt 1$, the
electromagnetic energy reached is $\sim 10\%$ of the rotational
kinetic energy. For such cases, however, the dissipation timescale is
quite long; in our present setting, it is longer than $\sim 1$\,s.
Since the timescale for launching the neutrino-driven outflow is
shorter than or as short as this timescale, the thermal energy
generated by the resistive dissipation is not likely to play a major
role for the evolution of the remnant neutron stars and mass ejection.

\subsubsection{Outcome after the winding}


Figure~\ref{fig4} displays the profiles for the density, temperature,
specific entropy, electron fraction (left for these four plots),
poloidal and toroidal magnetic-field strength, the ratio of the
rest-mass to magnetic energy density, and the ratio of the gas
pressure to the magnetic pressure (right for these four plots) after
the saturation of the electromagnetic energy is reached for models
Bm-R0-Y (ideal MHD model: top panels) and Bm-Rm-Y (moderately high
resistivity model: bottom panels). These plots show that the system is
composed of a massive neutron star surrounded by a disk and an outflow
driven from the polar region of the neutron star toward the
$z$-direction irrespective of the presence or absence of the
resistivity.

However, by comparing the profiles for the two models, two
quantitative differences are found. First, the magnetic-field strength
for Bm-R0-Y is stronger than for Bm-Rm-Y as we already mentioned in
Sec.~\ref{sec3-1b} (see Fig.~\ref{fig3}). This is clearly reflected
outside the neutron star in Fig.~\ref{fig4}.  Second, the property of
the outflow driven from the polar surface of the neutron star depends
on the strength of an MHD effect: In the presence of the strong MHD
pressure (i.e., the ideal MHD case), the matter density in the polar
region is smaller, and also, the velocity of the outflow is
higher. This modifies the electron fraction of the matter in the polar
region and ejecta: The electron fraction is lower for model Bm-R0-Y
for which the outflow velocity is higher and the effect of the
neutrino irradiation to the ejecta is weaker (see also
Sec.~\ref{sec3-2}).

\subsubsection{Remark: Comparison with previous work}

In the present work, we find that the massive neutron star does not
contract significantly and its structure is not modified essentially,
after the magnetic winding and braking processes.  This conclusion is
in contrast to previous work~\cite{Duez06a,Duez06b,Duez06c}, which
showed that differentially rotating neutron stars significantly
contract, and if they are hypermassive, they collapse to a black hole, 
after the magnetic winding effect and associated angular momentum
transport by the magnetic braking take place. The major reason for the
absence of the significant contraction is that in the present work, a
realistic angular velocity profile is determined by the merger
simulation (i.e., $d\Omega/d\varpi > 0$ in the central region of the
neutron star) while in Refs.~\cite{Duez06a,Duez06b,Duez06c}, they
assumed that the angular velocity decreases steeply with the
cylindrical radius.  For $d\Omega/d\varpi > 0$, the centrifugal force
in the central region of the neutron star increases after the magnetic
winding and subsequent magnetic braking proceed, whereas for
$d\Omega/d\varpi < 0$ (in the previous work), the centrifugal force in
the central region decreases by the magnetic-field effects, and hence,
the eventual collapse of the hypermassive neutron stars to a black
hole can occur. Moreover, the outward angular momentum transport
proceeds efficiently for $d\Omega/d\varpi < 0$, while for
$d\Omega/d\varpi > 0$ (the present case) an inward angular momentum
transport occurs.  Our present work indicates that the choice of the
angular velocity profile is an important factor for exploring a
realistic evolution process of the hypermassive neutron stars in
numerical simulations.

\subsection{Ejecta}\label{sec3-2}

\begin{figure}[t]
\includegraphics[width=88mm]{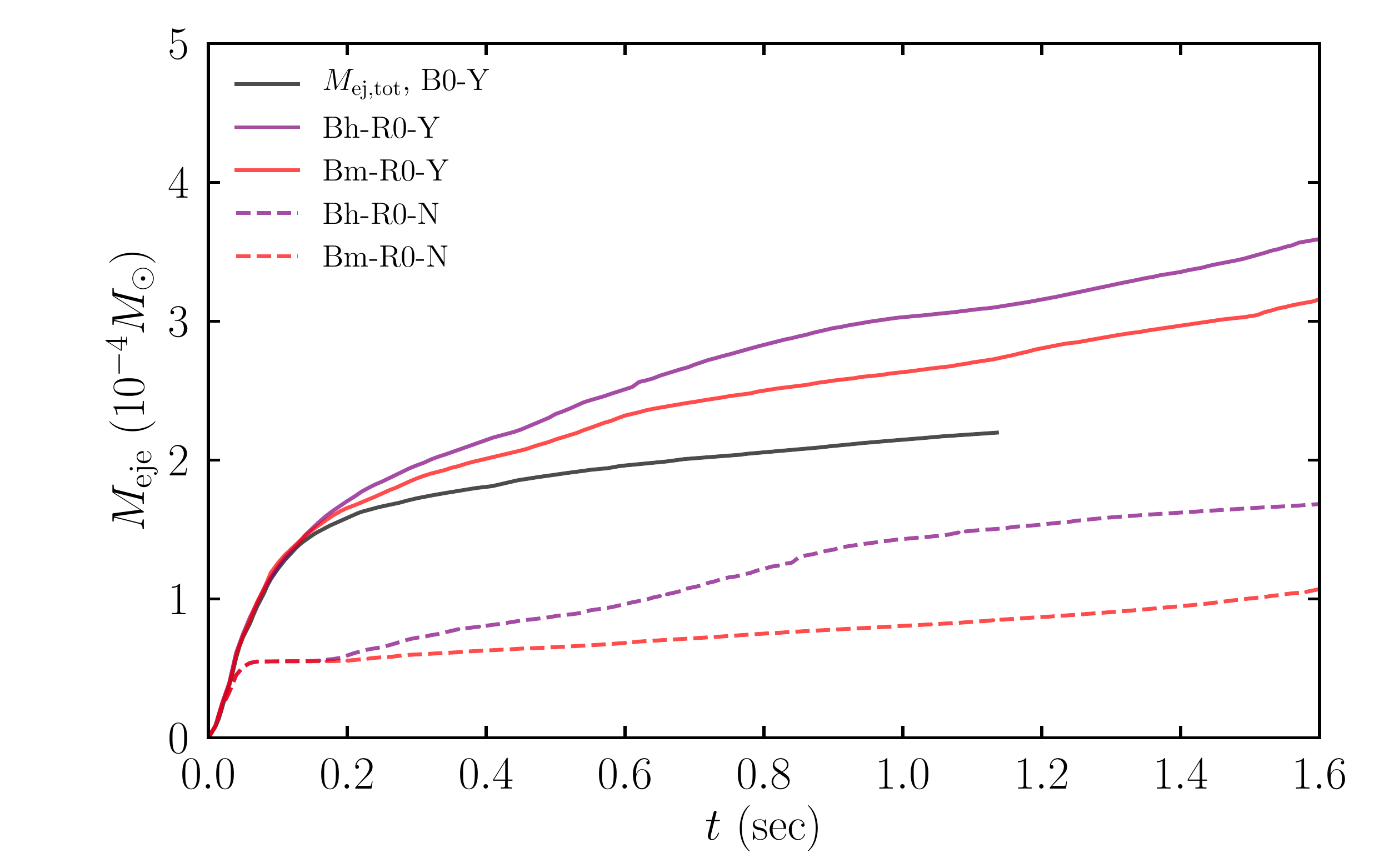}
\caption{Ejecta mass as a function of time for the ideal MHD cases.
  Model B0-Y refers to the result in the absence of the magnetic-field
  effect but with the neutrino effects. 
\label{fig5}}
\end{figure}

\begin{figure*}[t]
\includegraphics[width=88mm]{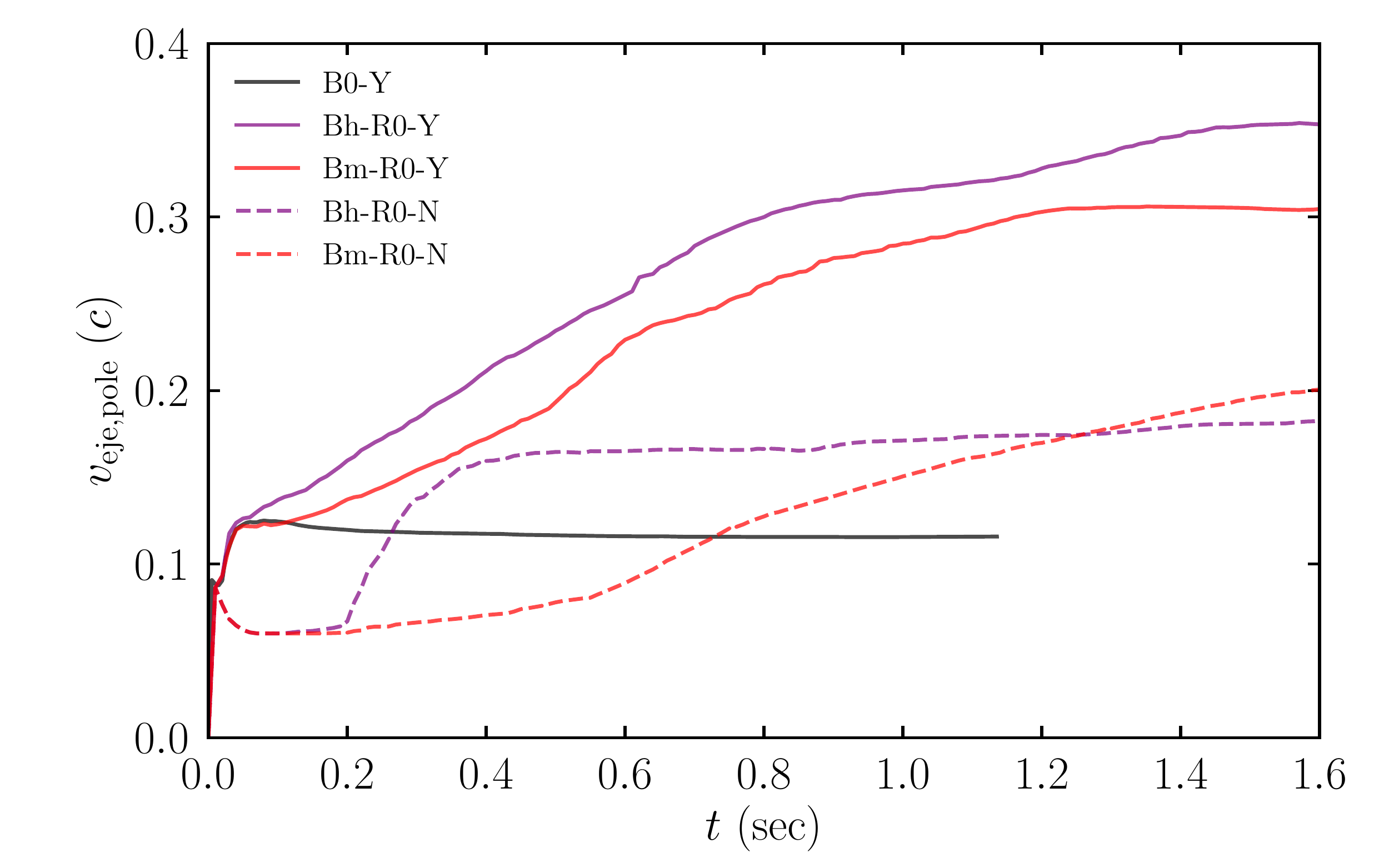}~~ 
\includegraphics[width=88mm]{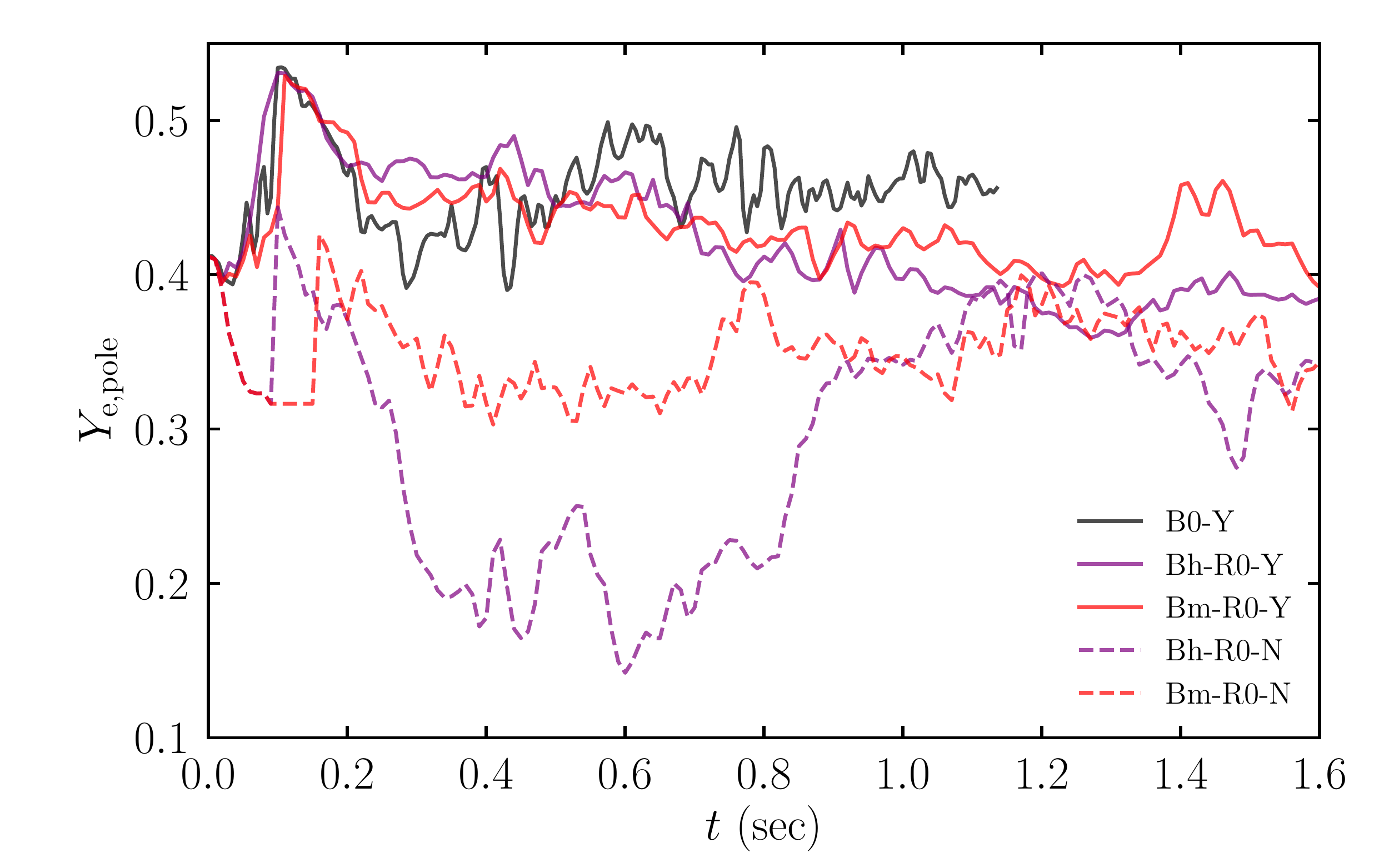} 
\caption{Average velocity (left panel) and electron fraction (right
  panel) of the ejecta as functions of time for the ideal MHD
  simulations and for one simulation without the MHD effect (B0-Y).
\label{fig6}}
\end{figure*}

As we reported in our previous
papers~\cite{Fujiba17,Fujiba18,Fujiba2020}, the mass ejection proceeds
from the remnant of binary neutron star mergers through the neutrino
irradiation/heating even in the absence of any other effects like the
MHD and viscous effects. The rest mass of this neutrino-irradiation
component is not very large as $\alt
10^{-3}M_\odot$~\cite{Fujiba17,Fujiba2020}.  In the context of the
present paper, the mass ejection may be enhanced by the MHD effect
that results primarily from the amplified toroidal magnetic field. In
this subsection, we pay attention to this enhancement.  Because we
finally find that the mass ejection by the magnetic pressure enhanced
by the magnetic winding in the neutron star is a minor effect, we here
pay attention to this topic focusing only on the ideal MHD results.

The ejecta component is determined using the same criterion as in
Refs.~\cite{Fujiba20,Fujiba2020}; we identify a matter component with
$|h u_t| > h_{\rm min}$ as the ejecta. Here $h_{\rm min}$ denotes
the minimum value of the specific enthalpy in the adopted
equation-of-state table, which is $\approx 0.9987c^2$.  For the matter
escaping from a sphere of $r=r_{\rm ext}$, we define the ejection
rates of the rest mass and energy (kinetic energy plus internal
energy) at a given radius and time by
\beqn
\dot M_{\rm eje}&:=&\oint \rho \sqrt{-g} u^i dS_i, \label{ejectrate}
\\
\dot E_{\rm eje}&:=&\oint \rho \hat e \sqrt{-g} u^i dS_i,
\label{ejectrate2}
\eeqn
where $\hat e:=h \alpha u^t-P/(\rho \alpha u^t)$.
The surface integral is performed at $r=r_{\rm ext}$ with
$dS_i=\delta_{ir}r_{\rm ext}^2 \sin\theta d\theta d\varphi$ for the ejecta
component. $r_{\rm ext}$ is chosen to be 1000\,km in this work. 

$\rho \sqrt{-g} u^t(=\rho_*)$ obeys the continuity equation for the
rest mass (see Eq.~(\ref{contin})), and thus, the time integration for
it is a conserved quantity.  Also in the absence of gravity and
magnetic-field effects, $\rho \hat e \sqrt{-g} u^t$ obeys the energy
conservation equation, and far from the central region, the sum of its
time integration and its gravitational potential energy are
approximately conserved (assuming that the electromagnetic energy is
much smaller than the kinetic energy).  Thus, the total rest mass and
energy (excluding the gravitational potential energy and
electromagnetic energy) of the ejecta (which escape away from a sphere
of $r=r_{\rm ext}$) are calculated by
\beqn
M_{\rm eje,esc}(t)&:=&\int^t \dot M_{\rm eje} dt, 
\\
E_{\rm eje,esc}(t)&:=&\int^t \dot E_{\rm eje} dt. 
\eeqn

Far from the central object, $E_{\rm eje,esc}$ is approximated by
\beq
E_{\rm eje,esc}\approx M_{\rm eje,esc}c^2 + U + T_{\rm kin}+{GM M_{\rm
  eje,esc} \over r_{\rm ext}}, \label{Eeje}
\eeq
where $U$ and $T_{\rm kin}$ are the values of the internal energy and
kinetic energy of the ejecta at $r_{\rm ext}\rightarrow \infty$,
respectively. The last term of Eq.~(\ref{Eeje}) approximately denotes
the contribution of the gravitational potential energy of the matter
at $r=r_{\rm ext}$ with $M$ being the total gravitational mass of the
system~\cite{Fujiba20}.  Since the ratio of the internal energy to the
kinetic energy of the ejecta decreases with its expansion, we may
approximate $U/T_{\rm kin} \approx 0$, and hence, $E_{\rm eje,esc}$ by
$E_{\rm eje,esc} \approx M_{\rm eje,esc}c^2 + T_{\rm kin}+GM M_{\rm
  eje,esc}/r_{\rm ext}$ for the region far from the central object.
We then define the average velocity of the ejecta (for the component
that escapes from a sphere of $r=r_{\rm ext}$) by
\beq
v_{\rm eje}:=\sqrt{{2(E_{\rm eje,esc}-M_{\rm eje,esc}c^2-GM M_{\rm
    eje,esc}/r_{\rm ext}) \over M_{\rm eje,esc}}}. 
\eeq

In the setting of the present paper, the mass outflow and resulting
ejection occur only from the polar region of the neutron star
(cf. Fig.~\ref{fig4}), because we do not take into account the
realistic evolution of the disk surrounding the neutron star, and mass
outflow does not occur from it. Thus, in the present work, we perform
the surface integral of Eqs.~(\ref{ejectrate}) and (\ref{ejectrate2})
only for $\theta \leq 15^\circ$ where $\theta$ denotes the polar
angle. Since the initial condition of our simulations is prepared
using the result of a merger simulation for binary neutron stars, the
dynamical ejecta component is present from the beginning in the
computational domain~\cite{Fujiba2020}. However, the dynamical ejecta
component is located primarily near the equatorial plane. Thus, with
the restriction for the surface integral of $\theta \leq 15^\circ$, we
can focus approximately on the post-merger mass ejection, which comes
primarily from the polar region of the neutron star in the present
context.

Figure~\ref{fig5} shows the rest mass of the post-merger ejecta
component as a function of time for four ideal MHD simulations.  For
comparison, we also plot the result in the absence of the magnetic
field (but with the neutrino effects:~model B0-Y)~\cite{Fujiba2020}.
Figure~\ref{fig5} shows that the mass ejection is driven primarily by
the neutrino irradiation/heating and pair-annihilation heating toward
the polar region, because the rest mass of the ejecta in the absence
of these neutrino effects is less than half of that in their presence
and also less than that in the absence of the magnetic-field effect
(model B0-Y). By comparing the results of models Bh-R0-Y and Rm-R0-Y
with that of model B0-Y, it is still found that the magnetic pressure
enhanced by the magnetic winding increases the ejecta mass.  However,
as obviously found from Fig.~\ref{fig5}, the rest mass of this ejecta
component is of the order of $10^{-4}M_\odot$ and by two orders of
magnitude smaller than that of the viscosity-driven ejecta, which
comes from disks (tori) around the neutron
star~\cite{Fujiba18,Fujiba2020}.
Therefore, the MHD effect that stems from the winding in the neutron
star does not contribute appreciably to increasing the ejecta
mass. This is likely to be due to the facts that (i) the MHD effect
can strip the material only in a thin polar surface layer of the
neutron star for which the total rest mass is tiny and (ii) the
gravitational potential near the neutron star is so deep that the mass
ejection cannot efficiently occur from its surface.

The present work indicates that the mass ejection by the MHD effect
from the neutron star is much less significant than the
viscosity-driven mass ejection from disks (tori) surrounding the
neutron star.  These results suggest that the major source of the mass
ejection from the remnant of binary neutron star mergers may be the
disk, not the remnant neutron star. However, this speculation should
be explored in the future by the simulations in which other important
effects such as dynamo effects are taken into account.


A noticeable MHD effect is found in the average velocity and electron
fraction of the ejecta (see Fig.~\ref{fig6}; cf.~Fig.~\ref{fig4} as
well).  In the absence of the MHD effect (model B0-Y), the average
velocity of the ejecta is at most $0.15c$.  However, in the presence
of the strong MHD effect (models Bh-R0-Y and Bm-R0-Y) the average
velocity of the ejecta is enhanced to be $\sim 0.3c$ at $t \sim 1$\,s.
The comparison of the results for models Bh-R0-Y and Bm-R0-Y with
Bh-R0-N and Bm-R0-N (models with no irradiation and pair-annihilation
of neutrinos) shows that the neutrino effects also enhance the
average velocity by the magnitude similar to that by the MHD effect.

Due to the acceleration by the MHD effect, the influence of the
irradiation/heating by neutrinos to the electron fraction of the
ejecta becomes weak: In the absence of the MHD effect, the electron
fraction of the ejecta becomes always high, $\agt 0.45$, due to the
irradiation both by electron neutrinos and anti electron neutrinos,
while in the presence of the MHD effect, the electron fraction becomes
slightly lower, $\sim 0.4$.  For example, for the ideal MHD models
Rh-B0-Y and Rm-B0-Y, the electron fraction in the polar region is as
high as $\agt 0.4$ before the magnetic field is amplified, but it
decreases below $0.4$ after the growth of the magnetic field and
associated onset of the accelerated mass outflow. This is due to the
fact that the neutron-rich nature of the matter outflowed originally
from the neutron-star surface is preserved stronger.  Thus, the MHD
effect in the neutron star does not contribute much to increasing the
ejecta mass, but still has the influence to modify the property of the
outflow and ejecta. However, this effect is not as strong as the
neutrino irradiation/heating effect (compare models Bh-R0-Y and
Bm-R0-Y with Bh-R0-N and Bm-R0-N) as already mentioned.

\section{Summary}\label{sec4}

We performed ideal and resistive MHD simulations for a remnant neutron
star of binary neutron star mergers in general relativity with
neutrino effects.  As a first step, we paid attention to the effect of
the magnetic winding for the evolution of the remnant neutron star and
resulting mass ejection.  The initial matter profile for the
simulations was obtained from a merger simulation. A seed poloidal
magnetic field, for which the electromagnetic energy is much smaller
than the rotational kinetic and internal energy of the system, was
initially superimposed inside the remnant neutron star, and we focused
only on the evolution of it; we did not pay attention to the MHD
evolution of the disk (torus) surrounding the neutron star.

Because of the magnetic winding effect, the toroidal magnetic field is
generated and amplified in this setting. Since the toroidal
magnetic-field strength increases linearly with time, the
electromagnetic-field energy increases as $\propto t^2$ in the early
growth stage. When the electromagnetic-field energy exceeds $\sim 3\%$
of the rotational kinetic energy, the magnetic braking plays an
important role for the redistribution of the angular momentum inside
the neutron star.  For the ideal MHD case, this always occurs and, as
a result of the angular momentum redistribution, the angular velocity,
which is initially an increase function of the cylindrical radius, is
enforced to approach a rigidly rotating state for the region in which
the magnetic braking works well.  The maximum electromagnetic energy
reached is $\sim 10\%$ of the rotational kinetic energy of the neutron
star, i.e., the maximum value of $E_{\rm B}$ is $\sim 10^{52}$\,erg.
By the angular-momentum redistribution, the rotational kinetic energy
of the neutron star is not significantly changed, while at the late
stage, the electromagnetic energy relaxes to $\sim 10^{51}$\,erg after
the magnetic braking is exerted, because the magnetic flux escapes
from the neutron star associated with the mass outflow.

For the resistive MHD simulation, the maximum electromagnetic energy
reached by the winding effect depends strongly on the magnitude of the
conductivity; specifically, the maximum value is determined by the
ratio, $R_\tau=\tau_{\rm dis}/\tau_{\rm wind}$. For $R_\tau \agt 1$,
the magnetic winding proceeds until the electromagnetic energy reaches
$\sim 10\%$ of the rotational kinetic energy as in the ideal MHD case.
On the other hand, for $R_\tau < 1$, the resistive dissipation plays a
role for suppressing the growth of the electromagnetic energy. If the
electromagnetic energy does not reach $\sim 3\%$ of the rotational
kinetic energy, the magnetic braking effect is weak, and hence, the
differentially rotating state of the neutron star is preserved; the
increase of the angular velocity near the central region is
suppressed.

It is also found that the magnetic field amplified by the winding in
the neutron star does not contribute much to enhancing the mass
ejection; for the mass ejection from the neutron star, the neutrino
effects such as neutrino irradiation/heating and pair-annihilation
heating are more important than the MHD effect.  Our interpretation
for this result is that (i) the MHD effect in the neutron star can
strip the material only in the thin polar surface region of the
neutron star for which the total rest mass is tiny and (ii) the
gravitational potential near the neutron star is so deep that the mass
outflow cannot be efficient.  The present result suggests that the
major source of the mass ejection from the remnant of binary neutron
star mergers may not be remnant neutron star but the disk surrounding
it. However, this speculation should be examined in more detail by the
simulations in which other important effects such as dynamo effects
are taken into account.


In the present work, we performed axisymmetric simulations.  This
implies that we might overlook important MHD effects.  One important
missing effect is the turbulence induced by Parker~\cite{Parker} and
Taylor instabilities~\cite{Taylor} (e.g., Ref.~\cite{Kiuchi11}).  As
we showed in this paper, the toroidal magnetic field is inevitably
amplified in the presence of a seed poloidal magnetic field.  It is
well-known that in the presence of the strong toroidal magnetic
fields, (non-axisymmetric) Parker and Taylor instabilities can induce
a turbulence, which could further induce the magnetic-field
amplification through the dynamo effect. It is not clear at all what
happens in such a situation. In addition, it is not very clear what
the final configuration of the magnetic field profile is after the
onset of the turbulence (e.g., Ref.~\cite{Duez}). To explore these
questions, we obviously need three-dimensional MHD simulations in the
subsequent work.  An alternatively phenomenological approach is to
incorporate a dynamo term to induce the turbulence
state~\cite{B2005,BDZ2013}. By this prescription, we do not need
three-dimensional simulations: An axisymmetric simulation would be
reasonable to obtain some insight on the effect of the turbulence. We
plan to perform simulations with a dynamo term in the subsequent work.

\acknowledgments

We thank Kenta Kiuchi and Federico Carrasco for helpful
discussions. MS and SF thank Yukawa Institute for Theoretical Physics,
Kyoto University for their hospitality during the first corona
pandemic time in Germany, in which this project was started.  This
work was in part supported by Grant-in-Aid for Scientific Research
(Grant Nos.~JP16H02183 and JP20H00158) of Japanese MEXT/JSPS.
Numerical computations were performed on Sakura and Cobra clusters at
Max Planck Computing and Data Facility.


\appendix

\section{Test-bed simulations for resistive MHD implementation}\label{app1}

We here present the results for a suit of test-bed simulations
performed with our resistive MHD implementation.  We employed several
test-bed problems introduced in Ref.~\cite{BDZ2013} (see also
Refs.~\cite{Komissarov,DZ09,PLRR2009,Qian2017}): the problems of
self-similar current sheet, resistive shock tube, and resistive
rotor. We also performed test simulations for the propagation of the
electromagnetic wave packet in the flat spacetime and for a dynamo
closure.

\subsection{Self-similar current sheet} 

\begin{figure}[t]
\includegraphics[width=86mm]{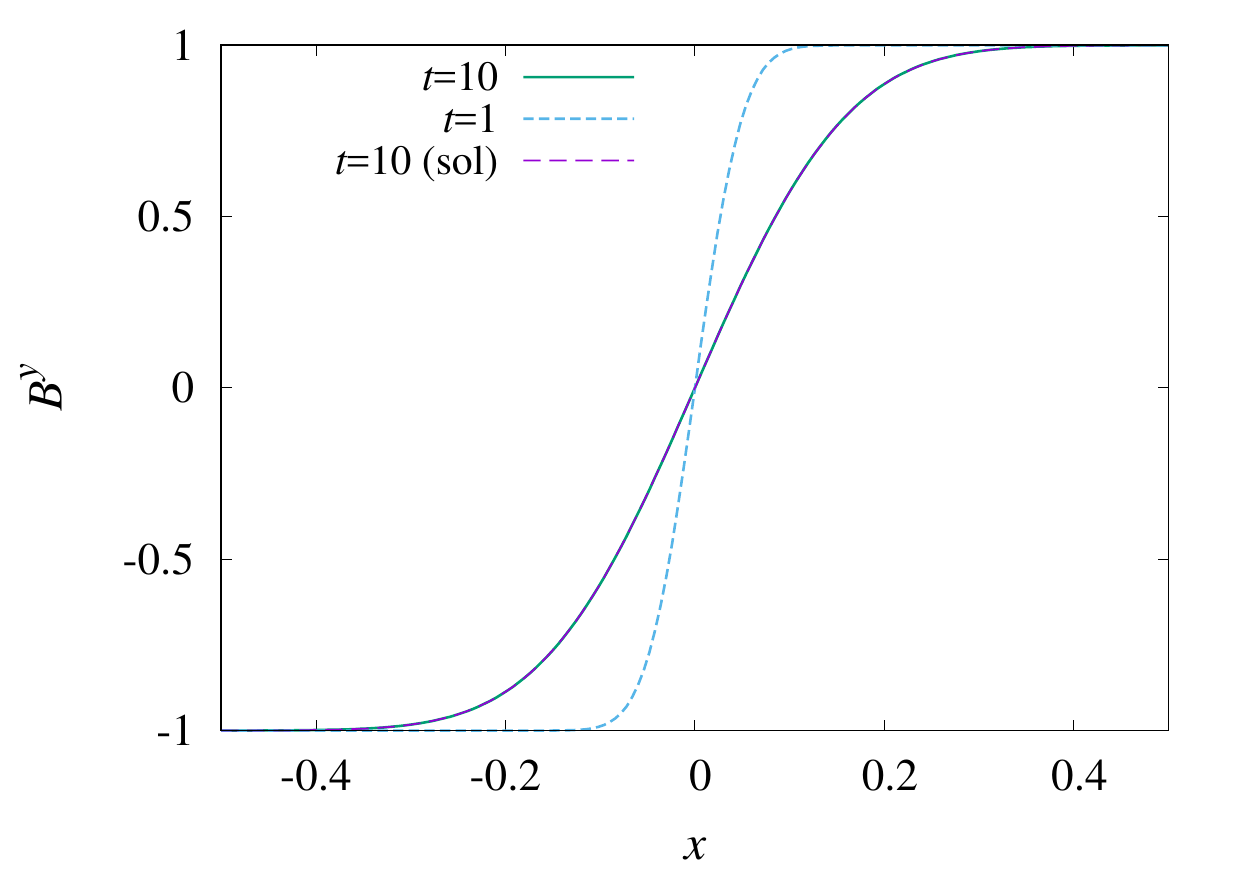} 
\caption{Numerical solution of $B^y$ at $t=1$ (dotted curve) and 10
  (solid curve) with the analytic solution of Eq.~(\ref{eqa1}) at
  $t=10$ (dashed curve).
\label{figa1}}
\end{figure}

This test-bed problem was first proposed in Ref.~\cite{Komissarov} and
subsequently employed in many
references~\cite{DZ09,PLRR2009,BDZ2013,Qian2017}.  This is the test to
check whether the magnetic field correctly diffuses out with the
special-relativistic resistive MHD implementation in the limit that
the gas pressure is much higher than the magnetic pressure. In the one
dimensional problem in which all the quantities depend only on the
$x$ coordinate, an approximate (nearly exact) solution is written,
e.g., in the following form:
\beq
B^y(x,t)=C\, {\rm erf}\left({\sqrt{\sigma_{\rm c}} x \over \sqrt{t}} \right),
\label{eqa1}
\eeq
where erf denotes the error function and $C$ is a constant with
$C^2/8\pi$ much smaller than the gas pressure. Here we set $C=1$ as
the choice of the units for simplicity.  Note that our notation for
the basic MHD equations is different from the previous papers because
of the presence of the factor $4\pi$ in front of $j^\mu$ in
Eq.~(\ref{eq:maxcell1}).

Following Ref.~\cite{BDZ2013}, we employ the initial condition at
$t=1$ as $\rho=1$, $p=50$, $E^i=1$, $u_i=0$, $B^x=B^z=0$, and
$B^y=B^y(x,t=1)$.  The $\Gamma$-law equation of state,
$p=(\Gamma-1)\rho\varep$ with $\Gamma=4/3$ is employed. $\sigma_{\rm
  c}$ is set to be $100$ (the resistivity is $1/(4\pi\sigma_{\rm
  c})$).  The computational domain is set up as $x=[-0.6:0.6]$ which
is covered by 201 uniform grid points.  The simulation is performed until
$t=10$. Figure~\ref{figa1} plots $B^y$ at $t=1$ and $10$. For $t=10$,
we also plot the functional form of $B^y$ described in
Eq.~(\ref{eqa1}). Figure~\ref{figa1} illustrates that our
implementation can reproduce the solution well. In this setting the
maximum relative error is $\approx 2.5 \times 10^{-4}$.

\subsection{Resistive shock tube} 

\begin{figure}[t]
\includegraphics[width=86mm]{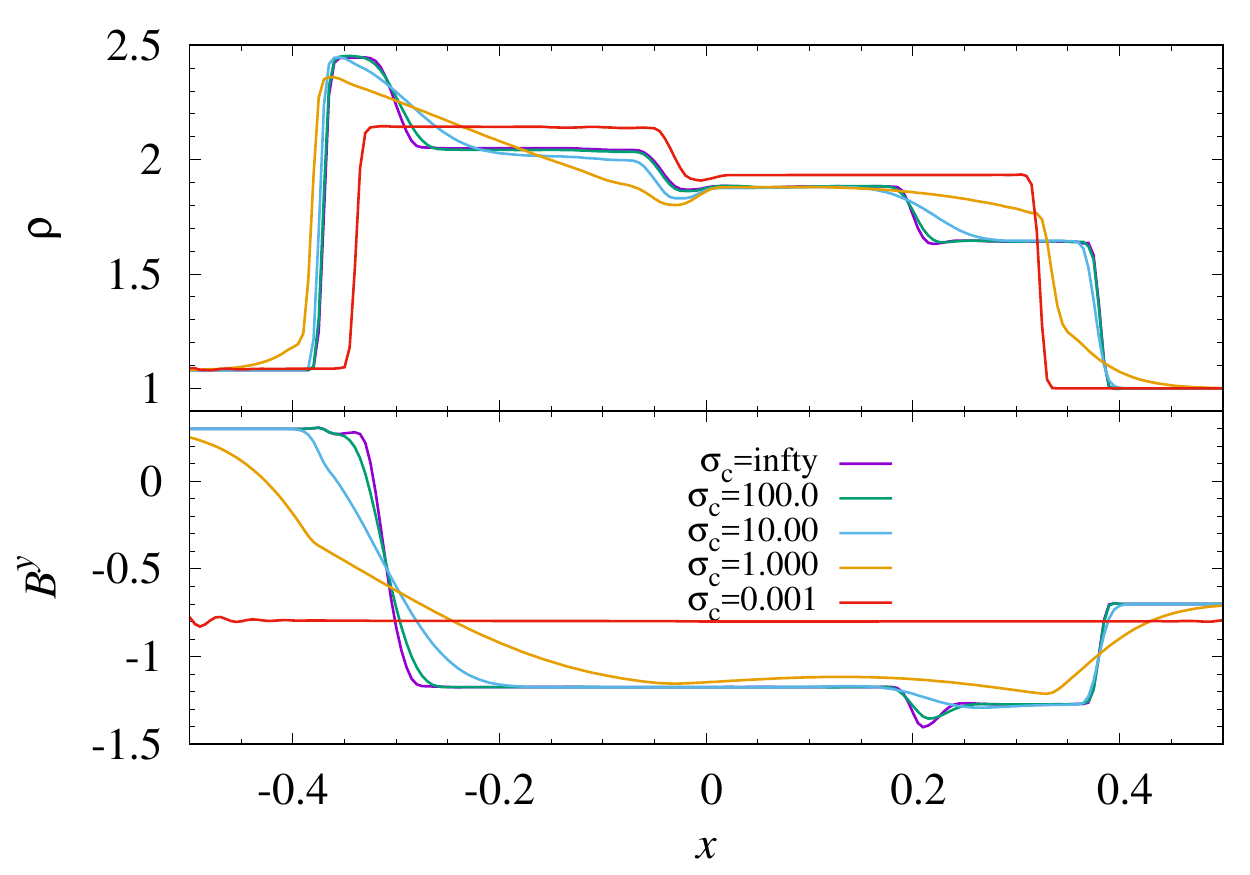} 
\caption{Numerical solutions of $\rho$ (upper panel) and $B^y$ (lower
  panel) for the shock tube test with $\sigma_{\rm c}=10^{10}$,
  $10^2$, 10, 1, and $10^{-3}$. The lower panel plots $B^y$ (not $\hat
  B^y$).
\label{figa2}}
\end{figure}

\begin{figure*}[t]
\vspace{-0.4cm}
\includegraphics[width=85mm]{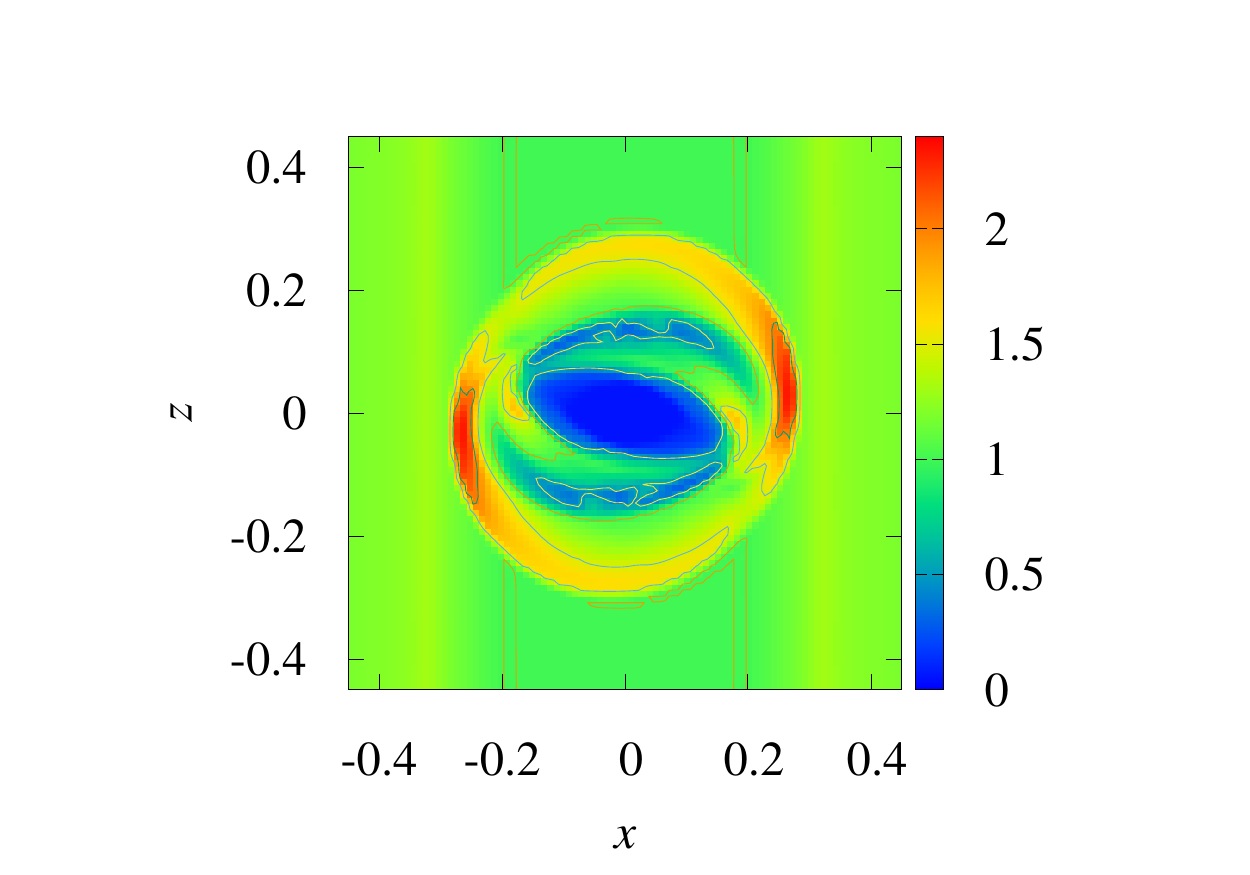} 
\hspace{-2.cm}
\includegraphics[width=85mm]{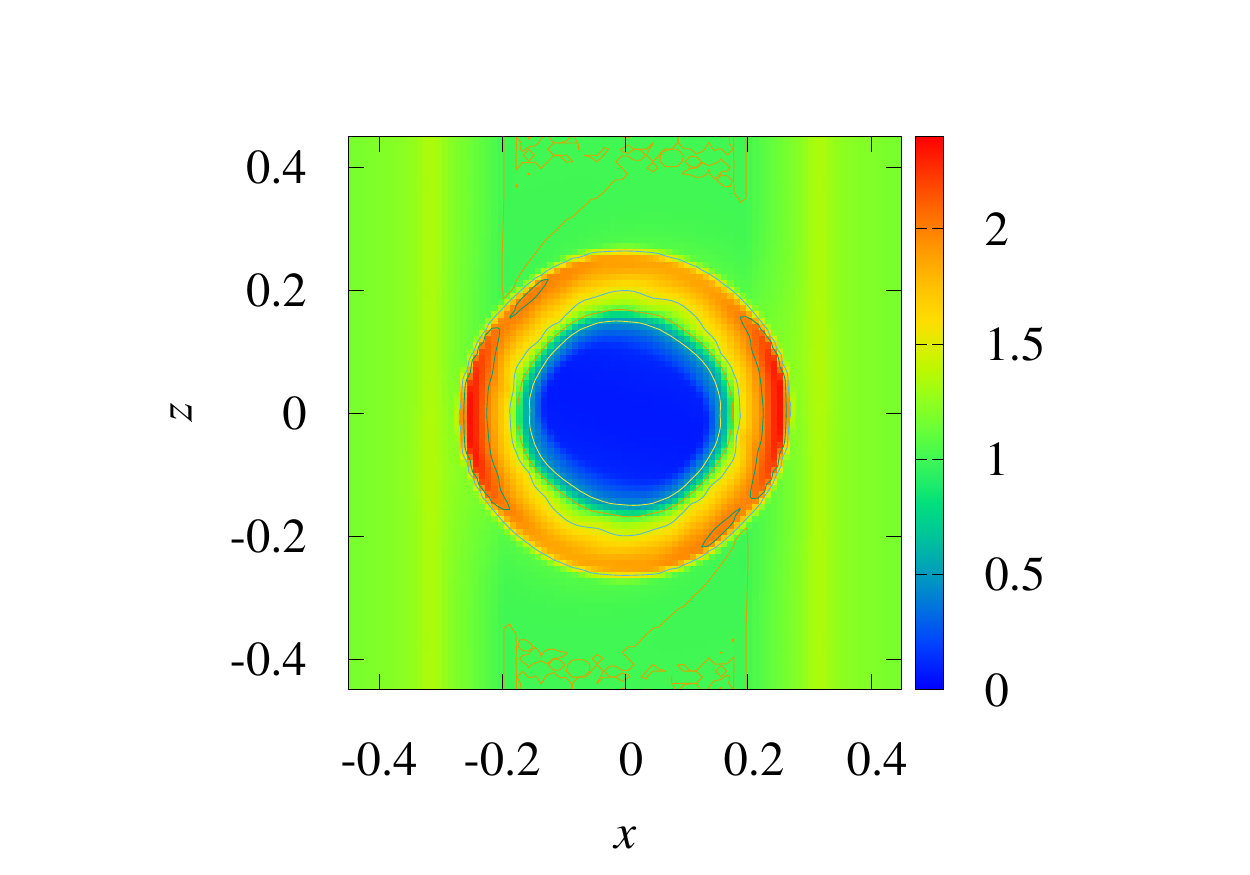} \\
\vspace{-0.7cm}
\includegraphics[width=85mm]{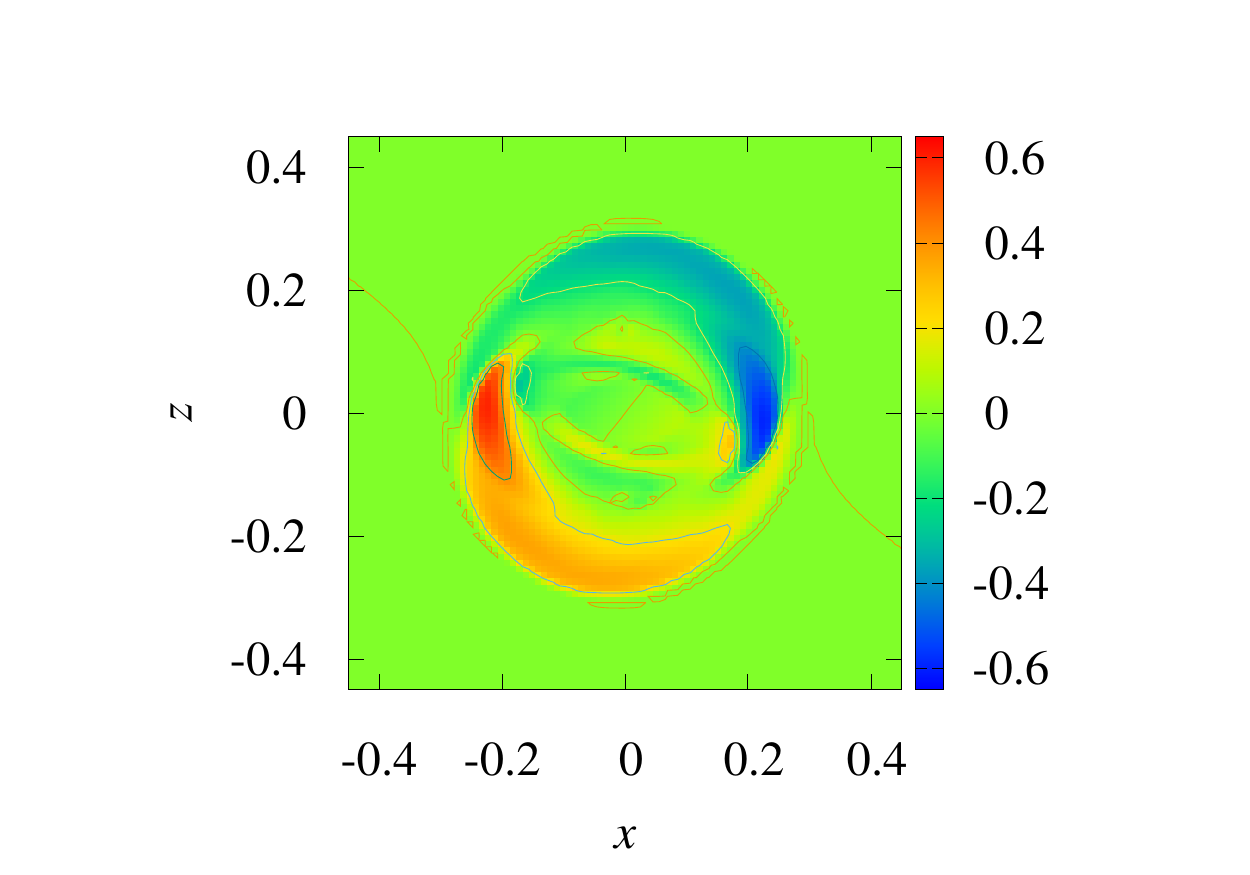} 
\hspace{-2.cm}
\includegraphics[width=85mm]{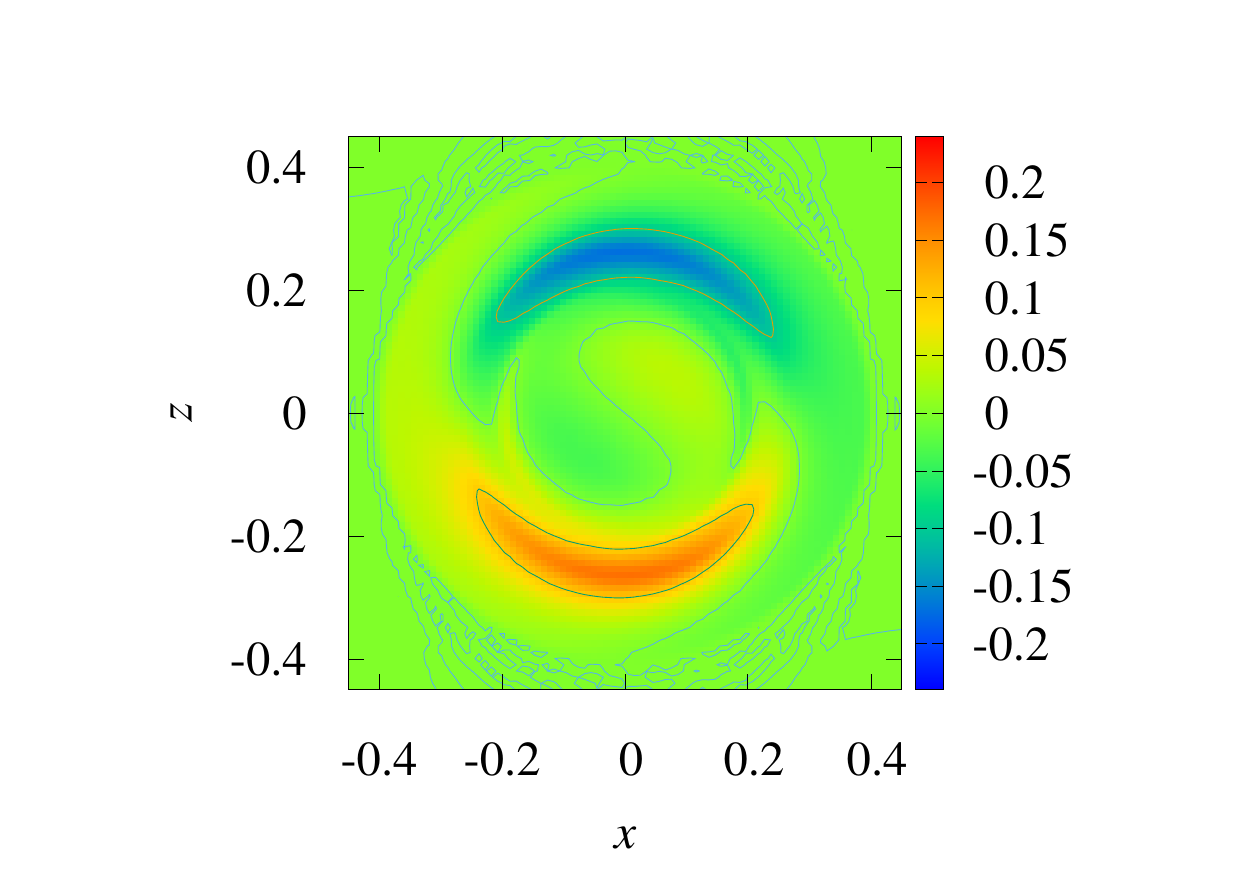} 
\vspace{-0.2cm}
\caption{The results of the resistive rotor problem for $\sigma_{\rm
    c}=10^{10}$ (left) and $1$ (right).  The upper and lower panels
  show $p$ and $E^z$, respectively.
\label{figa3}}
\end{figure*}

This problem was first proposed in Ref.~\cite{DZ09}.
The initial condition for this problem proposed in Ref.~\cite{BDZ2013} is
\beqn
&&(\rho, p, v^x, v^y, v^z, \hat B^x, \hat B^y. \hat B^z) \nonumber \\
&&=\left\{
\begin{array}{ll}
(1.08, 0.95, 0.4, 0.3, 0.2, 2.0, 0.3, 0.3) & {\rm for}~x < 0, \\
(1.0, 1.0, -0.45, -0.2, 0.2, 2.0, -0.7, 0.5) & {\rm for}~x \geq 0, \nonumber \\
\end{array}
\right.
\eeqn
where $\hat B^i=B^i/\sqrt{4\pi}$ (this renormalization is necessary to
align the unit with the previous work) and $v^i=u^i/u^t$. The initial
electric field is given by $E^i=-\epsilon^{ijk}v_j B_k$.  The
$\Gamma$-law equation of state with $\Gamma=5/3$ is employed.
$\sigma_{\rm c}$ is chosen to be $10^{-3}$, 1, 10, $10^2$, and
$10^{10}$.  The computation is performed from $t=0$ to $0.55$ covering
the computational domain of $x=[-0.5:0.5]$ by 401 uniform grid points.
The results are shown for $\rho$ and $B^y$ in Fig.~\ref{figa2} as in
Ref.~\cite{BDZ2013}. A small unphysical oscillation could be seen for
the curves of $\rho$ and $B^y$ in the absence of the Kreiss-Oliger
dissipation for evolving Maxwell's equations. However, this
dissipation term with a small coefficient of $1/640$ cures the
unphysical oscillation.  We find that the numerical solutions are
quite similar to those found in Ref.~\cite{BDZ2013}.

\subsection{Resistive rotor}

This is a two-dimensional problem with the coordinates $(x, z)$ and
the initial condition proposed in Ref.~\cite{BDZ2013} is given in the
following manner. Inside a radius of $r=\sqrt{x^2+z^2} \leq 0.1$, a
uniformly rotating medium with $\rho=10$ and the uniform angular
velocity of $\Omega=8.5$ is prepared (i.e., $v^x=-z \Omega$ and
$v^z=x\Omega$). Outside the circle of $r=0.1$, on the other hand, we
set $\rho=1$ and $\Omega=0$. For the entire region, the pressure and
the magnetic field are initially uniform as $p=1$ and $(\hat B^x, \hat
B^y, \hat B^z)=(1, 0, 0)$.  The initial electric field is again
determined by $E^i=-\epsilon^{ijk}v_j B_k$.  The $\Gamma$-law equation
of state with $\Gamma=4/3$ is employed, and the system is evolved from
$t=0$ to $0.3$.

The computational domain is chosen to be $x=[-0.5:0.5]$ and
$z=[-0.5:0.5]$ and is covered by a uniform grid of $401 \times 401$
points. This problem is numerically solved for $\sigma_{\rm
  c}=10^{10}$ (nearly ideal MHD case) and $1$. The results for the
profiles of the pressure and $z$-component of the electric field are
shown in Fig.~\ref{figa3}. By comparing these results with Fig.~3 of
Ref.~\cite{BDZ2013}, we find a good (qualitative/semi-quantitative)
agreement.

\begin{figure*}[t]
\includegraphics[width=88mm]{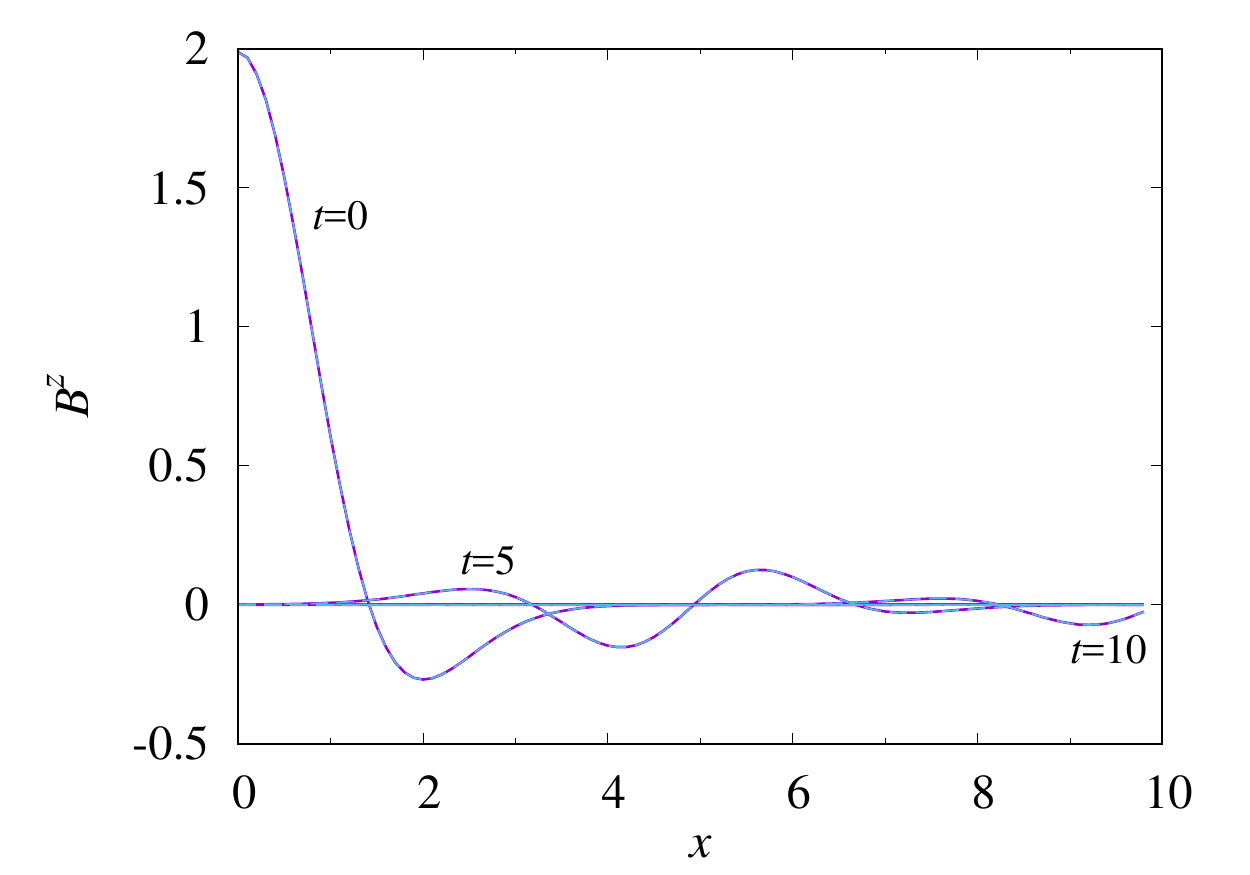} 
\includegraphics[width=88mm]{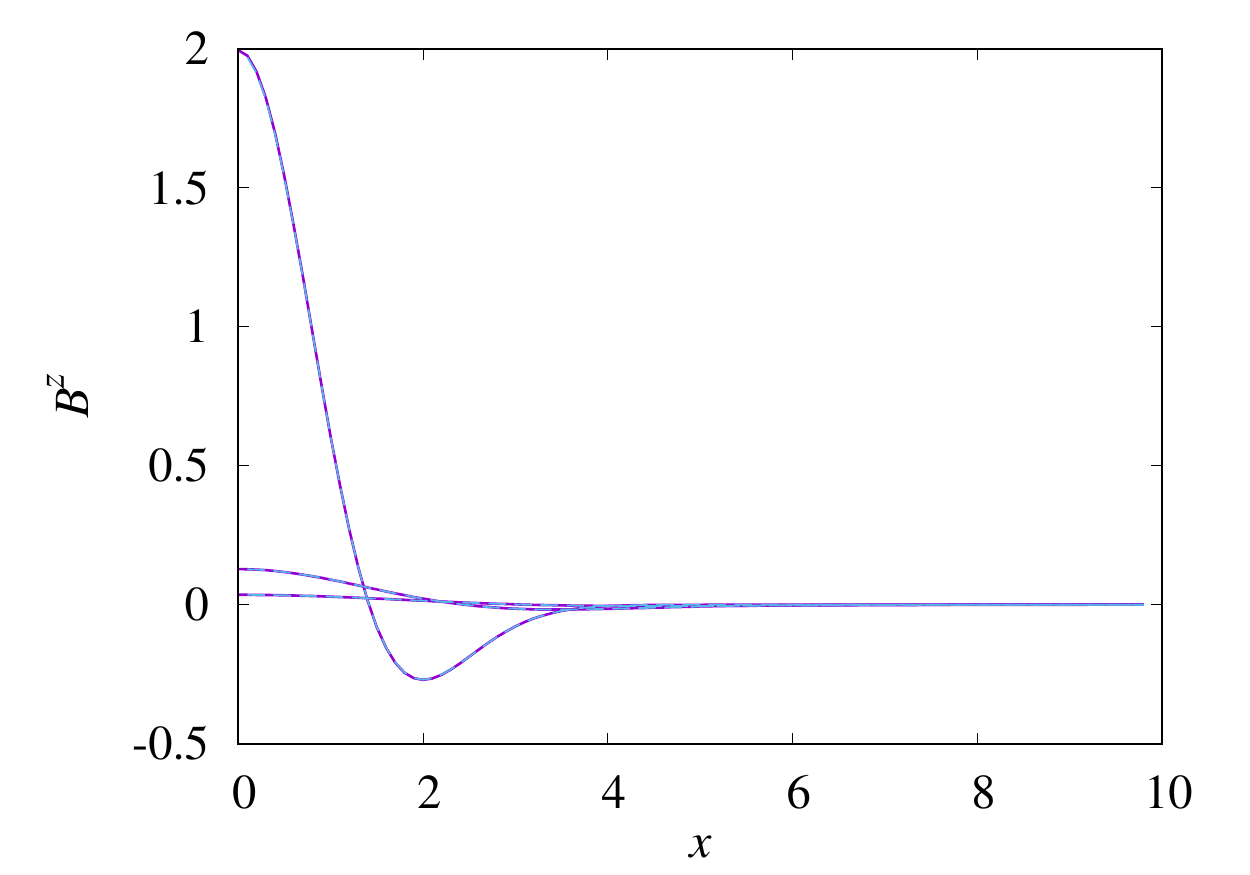} 
\caption{Left: Propagation of electromagnetic waves ($B^z$) in the
  flat spacetime for $\sigma_{\rm c}=0$ at $t=0$, 5, 10, and 15.  At
  $t=15$, the amplitude is much smaller than unity entirely.  The
  solid and dashed curves show the numerical and exact solutions,
  respectively.  Right: Dispersion of electromagnetic waves in the
  flat spacetime for $4\pi \sigma_{\rm c}=50$ at $t=0$, 50, and 100.
  The solid and dashed curves show the numerical and exact solutions,
  respectively (both curves approximately agree). For $x< 0$, the
  solution has the reflection symmetry for both panels (and we do not
  plot it).
\label{figa4}}
\end{figure*}

\subsection{Evolving electromagnetic fields}

To check that our implementation for evolving the electromagnetic
fields works well in a practical computational domain, we also
performed several test-bed simulations in three spatial dimensions
varying the value of $\sigma_{\rm c}$ from $0$ to high values.  For
$\sigma_{\rm c}=0$ and $\sigma_{\rm c} \rightarrow \infty$ in the flat
spacetime of $u_i=0$, it is easy to derive analytic solutions for the
electromagnetic equations, and thus, we compare the results of the
simulations with these analytic solutions.

For $\sigma_{\rm c}=0$, the electromagnetic fields obey wave equations
in vacuum, and general solutions for the basic equations are easily
derived. For example, an axisymmetric general solution for the dipole
radiation is written as
\beqn
B^r&=&{1 \over r}{\pa \over \pa r}\left({f(r+t)-f(r-t) \over r}\right)\cos\theta,\nonumber \\
B^\theta&=&-{1 \over 2r^2} {\pa \over \pa r}\left[
r{\pa \over \pa r}\left({f(r+t)-f(r-t) \over r}\right)\right]\sin\theta,~~
\eeqn
and $B^\varphi=0$. For this case, $E^\varphi$ is only non-zero
component of $E^i$ and it is derived straightforwardly from the field
equation.  Here, $f(u)$ is an arbitrary regular function, and for
$f(u)=- u e^{-u^2/2}$, the initial condition becomes
\beqn
B^x&=&x z e^{-r^2/2},\nonumber \\
B^y&=&y z e^{-r^2/2},\nonumber \\
B^z&=&(2-x^2-y^2) e^{-r^2/2}, \label{initB}
\eeqn
and $E^i=0$. Note here that we choose the width of the wave packet
(i.e., unity) as the unit of the length in this problem.

We then evolve this wave packet in the domain of $x=[-10:10]$,
$y=[-10:10]$, and $z=[0:10]$ with the grid spacing of $dx=0.1$.  The
reflection-symmetric boundary condition is imposed on the $z=0$ plane
and an outgoing wave boundary condition is imposed on the outer
boundaries. The left panel of Fig.~\ref{figa4} plots the numerical
profiles of $B^z$ along the $x$ axis at $t=0$, 5, 10, and 15 (solid
curves) together with the analytic solutions (dashed curves). Note
that at $t=15$, the wave packet already propagated away from the
computational domain. The relative error of the numerical solution to
the analytic one, e.g., at $t=5$, is of order $10^{-3}$ with this
setting (except for the outer boundaries).  This illustrates that our
implementation can well follow the propagation of the electromagnetic
waves.

For large values of $\sigma_{\rm c}$, on the other hand, the field
equations relax to parabolic ones. For the parabolic equations with
the initial condition of Eq.~(\ref{initB}), a solution for the dipole
magnetic field is written as
\beqn
B^r&=&2\left(1 + {t \over 2\pi \sigma_{\rm c}} \right)^{-5/2}
\exp\left(-{r^2 \over 2 + t/\pi\sigma_{\rm c}}\right) \cos\theta,
\nonumber \\
B^\theta&=&-{1 \over r} \left(1 + {t \over 2\pi \sigma_{\rm c}} \right)^{-5/2}
\left(2 - {r^2 \over 1+t/2\pi\sigma_{\rm c}}\right) \nonumber \\
&& \times \exp\left(-{r^2 \over 2 + t/\pi\sigma_{\rm c}}\right) \cos\theta,
\nonumber \\
B^\varphi&=&0. 
\eeqn

We again evolve this wave packet in the domain of $x=[-10:10]$,
$y=[-10:10]$, and $z=[0:10]$ with $dx=0.1$ and $4\pi \sigma_{\rm
  c}=50$.  The right panel of Fig.~\ref{figa4} plots the profiles of
$B^z$ along the $x$ axis at $t=0$, 50, and 100 (solid curves) together
with the analytic solutions (dashed curves). With the chosen grid
resolution, the maximum relative error size is $\approx 4\%$ at
$t=100$ (except for the region near the outer boundaries). The largest
error is always located near the region at which $B^z=0$, and for
other regions, the accuracy is much better.

\subsection{Steady dynamo}\label{app:dynamo}

\begin{figure}[t]
\includegraphics[width=88mm]{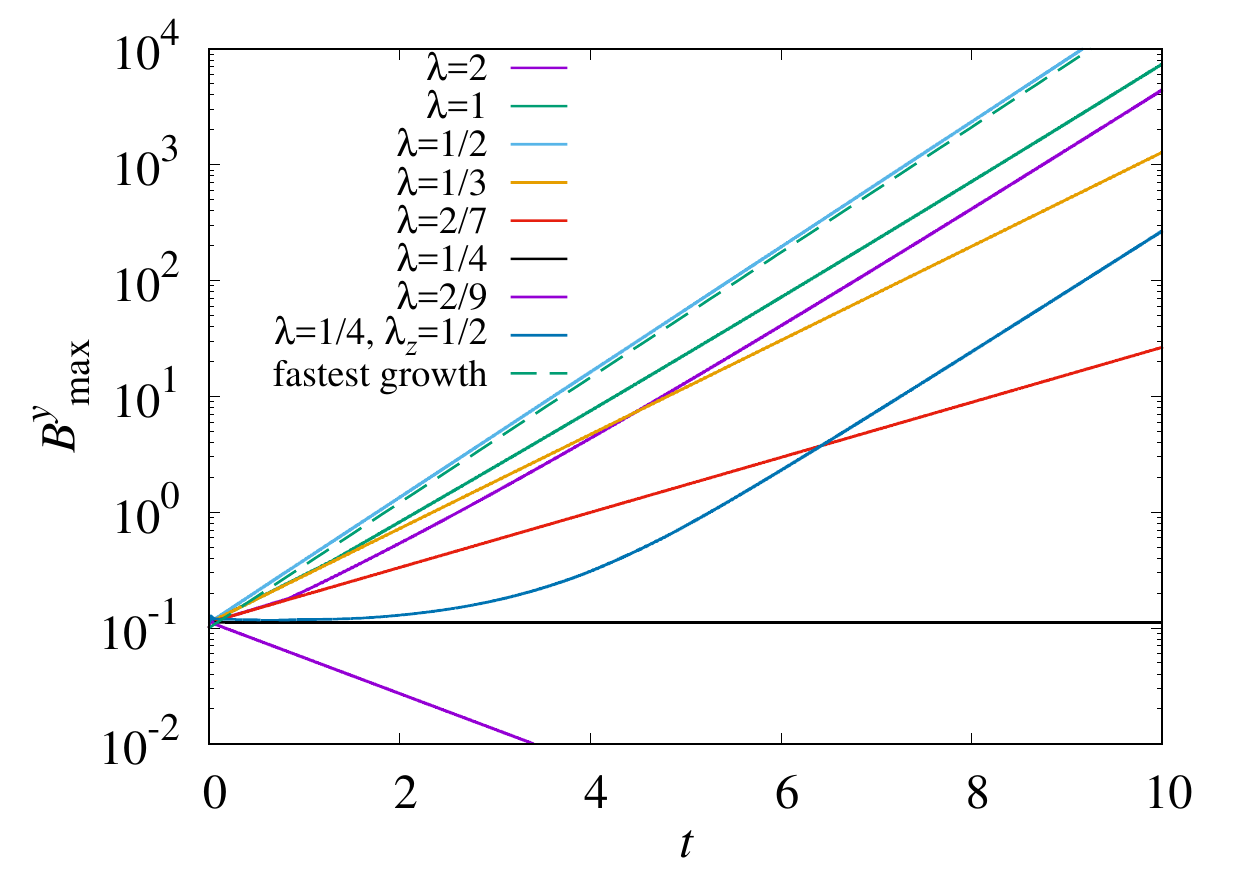} 
\caption{The maximum value of $B^y$ for the steady dynamo test for
  $\sigma_{\rm c}=10$ and $\alpha_{\rm d}=0.2$ with $\lambda_x=2$,
  $1$, $1/2$, $1/3$, $2/7$, $1/4$, $2/9$, $a=0.1$, and $b=0$ (no
  perturbation in the $z$ direction) and with $\lambda_x=1/4$,
  $\lambda_z=1/2$, $a=0.1$, and $b=0.01$. Note that the fastest
  growing mode has the wavelength of $1/2$ while for the marginally stable 
  mode it is $1/4$. $\lambda$ in the inset of the plot implies $\lambda_x$. 
\label{figa6}}
\end{figure}

Following Ref.~\cite{BDZ2013}, we also performed this simple test to
check that our implementation works well in the presence of a dynamo
term.  In this test-bed problem, again, we pay attention only to
solving the electromagnetic field setting $u_i=0$.

First we assume that $B^x=E^x=0$, and $B^y$, $B^z$, $E^y$, and $E^z$
are functions of
$\exp(i\omega t - ikx)$ where $\omega$ and $k$ are the angular
frequency and the wave number. Then, the following dispersion relation is
derived:
\beqn
\omega^2 - 4\pi i \sigma_{\rm c} \omega - k^2 \pm 4\pi\sigma_{\rm c} \alpha_{\rm d} k=0.
\eeqn
For this equation, we find that the exponential growth mode is present
for the case that $k < 4\pi \sigma_{\rm c}\alpha_{\rm d}$, and the
resultant expression of $\omega$ is
\beq
\omega= 2\pi i \sigma_{\rm c} \left( 1 - \sqrt{1 - (k/2\pi\sigma_{\rm c})^2 + k \alpha_{\rm d}/\pi\sigma_{\rm c}} \right),
\eeq
and for the fastest growing mode with $k=2\pi\sigma_{\rm c}\alpha_{\rm
  d}$, this becomes
\beq
\omega=-2\pi i \sigma_{\rm c} \left(\sqrt{1 + \alpha_{\rm d}^2} -1\right). 
\eeq

Because the basic equation is linear in $E^i$ and $B^i$, it is
straightforward to extend this analysis in the multidimensional case
as long as we focus only on the transverse component. That is, for the
case that we initially prepare two independent modes proportional to,
e.g., $\exp(i\omega t - i k_xx)$ and $\exp(i\omega t - i k_zz)$ where
$k_x$ and $k_z$ are the wave numbers, the stability of the system is
determined simply by analyzing the dispersion relations in the $x$ and
$z$ direction independently.  Taking into account this fact, we
perform two-dimensional simulations setting the region of $x=[-1:1]$
and $z=[-1:1]$ with the periodic boundary conditions in both
coordinates.

We then prepare the following initial condition:
\beqn
B^x&=&-b \cos(k_z z),\\
B^y&=& a \sin(k_x x) + b \sin(k_z z),\\
B^z&=&-a \cos(k_x x),
\eeqn
and $E^i$ is determined by the ideal MHD condition. Here, $a$ and $b$
are constants. In this setting, for the case that $2 \lambda_i
\sigma_{\rm c} \alpha_{\rm d} > 1$, the initial seed field should grow
exponentially with time, where $\lambda_i:=2\pi/k_i$ ($i$ is $x$ or
$z$) denotes the wave length.  Note that our setting of the
computational domain can follow the waves only with $\lambda_i \leq
2$.  Thus, the simulations are performed for $\sigma_{\rm c}=10$ and
$\alpha_{\rm d}=0.2$.  For this setting, the fastest growing mode has
$\lambda_i=1/2$ while the marginally stable mode has $\lambda_i=1/4$
(i.e., the mode with $\lambda_i > 1/4$ is unstable). 

Figure~\ref{figa6} plots the evolution of the maximum value of $B^y$
for $\sigma_{\rm c}=10$ and $\alpha_{\rm d}=0.2$ with $\lambda_x=2$,
$1$, $1/2$, $1/3$, $2/7$, $1/4$, $2/9$, $a=0.1$, and $b=0$ (no
perturbation in the $z$ direction) and with $\lambda_x=1/4$,
$\lambda_z=1/2$, $a=0.1$, and $b=0.01$. The growth rates of the
unstable modes are well captured in each simulation.  This figure
illustrates that our implementation can derive the predicted results
irrespective of the prepared initial conditions.


\section{Resistive evolution of the massive neutron star with purely toroidal magnetic field}\label{app2}

\begin{figure}[t]
\vspace{-2mm}
\includegraphics[width=88mm]{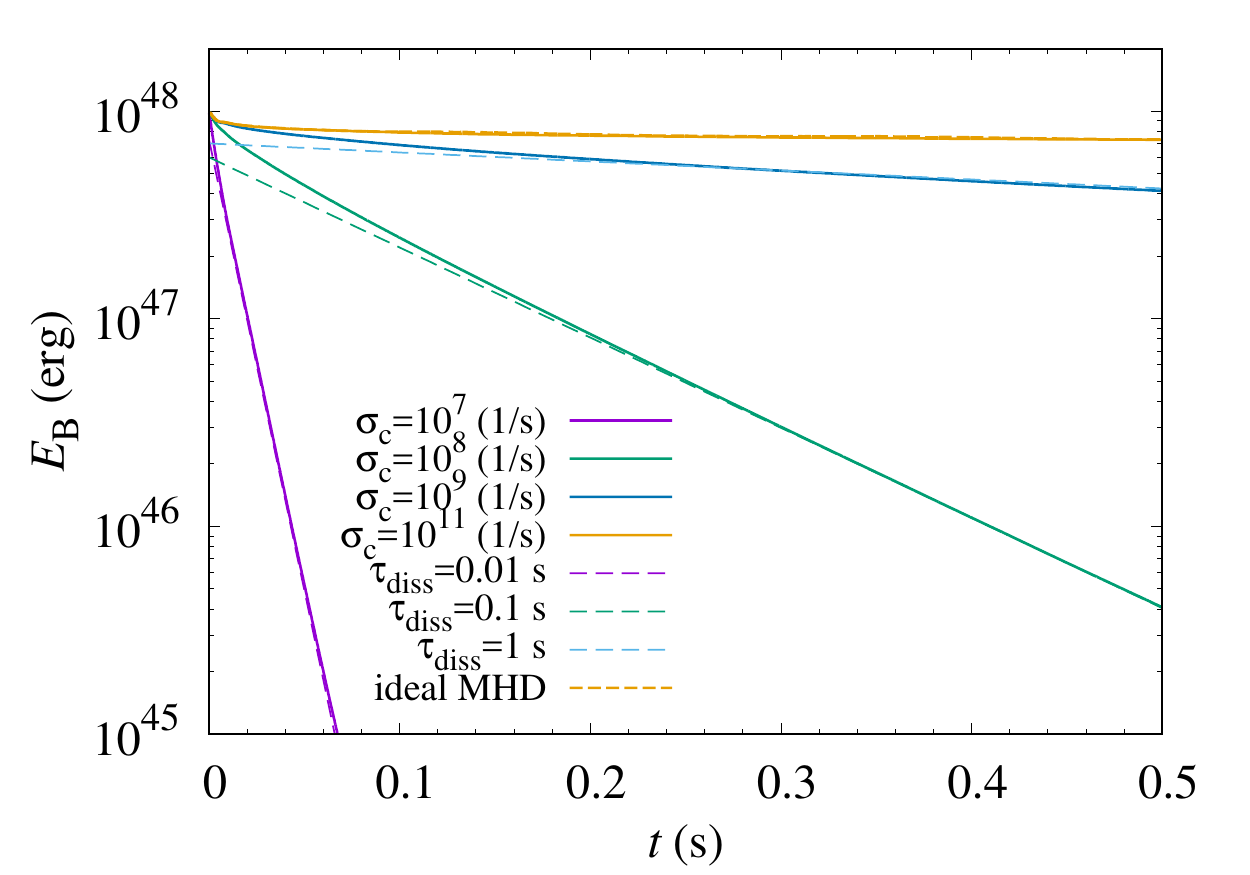} 
\caption{Evolution of the magnetic-field energy for the remnant
  massive neutron star initially with a purely toroidal magnetic field
  for $\sigma_{\rm c}=10^7$, $10^8$, $10^9$, and $10^{11}\,{\rm
    s}^{-1}$ (solid curves). The dashed lines denote $\propto
  \exp(-t/\tau_{\rm diss})$ with $\tau_{\rm diss}=0.01$, $0.1$, and
  $1.0$\,s for $\sigma_{\rm c}=10^7$, $10^8$, and $10^9$\,${\rm
    s}^{-1}$, which are consistent with Eq.~(\ref{taudis}). The result
  by the ideal MHD implementation is also shown by the dotted curve, which
  agrees approximately with the curve for $\sigma_{\rm c}=10^{11}\,{\rm s}^{-1}$. 
\label{figa7}}
\end{figure}

In this section, we present the results for the resistive MHD evolution
of the merger remnant neutron star with purely toroidal magnetic fields
and demonstrate that our implementation can follow the resistive
dissipation of the toroidal magnetic field successfully.

Figure~\ref{figa7} plots the evolution of the electromagnetic energy
for the remnant massive neutron star, for which a purely toroidal
magnetic field of Eq.~(\ref{initoro}) is superimposed at $t=0$, in the
axisymmetric resistive MHD simulations with $\sigma_{\rm c}=10^7$,
$10^8$, $10^9$, and $10^{11}\,{\rm s}^{-1}$ (solid curves).  The
result of an ideal MHD simulation is shown together (dashed curve).
Since the toroidal magnetic field is simply superimposed and thus the
initial condition is not exactly in an equilibrium state, the
electromagnetic energy initially decreases by 10--20\% even in the
absence of the resistivity during the early relaxation stage. The
subsequent long-term gradual decrease of the electromagnetic energy
for the ideal MHD case would be partly due to the numerical
dissipation or diffusion.\footnote{In these simulations the neutrino
  cooling is incorporated and the density and pressure profiles are
  modified during the simulation. This effect also affects the
  evolution of the magnetic field configuration. } Note again that
this evolution process is valid only in the assumption of the axial
symmetry. In general cases, non-axisymmetric instability such as
Parker and Taylor instability~\cite{Parker,Taylor} could occur and the
evolution process could be significantly modified.

For the resistive MHD case, the magnetic field decreases exponentially
with time after the early relaxation stage.
The short-dashed
slopes denote the relation of $\propto \exp(-t/\tau_{\rm diss})$ where
$\tau_{\rm diss}$ is the decay timescale, and are plotted for
approximately fitting the curves for the resistive MHD results. It is
found that the dissipation timescale of the magnetic field is $\approx
0.01$, $0.1$, and $1.0$\,s for $\sigma_{\rm c}=10^7$, $10^8$, and
$10^9\,{\rm s}^{-1}$, which are consistent with the timescales
estimated by Eq.~(\ref{taudis}).  For $\sigma_{\rm c}=10^{11}\,{\rm
  s}^{-1}$, the dissipation timescale would be $\sim 10^2$\,s, and
much longer than the simulation time.  Therefore, the result for this
case agrees approximately with that in the ideal MHD simulation: Our
ideal and resistive MHD implementations can derive approximately the
identical results.

\begin{figure}[t]
\vspace{-2mm}
\includegraphics[width=88mm]{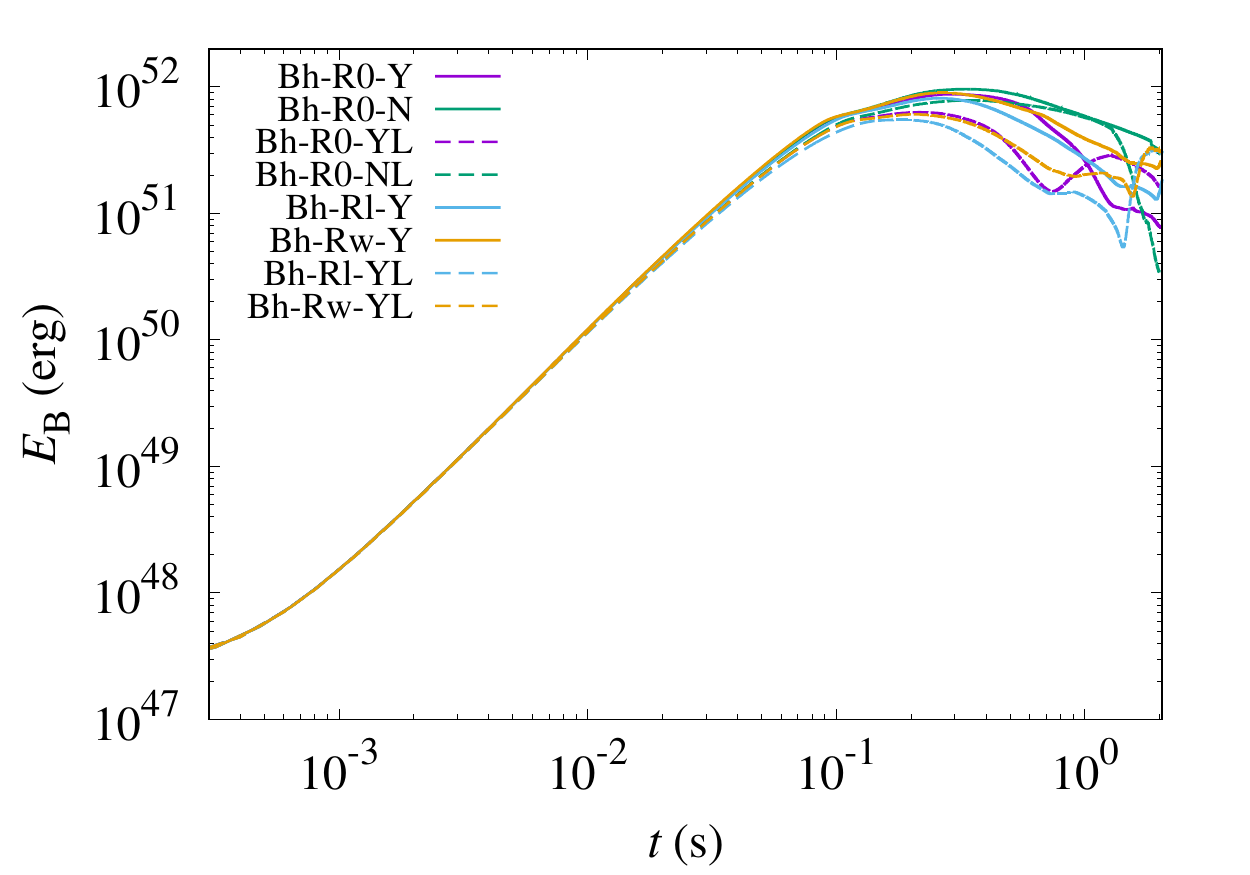} 
\caption{Evolution of the electromagnetic energy for two ideal MHD models
  Bh-R0-Y, Bh-R0-N and for two resistive MHD models Bh-Rw-Y and Bh-Rl-Y
  with two different grid resolutions. The solid and dashed curves
  show the results for $\varDelta x=200$\,m and $\varDelta x=250$\,m, respectively. 
\label{figapp2}}
\end{figure}

\section{Grid resolution dependence}\label{app3}

As a convergence test, we performed numerical simulations with two
lower grid resolutions for two ideal MHD (Bh-R0-Y and Bh-R0-N) and two
resistive MHD (Bh-Rw-Y and Bh-Rl-Y) cases. Figure~\ref{figapp2} is the
same as Fig.~\ref{fig1} but for the comparison in the evolution of the
electromagnetic energy for two different resolutions.  The solid and
dashed curves show the results for the cases that the neutron star is
resolved with the grid spacing of $\varDelta x=200$\,m and $\varDelta
x=250$\,m, respectively.

As found from Fig.~\ref{figapp2}, the growth of the magnetic-field
strength by the winding effect is followed with the lower grid
resolution fairly well.  However, the amplification of the
magnetic-field strength is suppressed for the lower resolution, and
thus, the maximum magnetic-field energy becomes smaller.  This effect
is not appreciable for the case that the effects of the neutrino
irradiation/heating and pair-annihilation are switched off.  However,
in the presence of these neutrino effects, the suppression is quite
large: The maximum magnetic-field strength for the lower resolution is
by a factor of $\sim 2$ smaller than for the higher one. It appears
that the mass ejection associated with these neutrino effects enhance
the magnetic-field dissipation or magnetic flux outflow from the
neutron star. Thus the poor convergence is likely to be
associated with a poor convergence in the neutrino transfer
(cf. Ref.~\cite{Fujiba17}). 

For the resistive MHD simulations, we also find the tendency similar
to that found for the ideal MHD case with neutrino effects. That is,
the dissipation or magnetic flux escape from the neutron star is
spuriously enhanced for the lower grid resolution. However, overall,
the evolution process of the neutron star is not modified
qualitatively by the grid resolution.


\end{document}